\documentclass[%
aip,
amsmath,amssymb,
reprint,%
]{revtex4-1}
\usepackage{braket}
\usepackage{graphicx}
\usepackage{dcolumn}
\usepackage{bm}

\usepackage[utf8]{inputenc}
\usepackage[T1]{fontenc}
\usepackage{mathptmx}
\usepackage{etoolbox}

\makeatletter
\def\@email#1#2{%
	\endgroup
	\patchcmd{\titleblock@produce}
	{\frontmatter@RRAPformat}
	{\frontmatter@RRAPformat{\produce@RRAP{*#1\href{mailto:#2}{#2}}}\frontmatter@RRAPformat}
	{}{}
}%
\makeatother

\begin{document}
	
	\preprint{AIP/123-QED}
	
	\title[Leveraging Off-the-Shelf Silicon Chips for Quantum Computing]{Leveraging Off-the-Shelf Silicon Chips for Quantum Computing}
	\author{J. Michniewicz}
	\affiliation{Blackett Laboratory, Imperial College London, SW7 2AZ, London, UK}
	\author{M.S. Kim}%
	\email{j.michniewicz23@imperial.ac.uk}
	\affiliation{Blackett Laboratory, Imperial College London, SW7 2AZ, London, UK}%

	\date{\today}
	
	\begin{abstract}
		There is a growing demand for quantum computing across various sectors, including finance, materials and studying chemical reactions. A promising implementation involves semiconductor qubits utilizing quantum dots within transistors. While academic research labs currently produce their own devices, scaling this process is challenging, requires expertise, and results in devices of varying quality. Some initiatives are exploring the use of commercial transistors, offering scalability, improved quality, affordability, and accessibility for researchers. This paper delves into potential realizations and the feasibility of employing off-the-shelf commercial devices for qubits. It addresses challenges such as noise, coherence, limited customizability in large industrial fabs, and scalability issues. The exploration includes discussions on potential manufacturing approaches for early versions of small qubit chips. The use of state-of-the-art transistors as hosts for quantum dots, incorporating readout techniques based on charge sensing or reflectometry, and methods like electron shuttling for qubit connectivity are examined. Additionally, more advanced designs, including 2D arrays and crossbar or DRAM-like access arrays, are considered for the path toward accessible quantum computing.
	\end{abstract}
	
	\maketitle
	
	
	
	\section{\label{sec1} Introduction}
	Quantum computing's potential in computation and simulations attracts significant investment \cite{intro1, intro2}, but current implementations, such as trapped ions and superconducting circuits, demand specialized hardware and expertise, limiting accessibility. Cloud-based services like IBM Quantum and Azure Quantum provide remote access but are constrained by hardware limitations, high costs, and a steep learning curve \cite{cloud_critism_1, cloud_criticism_2}. Leveraging mass-produced commercial transistors in silicon semiconductors offers cost-effective and accessible alternatives of superior quality \cite{intel_paper}, enabling easier experimentation with various qubit technologies. This study explores the feasibility and potential impact of this approach.
	
	\section{Transistors as hosts of qubits}\label{sec2}
\begin{figure}
		\centering
		\includegraphics[scale=0.16]{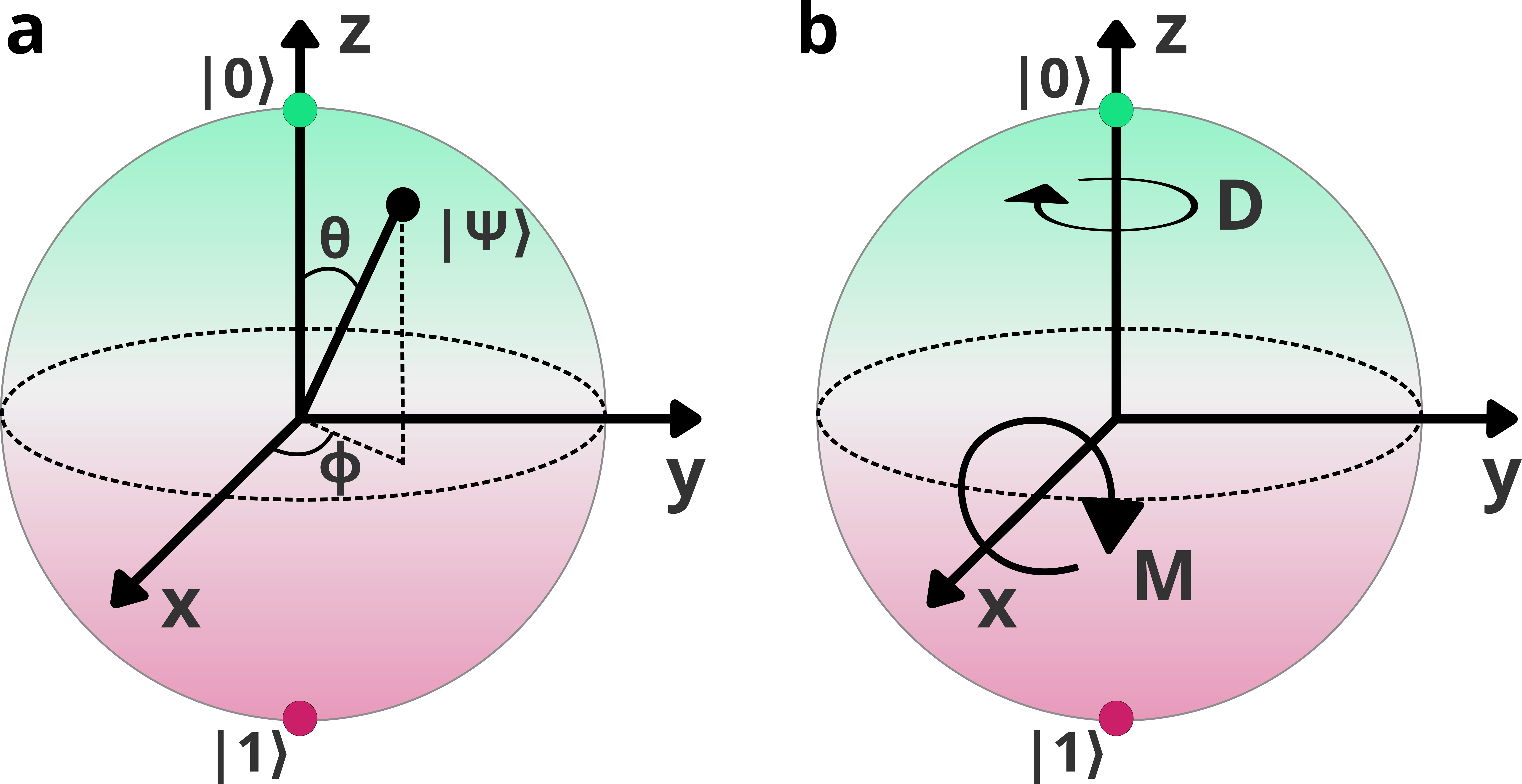}
		\caption{Bloch sphere schematic representation:\\
			\textbf{a} Pure states $\ket{0}$ and $\ket{1}$ are located at the poles. A general state $\ket{\Psi}$ is represented as a vector at an angle $\theta$ with the z-axis and $\phi$ with the x-axis.\\
			\textbf{b} Rotations about the axes are induced by the diagonal (D) and off-diagonal/mixing (M) terms of the Hamiltonian Eq. \ref{Hamiltonian}}
		\label{bloch}
	\end{figure}

	Qubits, representing two-level quantum systems, exhibit logical states $\ket{0}$ and $\ket{1}$, with their physical representation varying based on the implementation. Key qubit operations include initialization, manipulation (using the universal set of single qubit and two qubit controlled gates), and readout. The Bloch sphere abstractly represents states $\ket{0}$ and $\ket{1}$ at the poles along the z-axis, with a general state being:
	
	\begin{equation}
		\ket{\Psi} = \mathrm{cos \frac{\theta}{2}} \ket{0} + \mathrm{e^{i\phi}}\mathrm{sin \frac{\theta}{2}} \ket{1}
	\end{equation}
	
	\noindent
	where $\mathrm{\theta}$ and $\mathrm{\phi}$ can be thought of us the azimuthal and polar angles on the Bloch sphere respectively (Fig. \ref{bloch}).

	\noindent
	To construct the general Hamiltonian $\mathcal{H}$, $\ket{0}$ is represented $\big(\begin{smallmatrix}
		1\\
		0
	\end{smallmatrix}\big)$  and $\ket{1}$ as $\big(\begin{smallmatrix}
		0\\
		1
	\end{smallmatrix}\big)$, with energies assigned as D and $-$D respectively.
	Supplying energy 2D induces transitions between states. For a $\ket{0} - \ket{1}$ superposition, mixing terms M are added to the Hamiltonian.
	\begin{equation}
		{\mathcal{H}}=
		\begin{pmatrix}
			D & M \\
			M & -D
		\end{pmatrix}
		\label{Hamiltonian}
	\end{equation}\\
	\noindent
	Solving for eigenvalues and eigenstates gives:
	
	\begin{equation}
		E_{\pm} = \pm\sqrt{D^2+M^2}, \,\,\,\,\,\,\,\,\,\,\,\,\, \Psi{\pm}= \begin{pmatrix}
			(D  + E_{\pm})/M \\
			1
		\end{pmatrix}
		\label{hamiltonian}
	\end{equation}
	
	\noindent
	Effective qubit manipulation occurs when one term significantly dominates. In the limit of D/M $\to$ 0, $\ket{0}$ and $\ket{1}$ become energetically degenerate, facilitating rotation about the x-axis of the Bloch sphere, as in Fig. \ref{bloch}b. Similarly, in the limit of M/D $\to$ 0, ($\ket{0} + \ket{1}$)/$\sqrt{2}$ and ($\ket{0} - \ket{1}$)/$\sqrt{2}$ undergo mixing, resulting in rotation about the z-axis.

	\subsection{Quantum dots and charge carriers}
	Semiconductor qubits utilize charge carriers like electrons or holes localized within quantum dots. These single-charge spins are typically confined either at the semiconductor–dielectric interface in metal-oxide-semiconductor (Si-MOS) stacks or in heterostructures, typically SiGe, residing in strained quantum wells buried beneath the epitaxial Si/SiGe interface. Electrons have been extensively studied, while holes in Silicon CMOS devices were initally demonstrated in 2016 \cite{silvano_1}. Challenges remain in achieving single-hole functionality in silicon quantum dots for spin-qubit applications. Issues with spin properties, spin-orbit interaction, technological implementation, and noise susceptibility need careful consideration for effective use of holes in quantum computing \cite{silvano_2, hole_qubits_review}. Silicon transistors control charge flow through gates, forming quantum dots (QDs) due to electrostatic repulsion. Adding charges occurs at specific gate voltages (V$_G$) when the charging energy (E$_\mathrm{c}$) barrier is overcome, leading to periodic conductance peaks \cite{wiel}. Quantum dots' energy levels are quantized by gate voltage, causing charge localization when tunneling energy uncertainty is smaller than E$_\mathrm{c}$. Temperatures at 1K or lower are crucial to minimize energy contributions from thermal excitations.
	
	\subsection{Qubit realizations}
	This section describes key semiconductor spin-qubit implementations \cite{3electron, big_review, qds_spinqubits}.

	\subsubsection{Single spin qubit}
	The simplest qubit uses a single electron with up and down spin orientations. The spins precess at the Larmor frequency $\omega_{L}$ determined by the g factor. A static magnetic field B$_z$ induces an energy difference, favoring the $\ket{\downarrow}$ state. This gives rise to the diagonal term D in Eq. \ref{Hamiltonian}. That D is $\hbar\omega_{L} = g\mu_BB_z$ with Bohr magneton $\mu_B$. The mixing term M comes from applying an oscillating pulse B$_x$ in a perpendicular plane usually sent using micro-strips or antennas near the quantum dot.
	Electron qubits using ESR need antennas/magnets. Holes can be electrically controlled via EDSR due to strong spin-orbit coupling.
	The simplest qubit uses a single charge with up and down spin orientations. The spins precess at the Larmor frequency $\omega_{L}$ determined by the g factor. A static magnetic field B$_z$ induces an energy difference, favoring the $\ket{\downarrow}$ state. This is the diagonal term D in Eq. \ref{Hamiltonian}. That D is $\hbar\omega_{L} = g\mu_BB_z$ with Bohr magneton $\mu_B$. The mixing term M comes from an oscillating B$_x$.
	
	\subsubsection{Singlet-triplet \& flip-flop qubits} 
	The singlet-triplet (or flip-flop) qubit uses a two-dot system with configurations (1,1) and (0,2), representing charges in the left and right dots. Detuning $\epsilon$ (V$_\mathrm{G}$), a function of gate voltage (V$_\mathrm{G}$), reflects the energy difference between dots. The anti-symmetric singlet state is allowed in both configurations, while the symmetric triplet state is disallowed in the (0,2) ground state due to Pauli exclusion \cite{qds_spinqubits}.
	The energy difference J($\epsilon$) between singlet and triplet states increases during the transition from (1,1) to (0,2), with D = J($\epsilon$)/2. An external magnetic field lifts triplet state degeneracy by Zeeman energy. Mixing is induced by a small gradient in the effective magnetic field between dots ($\Delta B_z$), achieved through coupling to bulk nuclear moments or micromagnets.

	\subsubsection{Exchange only (EO) qubits} 
	Extending to three charges requires three gate voltages, with detuning parameters $\epsilon$ and $\epsilon_m$. The system offers various arrangements, quadruplet and two degenerate doublet states that form logical states are commonly chosen, Possible realization are a 'subsystem' with an external magnetic field or a 'subspace' without one. In both cases, sum and difference of the two exchange couplings, J$_{12}$($\epsilon$, $\epsilon_m$) and J$_{23}$($\epsilon$, $\epsilon_m$), contribute to the Hamiltonian's diagonal and mixing terms. Three-charge systems offer diverse realizations \cite{3electron, big_review}, relying on all-electrical control for rotations on the Bloch sphere.
	\noindent
	\subsection{Qubit operation}
	During initialization, the gate voltage is adjusted to position $\ket{0}$ below the energy of ohmic contacts, enabling selective tunneling of only that state. Readout is performed analogusly. For manipulation, different qubit types utilize distinct techniques. Single-spin qubits transition between $\ket{\uparrow}$ and $\ket{\downarrow}$ via effective B$_x$. For electrons, B$_x$ is applied using micro-strips or antennas, facilitating Electron Spin Resonance (ESR). Holes, however, do not require such pulses; instead, spin operations are achieved by modulating gate voltages, which change the confinement potential and create time-dependent g-factors via Electric-Dipole Spin Resonance (EDSR), effectively generating a fluctuating magnetic field \cite{silvano_4}. Singlet-triplet and EO qubits achieve mixing in various charge configurations via J$_{12}$ and J$_{23}$ interactions.

	\begin{figure*}
		\centering
		\includegraphics[scale=0.16]{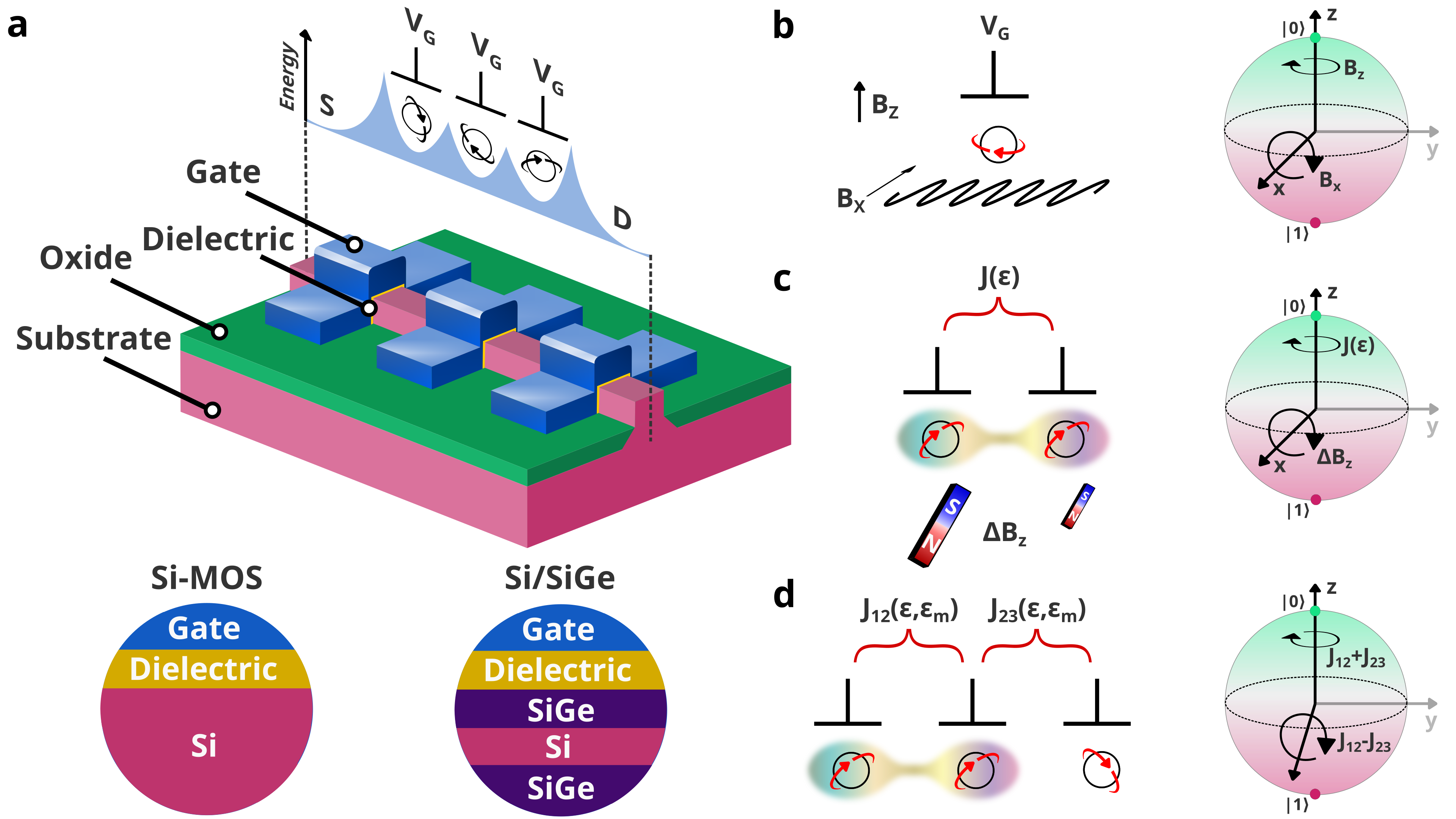}
		\caption{Qubit implementations in semiconductors:\\
			\textbf{a} Schematics of a transistor used for QD spin qubits, with additional gates and components omitted for clarity. Gate voltage V$_\mathrm{G}$ confines charges, as depicted on the energy diagram, where S is source and D is drain. Typical gate stack materials in heterostructures (SiGe) and Si-MOS shown below.\\
			\textbf{b} Single spin qubit. A global magnetic field B$_z$ favors one spin orientation. An oscillating B$_x$ transitions the charge between $\ket{\uparrow}$ and $\ket{\downarrow}$ \cite{pulse_eng}. Control is visualized on the Bloch sphere.\\
			\textbf{c} Singlet-triplet qubit. The color gradient connecting two trapped charges symbolizes entanglement in singlet/triplet states. Transitioning deep into the (1,1) charge configuration induces state mixing. Moving to a non-zero J($\epsilon$) region enables rotation on the Bloch sphere about the z-axis by an angle $\hbar \phi$ = J($\epsilon$)t$_e$, with t$_e$ as the time spent in that region. Rotation along the x-axis is induced by a slight gradient in the effective magnetic field between dots.\\
			\textbf{d} EO qubit. As with singlet-triplet implementation, the system can be maneuvered into configurations of J$_{12}$($\epsilon$, $\epsilon_m$) and J$_{23}$($\epsilon$, $\epsilon_m$) where $\ket{0}$ and $\ket{1}$ are degenerate, and by transitioning to non-degenerate regions, superposition states are mixed. Despite the 120-degree difference in rotation axes on the Bloch sphere, pulse sequences as detailed in \cite{120, 3electron} enable rotations necessary for single-qubit gates.}
		\label{qubits}
	\end{figure*}
	
	\section{Feasibility and challenges}\label{sec3}
	Commercial devices are efficient in mass production and precision but may lack the specificity required for quantum computing. The next section explores using commercial devices in quantum computing.
	
	\subsection{Fabrication considerations} 
	Large fabs handle essential qubit fabrication parameters well, like feature size, spacing rules and edge roughness, but some components and methods used may differ from academic approaches.
	
	\subsubsection{Dimensions}
	\noindent
	Practical qubit operation demands Plunger Gates (PGs) tens of nanometers long, with pitches below 100nm, and filled with Barrier Gates (BGs) of similar pitch to ensure large orbital energy spacing compared to thermal energy \cite{lab_to_industry_qubits}. Intel met these requirements with their 50nm finFET technology \cite{intel_50nm}, and CEA-Leti managed an 80nm pitch, with effective PG-BG pitch of 40nm, using DUV lithography \cite{leti_80nm}. TSMC announced 45nm minimum feature size \cite{tsmc_sota_pitch}, nearing the industry’s physical limit projected within a decade \cite{src_decadal_report}. Smaller feature sizes demand higher alignment precision, improving qubit performance. \cite{tsmc_sota_pitch}. Decreased distance between dots increases tunnel coupling, especially beneficial in SiGe systems \cite{intel_50nm}. Imec achieved a record 40nm separation in a single split gate with a spacer, albeit using e-beam lithography \cite{lab_to_industry_qubits}. Small quantum dots and pitches confine charges to smaller, less rough regions. Holes, requiring larger quantum dots, ease fabrication constraints but may encounter more disorder.

	\subsubsection{Lithography}
	\noindent
	In academia, e-beam lithography or lift-off techniques are preferred for their flexibility and affordability. Optical lithography, including EUV and DUV, is the industry standard \cite{nist_sota_euv} for its small features, high resolution, and speed, but faces hurdles in widespread adoption due to cost and complexity \cite{nist_sota_euv}. While 193nm DUV lithography is cheaper than EUV, its longer wavelength leads to larger feature sizes, partly mitigated by multi-patterning \cite{duv}. Nonetheless, DUV remains widely used \cite{duv_news}, offering a more cost-effective option for qubit device fabrication. Directed self-assembly (DSA) with EUV lithography can reduce roughness and disorder, though integration challenges may arise \cite{dsa_euv}. E-beam lithography has comparable resolution to EUV and significantly lower cost \cite{homemade_ebeam}, can be CMOS compatible \cite{ebeam_cmos_compatible}, and enables innovative applications like e-beam-enabled deposition for nanomagnets \cite{magnet_making_techniques} and EO qubits in Si-Ge \cite{fc_lc3}. However, its slower speed, deeper-penetrating localized electron interaction and scattering may favor optical lithography \cite{intel_paper, euv_for_si_qubits, ebeam_limitations, ebeam_optimization_challenges}.

	\subsubsection{Additional components}
	\noindent
	Electron qubits using ESR need antennas/magnets. Holes can be electrically controlled via EDSR due to strong spin-orbit coupling.  Intel demonstrated ESR with a copper coplanar stripline with a CMOS-compatible fabrication \cite{intel_paper}. QuTech fabricated antennas from Al or NbTin \cite{si_sige_array} and magnets from Cobalt in an adjacent, though not fully CMOS-compatible process \cite{si_6qubits}. Although large fabs have the precision to deposit cobalt layers and techniques for creating nanomagnets exist \cite{magnet_making_techniques}, investing in such niche applications may not be economically viable for them. Additionally, integrating magnets requires precise tuning and simulation, and surface oxide can unpredictably interfere with magnet properties, complicating the process.

	\subsection{Decoherence and noise}	
	Disorder affects qubit stability and coherence. Dynamic disorder arises from environmental fluctuations, leading to charge and spin noise \cite{fc_cn1}. Static disorder from imperfections, irregularities, and interface roughness makes control of many qubits harder.

	\subsubsection{Structural choice}
	\noindent 
	While silicon-28 offers lower spin noise, certain growth techniques for Si-28 may not align with industrial processes \cite{material_integration_challenges_si_qubits}, leading to the use of natural silicon with multiple isotopes. 
	
	\noindent
	Intrinsic semiconductor characteristics contribute to complex charge noise \cite{fc_cn2}, with fluctuating electric fields resulting from charge defects in semiconductor locations such as the quantum well, barrier, interface, and dielectric layers \cite{fc_cn3}. 
	
	\noindent
	Silicon crystals have six degenerate minima in the conduction band \cite{fon, sqe}, which quantum confinement and external factors like strain can lift. While generally viewed as a drawback \cite{ch2_valleys_bad}, these valleys can be advantageous in certain qubit systems, especially regarding spin and valley coupling \cite{ch2_valley_qubit1, ch2_valley_qubit2, ch2_valley_qubit3}.
	
	\noindent
	Si-MOS is widely available and aligns with industrial standards, unlike Si/Ge, which, although not incompatible, is less common. Si-MOS structures offer larger valley splitting but face higher charge noise from the dielectric interface (optimizing gate stack can help \cite{asser}). Interface roughness is a  disorder bottleneck in Si/SiO2 \cite{si_variability}. Si/Ge structures have reduced disorder and charge noise but may exhibit more device-to-device non-uniformity due to strain and compositional fluctuations \cite{modeling_sige_qd_variability}, and have limited and variable valley splitting \cite{modeling_sige_qd_variability, material_challenges_for_qc} (both can be partially mitigated \cite{sige_disorder_valley_mitigation}). Additionally, Si-MOS structures have lower capacitive cross-talk between QDs \cite{si_sige_array}.
	
	\noindent
	In EO qubits, noise reduction is possible through symmetric operations and operating at the sweet spot \cite{sweet_eo}, a strategy that can also minimize charge sensitivity in holes, approaching electron-like coherence levels \cite{silvano_6}. However, implementing reliable qubit gates remains a challenge \cite{material_challenges_for_qc}.

	\subsubsection{Geometry and materials}
	\noindent		
	FinFETs introduce structural confinement, improving qubit control, but non-planar structures can increase disorder due to rougher interfaces and defects \cite{lab_to_industry_qubits}. Nevertheless, FinFETs, and Gate-All-Around Field-Effect Transistors (GAA-FETs), are preferred for their smaller dimensions, lower leakage, and reduced power consumption, which minimizes thermal noise and enhances cooling efficiency, potentially benefitting qubit technology.
	Gate material choice greatly affects disorder \cite{gate_stack_disorder}. Polysilicon and SiO2, once common, are replaced by metals like TiN, Al, Cu, and ruthenium (used by Intel, IBM, Samsung) in gate stacks \cite{no_more_poly, ruthenium, gate_stack_sota_materials}.
	Polysilicon offers higher mobility, potentially reducing disorder, while TiN's superconductivity and low resistance at low temperatures is adavantegous for high frequency operation \cite{gate_stack_disorder}. TiN disorder results from oxygen scavenging during deposition, when combining with SiO2 \cite{gate_stack_disorder}. Despite SiO2's reduced use, alternatives like HfO2 also introduce defects and oxygen vacancies \cite{no_more_poly, dielectric_and_metal_materials}.
	
	\noindent
	Cobalt magnets require precise deposition as thin films, often resulting in low yield. Even successful deposition can introduce roughness and contamination, possibly exceeding cleanroom standards. Patterning thin films for single-electron/atom control is challenging, and creating small magnets for high field generation may be infeasible \cite{qutech_whitepaper}. Optimizing parameters like shape and distance to quantum dots is crucial for minimizing dephasing \cite{optimizing_magnets}.
	\newline
	
	\begin{figure*}
		\centering
		\includegraphics[scale=0.16]{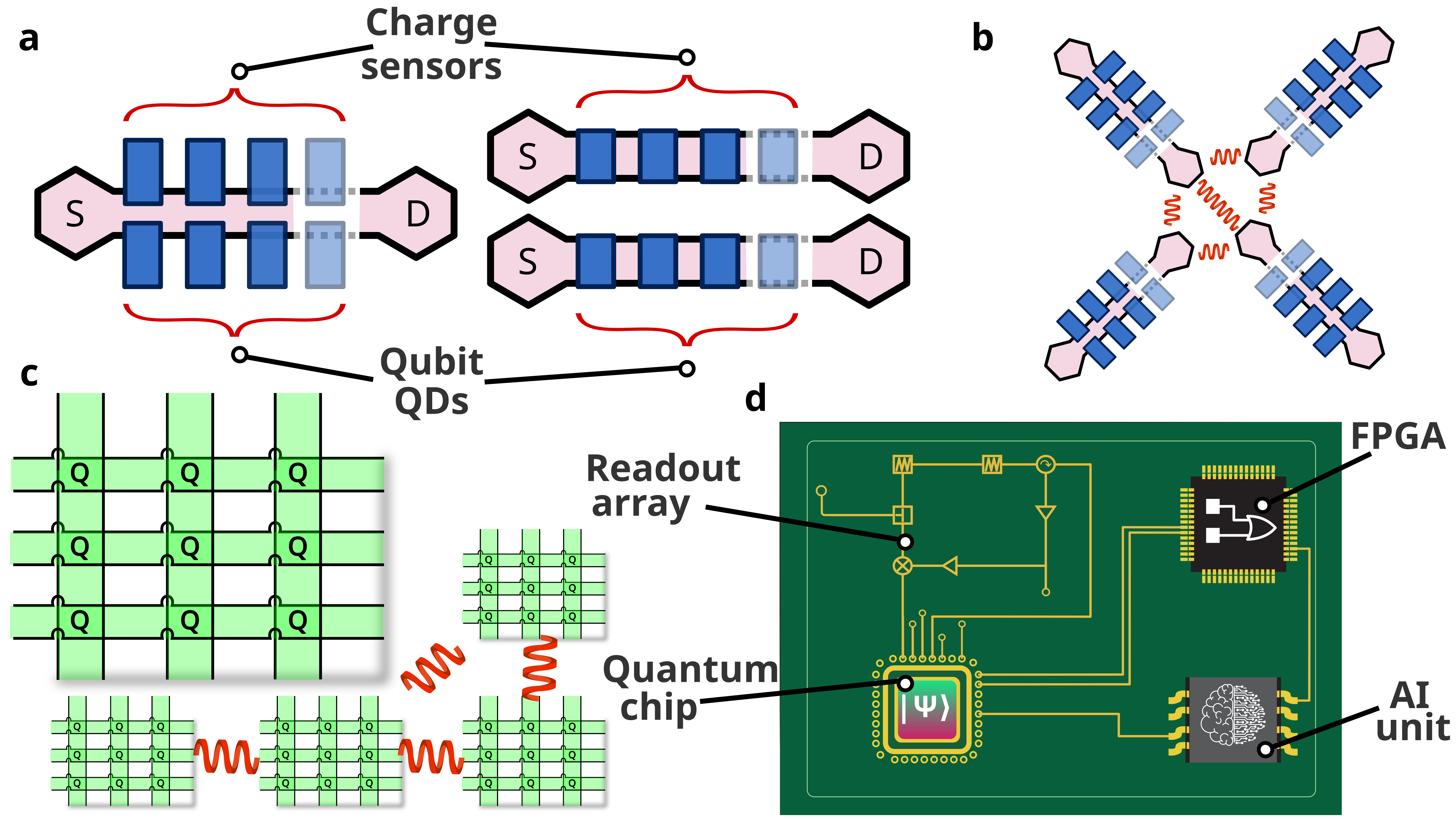}
		\caption{Potential designs of qubit systems using off-the-shelf semiconductors:\\
			\textbf{a} Two transistor designs: split-gates on the left, regular gates on the right. Bottom row hosts qubit QDs, top repurposes QDs as charge sensors. Designs extend indefinitely, one sensor can monitor multiple QD transitions.\\
			\textbf{b} Either design can be combined with other transistors. Coupling, depicted by a red line, can occur through electron shuttling, photon mediation, or other methods, offering various interconnectivity modes, although not with all qubits directly connected. The example shown represents just one of many possible layouts.\\
			\textbf{c} A 2D qubit array, with DRAM-like or staircase layout employs line and row control at each intersection with a qubit (Q), reducing wire count. Visualizations on the right illustrate interconnected smaller arrays with varying connection complexities.\\
			\textbf{d} An integrated wafer minimizes external communication needs by incorporating a readout array, FPGA/ASIC, and AI unit for error correction. Modular setup allows integration of a quantum chip, facilitating experimentation and debugging.}
		\label{fc}
	\end{figure*}

	\subsection{Scalability}
	As scale increases, fabricating large qubit arrays with precision is crucial for efficiently addressing and controlling individual qubits.

	\subsubsection{Readout}
	\noindent
	Readout options include using a Quantum Point Contact (QPC) or a Single-Electron Transistor (SET) \cite{fc_op_guide_simos_qubits}.
	
	\noindent
	Dispersive readout allows for faster readout times in both SiGe \cite{sige_dispersive_readout} and Si-MOS \cite{fc_s13, mine_many}, with Si-MOS demonstrating superior reflectometry capabilities thanks to its significantly higher lever-arm \cite{dispersive_lever-arm_theory, dispersive_platform_comparison}, giving it a notable advantage in load-aware comparisons \cite{dispersive_platform_comparison}.
	
	\noindent
	Industrial foundries have faced challenges in placing SETs close to QDs, but recent advancements have resolved this in both dual and single-channel layouts \cite{lab_to_industry_qubits}. QDs themselves can serve as charge sensors, especially when using dispersive readout \cite{silvano_5}, showcased by Intel's dual-channel system \cite{fc_lc4}. Flexible designs like the ones depicted in Fig. \ref{fc}a allow for row selection for sensing and qubit quantum dots, offering advantages such as the ability to swap rows in case of misalignment, resulting in a significantly higher signal-to-noise ratio when using reflectometry \cite{lab_to_industry_qubits}. Also, multiple transistors can be read by a single sensor without assigning one sensing per qubit dot \cite{si_6qubits}.

	\subsubsection{Addresability and power management}
	\noindent
	Maintaining millikelvin temperatures minimizes thermal noise, and various devices and components are operational at these temperatures \cite{ch6_lr_cryodem, ch6_lr_cryodem2, cryo2}. However, the standard one-input-per-qubit control method will overburden dilution refrigerators due to the rising number of wires. Alternative solutions like crossbar-like DRAM arrays \cite{mine_ieee,mine_ieee2, my_nat_true} and staircase configurations with integrated crossbar structures \cite{fc_s12} offer potential to reduce required input lines for qubit control. Coupling with superconductors from another plane of the dilution refrigerator is also worth exploring \cite{cryo1,cryo2}.

	\noindent
	Commercial devices may face addresability issues due to their uniform design. Implementations that require precise control over spins rely on small differences in g factors for individual qubit addressing \cite{gfactor}. Intentionally introducing these defects is an option \cite{fc_lc2}, but it is not feasible in industrial settings.

	\subsubsection{Interconnectivity}
	\noindent
	Interconnectivity solutions involve tunneling and charge swapping between dots, aided by additional tunneling gates to control charge flow direction. Proposed methods include adjusting gate voltages with pulses in linear arrays \cite{fc_s1}, T-shaped linear arrays, and pumping \cite{fc_s2}. Another approach involves conveyor belt-like shuttling using industry-standard methods, which works consistently across varying channel lengths and is easy to tune \cite{fc_s3}.
	\noindent
	Transitioning from linear transistors to more future-proof 2D designs is feasible for industrial processes but challenging. Efficient error correction and mitigation are also favored in this geometry \cite{2d_better1, 2d_better2}, and even 3D logical layouts could be achieved through shuttling \cite{fc_s4}. Implementing 2D structures often requires 3D manufacturing of layered lines for control \cite{fc_s5}, a complex task with current technologies. Moreover, using silicon-on-insulator (SOI) \cite{material_integration_challenges_si_qubits} or integrating magnets into 2D arrays may be nontrivial \cite{optimizing_magnets}.
	\noindent	
	Demonstrations of 2D quantum dot arrays include a split-gate device with two arrays of dots \cite{fc_s6}, and another, somewhat limited, approach towards a 2D arrangement \cite{fc_s7}. Further efforts involve a 2D design with four qubits in Si-Ge \cite{fc_s8}.
	\noindent	
	To simplify creating large 2D arrays, smaller islands can be interconnected over distance using shuttling \cite{fc_s9}, microwaves \cite{fc_s10}, or resonant coupling \cite{fc_s11}. See Fig. \ref{fc}c for illustrations. Interconnected components can show significant improvement compared to monolithic designs \cite{chiplets, fc_s10}.

	\subsubsection{Integration and modularity}
	\noindent
	Modular designs facilitate scalability and enhance experimentation and adaptibility across research groups. Integration with various CMOS components is feasible \cite{all_stack1, all_stack2, nawa_myphd, nawa_fpga, nawa_fullstack}, as shown in Fig. \ref{fc}d. Additionally, bringing components close together enables faster communication with the quantum chip, can ease wiring and power management and enhance sensitivity, for instance while incorporating resonators on the same wafer for dispersive readout \cite{dispersive_platform_comparison}.
	
	\noindent
	Determining crucial parameters for optimal operation is time-consuming and challenging. As the scale increases, implementing automatic tuning methods becomes crucial, significantly simplifying the task \cite{nawa_ares_ml, zwolak}.
	
	\section{Conclusions}\label{sec5}
	Large fabs offer precise fabrication and integration with CMOS components. However, certain components like micromagnets may have limited viability in industrial settings. All-electrical qubits, such as EO electron qubits (yet to be demonstrated in CMOS \cite{spinqubit_database}) or holes, may align better with industrial processes. The feasibility of using industrial devices for quantum computing depends on big fabs' interests in materials and technologies, as they impact disorder and noise levels significantly. Nonetheless, there is potential for implementing certain qubit types even with current processes.

	\section*{Acknowledgments}
	We thank the UK EPSRC (EP/Y004752/1, EP/W032643/1) and the Samsung GRC grant. We acknowledge input from Prof. Silvano de Franceschi during discussions.

	\section*{Author declarations}
	\subsection*{Conflict of interest}
	\noindent
	The authors have no conflicts to disclose.
	\subsection*{Author contributions}
	\noindent
	\textbf{John Michniewicz}: Conceptualization; Writing; Review \& Editing. \\
	\noindent
	\textbf{Myungshik Kim}: Conceptualization (supporting); Review (supporting); Project administration; Supervision.
	
	\section*{Data Availability Statement}
	Data sharing is not applicable to this article as no new data were created or analyzed in this study.

	\bibliography{manuscript_bibliography}

\begin{thebibliography}{95}%
\makeatletter
\providecommand \@ifxundefined [1]{%
 \@ifx{#1\undefined}
}%
\providecommand \@ifnum [1]{%
 \ifnum #1\expandafter \@firstoftwo
 \else \expandafter \@secondoftwo
 \fi
}%
\providecommand \@ifx [1]{%
 \ifx #1\expandafter \@firstoftwo
 \else \expandafter \@secondoftwo
 \fi
}%
\providecommand \natexlab [1]{#1}%
\providecommand \enquote  [1]{``#1''}%
\providecommand \bibnamefont  [1]{#1}%
\providecommand \bibfnamefont [1]{#1}%
\providecommand \citenamefont [1]{#1}%
\providecommand \href@noop [0]{\@secondoftwo}%
\providecommand \href [0]{\begingroup \@sanitize@url \@href}%
\providecommand \@href[1]{\@@startlink{#1}\@@href}%
\providecommand \@@href[1]{\endgroup#1\@@endlink}%
\providecommand \@sanitize@url [0]{\catcode `\\12\catcode `\$12\catcode
  `\&12\catcode `\#12\catcode `\^12\catcode `\_12\catcode `\%12\relax}%
\providecommand \@@startlink[1]{}%
\providecommand \@@endlink[0]{}%
\providecommand \url  [0]{\begingroup\@sanitize@url \@url }%
\providecommand \@url [1]{\endgroup\@href {#1}{\urlprefix }}%
\providecommand \urlprefix  [0]{URL }%
\providecommand \Eprint [0]{\href }%
\providecommand \doibase [0]{http://dx.doi.org/}%
\providecommand \selectlanguage [0]{\@gobble}%
\providecommand \bibinfo  [0]{\@secondoftwo}%
\providecommand \bibfield  [0]{\@secondoftwo}%
\providecommand \translation [1]{[#1]}%
\providecommand \BibitemOpen [0]{}%
\providecommand \bibitemStop [0]{}%
\providecommand \bibitemNoStop [0]{.\EOS\space}%
\providecommand \EOS [0]{\spacefactor3000\relax}%
\providecommand \BibitemShut  [1]{\csname bibitem#1\endcsname}%
\let\auto@bib@innerbib\@empty
\bibitem [{\citenamefont {Cartlidge}(2023)}]{intro1}%
  \BibitemOpen
  \bibfield  {author} {\bibinfo {author} {\bibfnamefont {E.}~\bibnamefont
  {Cartlidge}},\ }\bibfield  {title} {\enquote {\bibinfo {title}
  {Commercializing quantum computers step by step},}\ }\href
  {https://www.nature.com/articles/d41586-023-01644-3} {\bibfield  {journal}
  {\bibinfo  {journal} {Nature}\ } (\bibinfo {year} {2023})}\BibitemShut
  {NoStop}%
\bibitem [{int(2023)}]{intro2}%
  \BibitemOpen
  \href
  {https://www.mckinsey.com/capabilities/mckinsey-digital/our-insights/quantum-technology-sees-record-investments-progress-on-talent-gap}
  {\enquote {\bibinfo {title} {Record investments in quantum technology
  {\textbar} {McKinsey}},}\ } (\bibinfo {year} {2023})\BibitemShut {NoStop}%
\bibitem [{\citenamefont {Zwerver}\ \emph {et~al.}(2022)\citenamefont
  {Zwerver}, \citenamefont {Krahenmann}, \citenamefont {Watson}, \citenamefont
  {Lampert}, \citenamefont {George}, \citenamefont {Pillarisetty},
  \citenamefont {Bojarski}, \citenamefont {Amin}, \citenamefont {Amitonov},
  \citenamefont {Boter}, \citenamefont {Caudillo}, \citenamefont
  {Correas-Serrano}, \citenamefont {Dehollain}, \citenamefont {Droulers},
  \citenamefont {Henry}, \citenamefont {Kotlyar}, \citenamefont {Lodari},
  \citenamefont {Luthi}, \citenamefont {Michalak}, \citenamefont {Mueller},
  \citenamefont {Neyens}, \citenamefont {Roberts}, \citenamefont {Samkharadze},
  \citenamefont {Zheng}, \citenamefont {Zietz}, \citenamefont {Scappucci},
  \citenamefont {Veldhorst}, \citenamefont {Vandersypen},\ and\ \citenamefont
  {Clarke}}]{intel_paper}%
  \BibitemOpen
  \bibfield  {author} {\bibinfo {author} {\bibfnamefont {A.~M.~J.}\
  \bibnamefont {Zwerver}}, \bibinfo {author} {\bibfnamefont {T.}~\bibnamefont
  {Krahenmann}}, \bibinfo {author} {\bibfnamefont {T.~F.}\ \bibnamefont
  {Watson}}, \bibinfo {author} {\bibfnamefont {L.}~\bibnamefont {Lampert}},
  \bibinfo {author} {\bibfnamefont {H.~C.}\ \bibnamefont {George}}, \bibinfo
  {author} {\bibfnamefont {R.}~\bibnamefont {Pillarisetty}}, \bibinfo {author}
  {\bibfnamefont {S.~A.}\ \bibnamefont {Bojarski}}, \bibinfo {author}
  {\bibfnamefont {P.}~\bibnamefont {Amin}}, \bibinfo {author} {\bibfnamefont
  {S.~V.}\ \bibnamefont {Amitonov}}, \bibinfo {author} {\bibfnamefont {J.~M.}\
  \bibnamefont {Boter}}, \bibinfo {author} {\bibfnamefont {R.}~\bibnamefont
  {Caudillo}}, \bibinfo {author} {\bibfnamefont {D.}~\bibnamefont
  {Correas-Serrano}}, \bibinfo {author} {\bibfnamefont {J.~P.}\ \bibnamefont
  {Dehollain}}, \bibinfo {author} {\bibfnamefont {G.}~\bibnamefont {Droulers}},
  \bibinfo {author} {\bibfnamefont {E.~M.}\ \bibnamefont {Henry}}, \bibinfo
  {author} {\bibfnamefont {R.}~\bibnamefont {Kotlyar}}, \bibinfo {author}
  {\bibfnamefont {M.}~\bibnamefont {Lodari}}, \bibinfo {author} {\bibfnamefont
  {F.}~\bibnamefont {Luthi}}, \bibinfo {author} {\bibfnamefont {D.~J.}\
  \bibnamefont {Michalak}}, \bibinfo {author} {\bibfnamefont {B.~K.}\
  \bibnamefont {Mueller}}, \bibinfo {author} {\bibfnamefont {S.}~\bibnamefont
  {Neyens}}, \bibinfo {author} {\bibfnamefont {J.}~\bibnamefont {Roberts}},
  \bibinfo {author} {\bibfnamefont {N.}~\bibnamefont {Samkharadze}}, \bibinfo
  {author} {\bibfnamefont {G.}~\bibnamefont {Zheng}}, \bibinfo {author}
  {\bibfnamefont {O.~K.}\ \bibnamefont {Zietz}}, \bibinfo {author}
  {\bibfnamefont {G.}~\bibnamefont {Scappucci}}, \bibinfo {author}
  {\bibfnamefont {M.}~\bibnamefont {Veldhorst}}, \bibinfo {author}
  {\bibfnamefont {L.~M.~K.}\ \bibnamefont {Vandersypen}}, \ and\ \bibinfo
  {author} {\bibfnamefont {J.~S.}\ \bibnamefont {Clarke}},\ }\bibfield  {title}
  {\enquote {\bibinfo {title} {Qubits made by advanced semiconductor
  manufacturing},}\ }\href {\doibase 10.1038/s41928-022-00727-9} {\bibfield
  {journal} {\bibinfo  {journal} {Nature Electronics}\ }\textbf {\bibinfo
  {volume} {5}},\ \bibinfo {pages} {184--190} (\bibinfo {year}
  {2022})}\BibitemShut {NoStop}%
\bibitem [{\citenamefont {Maurand}\ \emph {et~al.}(2016)\citenamefont
  {Maurand}, \citenamefont {Jehl}, \citenamefont {Kotekar-Patil}, \citenamefont
  {Corna}, \citenamefont {Bohuslavskyi}, \citenamefont {Laviéville},
  \citenamefont {Hutin}, \citenamefont {Barraud}, \citenamefont {Vinet},
  \citenamefont {Sanquer},\ and\ \citenamefont {De~Franceschi}}]{silvano_1}%
  \BibitemOpen
  \bibfield  {author} {\bibinfo {author} {\bibfnamefont {R.}~\bibnamefont
  {Maurand}}, \bibinfo {author} {\bibfnamefont {X.}~\bibnamefont {Jehl}},
  \bibinfo {author} {\bibfnamefont {D.}~\bibnamefont {Kotekar-Patil}}, \bibinfo
  {author} {\bibfnamefont {A.}~\bibnamefont {Corna}}, \bibinfo {author}
  {\bibfnamefont {H.}~\bibnamefont {Bohuslavskyi}}, \bibinfo {author}
  {\bibfnamefont {R.}~\bibnamefont {Laviéville}}, \bibinfo {author}
  {\bibfnamefont {L.}~\bibnamefont {Hutin}}, \bibinfo {author} {\bibfnamefont
  {S.}~\bibnamefont {Barraud}}, \bibinfo {author} {\bibfnamefont
  {M.}~\bibnamefont {Vinet}}, \bibinfo {author} {\bibfnamefont
  {M.}~\bibnamefont {Sanquer}}, \ and\ \bibinfo {author} {\bibfnamefont
  {S.}~\bibnamefont {De~Franceschi}},\ }\bibfield  {title} {\enquote {\bibinfo
  {title} {A {CMOS} silicon spin qubit},}\ }\href {\doibase
  10.1038/ncomms13575} {\bibfield  {journal} {\bibinfo  {journal} {Nature
  Communications}\ }\textbf {\bibinfo {volume} {7}},\ \bibinfo {pages} {13575}
  (\bibinfo {year} {2016})}\BibitemShut {NoStop}%
\bibitem [{\citenamefont {Liles}\ \emph {et~al.}(2018)\citenamefont {Liles},
  \citenamefont {Li}, \citenamefont {Yang}, \citenamefont {Hudson},
  \citenamefont {Veldhorst}, \citenamefont {Dzurak},\ and\ \citenamefont
  {Hamilton}}]{silvano_2}%
  \BibitemOpen
  \bibfield  {author} {\bibinfo {author} {\bibfnamefont {S.~D.}\ \bibnamefont
  {Liles}}, \bibinfo {author} {\bibfnamefont {R.}~\bibnamefont {Li}}, \bibinfo
  {author} {\bibfnamefont {C.~H.}\ \bibnamefont {Yang}}, \bibinfo {author}
  {\bibfnamefont {F.~E.}\ \bibnamefont {Hudson}}, \bibinfo {author}
  {\bibfnamefont {M.}~\bibnamefont {Veldhorst}}, \bibinfo {author}
  {\bibfnamefont {A.~S.}\ \bibnamefont {Dzurak}}, \ and\ \bibinfo {author}
  {\bibfnamefont {A.~R.}\ \bibnamefont {Hamilton}},\ }\bibfield  {title}
  {\enquote {\bibinfo {title} {Spin and orbital structure of the first six
  holes in a silicon metal-oxide-semiconductor quantum dot},}\ }\href {\doibase
  10.1038/s41467-018-05700-9} {\bibfield  {journal} {\bibinfo  {journal}
  {Nature Communications}\ }\textbf {\bibinfo {volume} {9}},\ \bibinfo {pages}
  {3255} (\bibinfo {year} {2018})}\BibitemShut {NoStop}%
\bibitem [{\citenamefont {Fang}\ \emph {et~al.}(2023)\citenamefont {Fang},
  \citenamefont {Philippopoulos}, \citenamefont {Culcer}, \citenamefont
  {Coish},\ and\ \citenamefont {Chesi}}]{hole_qubits_review}%
  \BibitemOpen
  \bibfield  {author} {\bibinfo {author} {\bibfnamefont {Y.}~\bibnamefont
  {Fang}}, \bibinfo {author} {\bibfnamefont {P.}~\bibnamefont
  {Philippopoulos}}, \bibinfo {author} {\bibfnamefont {D.}~\bibnamefont
  {Culcer}}, \bibinfo {author} {\bibfnamefont {W.~A.}\ \bibnamefont {Coish}}, \
  and\ \bibinfo {author} {\bibfnamefont {S.}~\bibnamefont {Chesi}},\ }\bibfield
   {title} {\enquote {\bibinfo {title} {Recent advances in hole-spin qubits},}\
  }\href {\doibase 10.1088/2633-4356/acb87e} {\bibfield  {journal} {\bibinfo
  {journal} {Materials for Quantum Technology}\ }\textbf {\bibinfo {volume}
  {3}},\ \bibinfo {pages} {012003} (\bibinfo {year} {2023})}\BibitemShut
  {NoStop}%
\bibitem [{\citenamefont {van~der Wiel}\ \emph {et~al.}(2002)\citenamefont
  {van~der Wiel}, \citenamefont {De~Franceschi}, \citenamefont {Elzerman},
  \citenamefont {Fujisawa}, \citenamefont {Tarucha},\ and\ \citenamefont
  {Kouwenhoven}}]{wiel}%
  \BibitemOpen
  \bibfield  {author} {\bibinfo {author} {\bibfnamefont {W.~G.}\ \bibnamefont
  {van~der Wiel}}, \bibinfo {author} {\bibfnamefont {S.}~\bibnamefont
  {De~Franceschi}}, \bibinfo {author} {\bibfnamefont {J.~M.}\ \bibnamefont
  {Elzerman}}, \bibinfo {author} {\bibfnamefont {T.}~\bibnamefont {Fujisawa}},
  \bibinfo {author} {\bibfnamefont {S.}~\bibnamefont {Tarucha}}, \ and\
  \bibinfo {author} {\bibfnamefont {L.~P.}\ \bibnamefont {Kouwenhoven}},\
  }\bibfield  {title} {\enquote {\bibinfo {title} {Electron transport through
  double quantum dots},}\ }\href {\doibase 10.1103/RevModPhys.75.1} {\bibfield
  {journal} {\bibinfo  {journal} {Reviews of Modern Physics}\ }\textbf
  {\bibinfo {volume} {75}},\ \bibinfo {pages} {1--22} (\bibinfo {year}
  {2002})},\ \bibinfo {note} {publisher: American Physical Society}\BibitemShut
  {NoStop}%
\bibitem [{\citenamefont {Russ}\ and\ \citenamefont
  {Burkard}(2017)}]{3electron}%
  \BibitemOpen
  \bibfield  {author} {\bibinfo {author} {\bibfnamefont {M.}~\bibnamefont
  {Russ}}\ and\ \bibinfo {author} {\bibfnamefont {G.}~\bibnamefont {Burkard}},\
  }\bibfield  {title} {\enquote {\bibinfo {title} {Three-electron spin
  qubits},}\ }\href {\doibase 10.1088/1361-648X/aa761f} {\bibfield  {journal}
  {\bibinfo  {journal} {Journal of Physics: Condensed Matter}\ }\textbf
  {\bibinfo {volume} {29}},\ \bibinfo {pages} {393001} (\bibinfo {year}
  {2017})}\BibitemShut {NoStop}%
\bibitem [{\citenamefont {Burkard}\ \emph {et~al.}(2023)\citenamefont
  {Burkard}, \citenamefont {Ladd}, \citenamefont {Pan}, \citenamefont
  {Nichol},\ and\ \citenamefont {Petta}}]{big_review}%
  \BibitemOpen
  \bibfield  {author} {\bibinfo {author} {\bibfnamefont {G.}~\bibnamefont
  {Burkard}}, \bibinfo {author} {\bibfnamefont {T.~D.}\ \bibnamefont {Ladd}},
  \bibinfo {author} {\bibfnamefont {A.}~\bibnamefont {Pan}}, \bibinfo {author}
  {\bibfnamefont {J.~M.}\ \bibnamefont {Nichol}}, \ and\ \bibinfo {author}
  {\bibfnamefont {J.~R.}\ \bibnamefont {Petta}},\ }\bibfield  {title} {\enquote
  {\bibinfo {title} {Semiconductor spin qubits},}\ }\href {\doibase
  10.1103/RevModPhys.95.025003} {\bibfield  {journal} {\bibinfo  {journal}
  {Reviews of Modern Physics}\ }\textbf {\bibinfo {volume} {95}},\ \bibinfo
  {pages} {025003} (\bibinfo {year} {2023})}\BibitemShut {NoStop}%
\bibitem [{\citenamefont {Harvey}(2022)}]{qds_spinqubits}%
  \BibitemOpen
  \bibfield  {author} {\bibinfo {author} {\bibfnamefont {S.}~\bibnamefont
  {Harvey}},\ }\bibfield  {title} {\enquote {\bibinfo {title} {Quantum {Dots} /
  {Spin} {Qubits}},}\ \ }(\bibinfo {year} {2022})\ \bibinfo {note}
  {arXiv:2204.04261 [cond-mat, physics:quant-ph]}\BibitemShut {NoStop}%
\bibitem [{\citenamefont {Crippa}\ \emph {et~al.}(2018)\citenamefont {Crippa},
  \citenamefont {Maurand}, \citenamefont {Bourdet}, \citenamefont
  {Kotekar-Patil}, \citenamefont {Amisse}, \citenamefont {Jehl}, \citenamefont
  {Sanquer}, \citenamefont {Laviéville}, \citenamefont {Bohuslavskyi},
  \citenamefont {Hutin}, \citenamefont {Barraud}, \citenamefont {Vinet},
  \citenamefont {Niquet},\ and\ \citenamefont {De~Franceschi}}]{silvano_4}%
  \BibitemOpen
  \bibfield  {author} {\bibinfo {author} {\bibfnamefont {A.}~\bibnamefont
  {Crippa}}, \bibinfo {author} {\bibfnamefont {R.}~\bibnamefont {Maurand}},
  \bibinfo {author} {\bibfnamefont {L.}~\bibnamefont {Bourdet}}, \bibinfo
  {author} {\bibfnamefont {D.}~\bibnamefont {Kotekar-Patil}}, \bibinfo {author}
  {\bibfnamefont {A.}~\bibnamefont {Amisse}}, \bibinfo {author} {\bibfnamefont
  {X.}~\bibnamefont {Jehl}}, \bibinfo {author} {\bibfnamefont {M.}~\bibnamefont
  {Sanquer}}, \bibinfo {author} {\bibfnamefont {R.}~\bibnamefont
  {Laviéville}}, \bibinfo {author} {\bibfnamefont {H.}~\bibnamefont
  {Bohuslavskyi}}, \bibinfo {author} {\bibfnamefont {L.}~\bibnamefont {Hutin}},
  \bibinfo {author} {\bibfnamefont {S.}~\bibnamefont {Barraud}}, \bibinfo
  {author} {\bibfnamefont {M.}~\bibnamefont {Vinet}}, \bibinfo {author}
  {\bibfnamefont {Y.-M.}\ \bibnamefont {Niquet}}, \ and\ \bibinfo {author}
  {\bibfnamefont {S.}~\bibnamefont {De~Franceschi}},\ }\bibfield  {title}
  {\enquote {\bibinfo {title} {Electrical {Spin} {Driving} by g-{Matrix}
  {Modulation} in {Spin}-{Orbit} {Qubits}},}\ }\href {\doibase
  10.1103/PhysRevLett.120.137702} {\bibfield  {journal} {\bibinfo  {journal}
  {Physical Review Letters}\ }\textbf {\bibinfo {volume} {120}},\ \bibinfo
  {pages} {137702} (\bibinfo {year} {2018})}\BibitemShut {NoStop}%
\bibitem [{\citenamefont {Wolfowicz}\ and\ \citenamefont
  {Morton}(2016)}]{pulse_eng}%
  \BibitemOpen
  \bibfield  {author} {\bibinfo {author} {\bibfnamefont {G.}~\bibnamefont
  {Wolfowicz}}\ and\ \bibinfo {author} {\bibfnamefont {J.~J.}\ \bibnamefont
  {Morton}},\ }\bibfield  {title} {\enquote {\bibinfo {title} {Pulse
  {Techniques} for {Quantum} {Information} {Processing}},}\ }in\ \href
  {\doibase 10.1002/9780470034590.emrstm1521} {\emph {\bibinfo {booktitle}
  {{eMagRes}}}},\ \bibinfo {editor} {edited by\ \bibinfo {editor}
  {\bibfnamefont {R.~K.}\ \bibnamefont {Harris}}\ and\ \bibinfo {editor}
  {\bibfnamefont {R.~L.}\ \bibnamefont {Wasylishen}}}\ (\bibinfo  {publisher}
  {John Wiley \& Sons, Ltd},\ \bibinfo {address} {Chichester, UK},\ \bibinfo
  {year} {2016})\ pp.\ \bibinfo {pages} {1515--1528}\BibitemShut {NoStop}%
\bibitem [{\citenamefont {DiVincenzo}\ \emph {et~al.}(2000)\citenamefont
  {DiVincenzo}, \citenamefont {Bacon}, \citenamefont {Kempe}, \citenamefont
  {Burkard},\ and\ \citenamefont {Whaley}}]{120}%
  \BibitemOpen
  \bibfield  {author} {\bibinfo {author} {\bibfnamefont {D.~P.}\ \bibnamefont
  {DiVincenzo}}, \bibinfo {author} {\bibfnamefont {D.}~\bibnamefont {Bacon}},
  \bibinfo {author} {\bibfnamefont {J.}~\bibnamefont {Kempe}}, \bibinfo
  {author} {\bibfnamefont {G.}~\bibnamefont {Burkard}}, \ and\ \bibinfo
  {author} {\bibfnamefont {K.~B.}\ \bibnamefont {Whaley}},\ }\bibfield  {title}
  {\enquote {\bibinfo {title} {Universal quantum computation with the exchange
  interaction},}\ }\href {\doibase 10.1038/35042541} {\bibfield  {journal}
  {\bibinfo  {journal} {Nature}\ }\textbf {\bibinfo {volume} {408}},\ \bibinfo
  {pages} {339--342} (\bibinfo {year} {2000})},\ \bibinfo {note} {number: 6810
  Publisher: Nature Publishing Group}\BibitemShut {NoStop}%
\bibitem [{\citenamefont {Michielis}\ \emph {et~al.}(2023)\citenamefont
  {Michielis}, \citenamefont {Ferraro}, \citenamefont {Prati}, \citenamefont
  {Hutin}, \citenamefont {Bertrand}, \citenamefont {Charbon}, \citenamefont
  {Ibberson},\ and\ \citenamefont {Gonzalez-Zalba}}]{lab_to_industry_qubits}%
  \BibitemOpen
  \bibfield  {author} {\bibinfo {author} {\bibfnamefont {M.~D.}\ \bibnamefont
  {Michielis}}, \bibinfo {author} {\bibfnamefont {E.}~\bibnamefont {Ferraro}},
  \bibinfo {author} {\bibfnamefont {E.}~\bibnamefont {Prati}}, \bibinfo
  {author} {\bibfnamefont {L.}~\bibnamefont {Hutin}}, \bibinfo {author}
  {\bibfnamefont {B.}~\bibnamefont {Bertrand}}, \bibinfo {author}
  {\bibfnamefont {E.}~\bibnamefont {Charbon}}, \bibinfo {author} {\bibfnamefont
  {D.~J.}\ \bibnamefont {Ibberson}}, \ and\ \bibinfo {author} {\bibfnamefont
  {M.~F.}\ \bibnamefont {Gonzalez-Zalba}},\ }\bibfield  {title} {\enquote
  {\bibinfo {title} {Silicon spin qubits from laboratory to industry},}\ }\href
  {\doibase 10.1088/1361-6463/acd8c7} {\bibfield  {journal} {\bibinfo
  {journal} {Journal of Physics D: Applied Physics}\ }\textbf {\bibinfo
  {volume} {56}},\ \bibinfo {pages} {363001} (\bibinfo {year}
  {2023})}\BibitemShut {NoStop}%
\bibitem [{\citenamefont {Pillarisetty}\ \emph {et~al.}(2021)\citenamefont
  {Pillarisetty}, \citenamefont {Watson}, \citenamefont {Mueller},
  \citenamefont {Henry}, \citenamefont {George}, \citenamefont {Bojarski},
  \citenamefont {Lampert}, \citenamefont {Luthi}, \citenamefont {Kotlyar},
  \citenamefont {Zietz}, \citenamefont {Neyens}, \citenamefont {Borjans},
  \citenamefont {Caudillo}, \citenamefont {Michalak}, \citenamefont {Nahm},
  \citenamefont {Park}, \citenamefont {Ramsey}, \citenamefont {Roberts},
  \citenamefont {Schaal}, \citenamefont {Zheng}, \citenamefont {Krähenmann},
  \citenamefont {Lodari}, \citenamefont {Zwerver}, \citenamefont {Veldhorst},
  \citenamefont {Scappucci}, \citenamefont {Vandersvpen},\ and\ \citenamefont
  {Clarke}}]{intel_50nm}%
  \BibitemOpen
  \bibfield  {author} {\bibinfo {author} {\bibfnamefont {R.}~\bibnamefont
  {Pillarisetty}}, \bibinfo {author} {\bibfnamefont {T.}~\bibnamefont
  {Watson}}, \bibinfo {author} {\bibfnamefont {B.}~\bibnamefont {Mueller}},
  \bibinfo {author} {\bibfnamefont {E.}~\bibnamefont {Henry}}, \bibinfo
  {author} {\bibfnamefont {H.}~\bibnamefont {George}}, \bibinfo {author}
  {\bibfnamefont {S.}~\bibnamefont {Bojarski}}, \bibinfo {author}
  {\bibfnamefont {L.}~\bibnamefont {Lampert}}, \bibinfo {author} {\bibfnamefont
  {F.}~\bibnamefont {Luthi}}, \bibinfo {author} {\bibfnamefont
  {R.}~\bibnamefont {Kotlyar}}, \bibinfo {author} {\bibfnamefont
  {O.}~\bibnamefont {Zietz}}, \bibinfo {author} {\bibfnamefont
  {S.}~\bibnamefont {Neyens}}, \bibinfo {author} {\bibfnamefont
  {F.}~\bibnamefont {Borjans}}, \bibinfo {author} {\bibfnamefont
  {R.}~\bibnamefont {Caudillo}}, \bibinfo {author} {\bibfnamefont
  {D.}~\bibnamefont {Michalak}}, \bibinfo {author} {\bibfnamefont
  {R.}~\bibnamefont {Nahm}}, \bibinfo {author} {\bibfnamefont {J.}~\bibnamefont
  {Park}}, \bibinfo {author} {\bibfnamefont {M.}~\bibnamefont {Ramsey}},
  \bibinfo {author} {\bibfnamefont {J.}~\bibnamefont {Roberts}}, \bibinfo
  {author} {\bibfnamefont {S.}~\bibnamefont {Schaal}}, \bibinfo {author}
  {\bibfnamefont {G.}~\bibnamefont {Zheng}}, \bibinfo {author} {\bibfnamefont
  {T.}~\bibnamefont {Krähenmann}}, \bibinfo {author} {\bibfnamefont
  {M.}~\bibnamefont {Lodari}}, \bibinfo {author} {\bibfnamefont
  {A.}~\bibnamefont {Zwerver}}, \bibinfo {author} {\bibfnamefont
  {M.}~\bibnamefont {Veldhorst}}, \bibinfo {author} {\bibfnamefont
  {G.}~\bibnamefont {Scappucci}}, \bibinfo {author} {\bibfnamefont
  {L.}~\bibnamefont {Vandersvpen}}, \ and\ \bibinfo {author} {\bibfnamefont
  {J.}~\bibnamefont {Clarke}},\ }\bibfield  {title} {\enquote {\bibinfo {title}
  {Si {MOS} and {Si}/{SiGe} quantum well spin qubit platforms for scalable
  quantum computing},}\ }in\ \href {\doibase 10.1109/IEDM19574.2021.9720567}
  {\emph {\bibinfo {booktitle} {2021 {IEEE} {International} {Electron}
  {Devices} {Meeting} ({IEDM})}}}\ (\bibinfo {year} {2021})\ pp.\ \bibinfo
  {pages} {14.1.1--14.1.4}\BibitemShut {NoStop}%
\bibitem [{\citenamefont {Bédécarrats}\ \emph {et~al.}(2021)\citenamefont
  {Bédécarrats}, \citenamefont {Paz}, \citenamefont {Diaz}, \citenamefont
  {Niebojewski}, \citenamefont {Bertrand}, \citenamefont {Rambal},
  \citenamefont {Comboroure}, \citenamefont {Sarrazin}, \citenamefont
  {Boulard}, \citenamefont {Guyez}, \citenamefont {Hartmann}, \citenamefont
  {Morand}, \citenamefont {Magalhaes-Lucas}, \citenamefont {Nowak},
  \citenamefont {Catapano}, \citenamefont {Cassé}, \citenamefont
  {Urdampilleta}, \citenamefont {Niquet}, \citenamefont {Gaillard},
  \citenamefont {De~Franceschi}, \citenamefont {Meunier},\ and\ \citenamefont
  {Vinet}}]{leti_80nm}%
  \BibitemOpen
  \bibfield  {author} {\bibinfo {author} {\bibfnamefont {T.}~\bibnamefont
  {Bédécarrats}}, \bibinfo {author} {\bibfnamefont {B.~C.}\ \bibnamefont
  {Paz}}, \bibinfo {author} {\bibfnamefont {B.~M.}\ \bibnamefont {Diaz}},
  \bibinfo {author} {\bibfnamefont {H.}~\bibnamefont {Niebojewski}}, \bibinfo
  {author} {\bibfnamefont {B.}~\bibnamefont {Bertrand}}, \bibinfo {author}
  {\bibfnamefont {N.}~\bibnamefont {Rambal}}, \bibinfo {author} {\bibfnamefont
  {C.}~\bibnamefont {Comboroure}}, \bibinfo {author} {\bibfnamefont
  {A.}~\bibnamefont {Sarrazin}}, \bibinfo {author} {\bibfnamefont
  {F.}~\bibnamefont {Boulard}}, \bibinfo {author} {\bibfnamefont
  {E.}~\bibnamefont {Guyez}}, \bibinfo {author} {\bibfnamefont {J.-M.}\
  \bibnamefont {Hartmann}}, \bibinfo {author} {\bibfnamefont {Y.}~\bibnamefont
  {Morand}}, \bibinfo {author} {\bibfnamefont {A.}~\bibnamefont
  {Magalhaes-Lucas}}, \bibinfo {author} {\bibfnamefont {E.}~\bibnamefont
  {Nowak}}, \bibinfo {author} {\bibfnamefont {E.}~\bibnamefont {Catapano}},
  \bibinfo {author} {\bibfnamefont {M.}~\bibnamefont {Cassé}}, \bibinfo
  {author} {\bibfnamefont {M.}~\bibnamefont {Urdampilleta}}, \bibinfo {author}
  {\bibfnamefont {Y.-M.}\ \bibnamefont {Niquet}}, \bibinfo {author}
  {\bibfnamefont {F.}~\bibnamefont {Gaillard}}, \bibinfo {author}
  {\bibfnamefont {S.}~\bibnamefont {De~Franceschi}}, \bibinfo {author}
  {\bibfnamefont {T.}~\bibnamefont {Meunier}}, \ and\ \bibinfo {author}
  {\bibfnamefont {M.}~\bibnamefont {Vinet}},\ }\bibfield  {title} {\enquote
  {\bibinfo {title} {A new {FDSOI} spin qubit platform with 40nm effective
  control pitch},}\ }in\ \href {\doibase 10.1109/IEDM19574.2021.9720497} {\emph
  {\bibinfo {booktitle} {2021 {IEEE} {International} {Electron} {Devices}
  {Meeting} ({IEDM})}}}\ (\bibinfo {year} {2021})\ pp.\ \bibinfo {pages}
  {1--4}\BibitemShut {NoStop}%
\bibitem [{\citenamefont {Chang}\ \emph {et~al.}(2022)\citenamefont {Chang},
  \citenamefont {Chang}, \citenamefont {Pan}, \citenamefont {Lai},
  \citenamefont {Lu}, \citenamefont {Ng}, \citenamefont {Chen}, \citenamefont
  {Wu}, \citenamefont {Lin}, \citenamefont {Liang}, \citenamefont {Tsao},
  \citenamefont {Mor}, \citenamefont {Li}, \citenamefont {Lin}, \citenamefont
  {Hsieh}, \citenamefont {Chen}, \citenamefont {Hsu}, \citenamefont {Chen},
  \citenamefont {Chen}, \citenamefont {Yeh}, \citenamefont {Chiang},
  \citenamefont {Lin}, \citenamefont {Liaw}, \citenamefont {Wang},
  \citenamefont {Lee}, \citenamefont {Chen}, \citenamefont {Lin}, \citenamefont
  {Chen}, \citenamefont {Chen}, \citenamefont {Chui}, \citenamefont {Yeo},
  \citenamefont {Huang}, \citenamefont {Lee}, \citenamefont {Tsai},
  \citenamefont {Chen}, \citenamefont {Lu}, \citenamefont {Jang},\ and\
  \citenamefont {Wu}}]{tsmc_sota_pitch}%
  \BibitemOpen
  \bibfield  {author} {\bibinfo {author} {\bibfnamefont {C.-H.}\ \bibnamefont
  {Chang}}, \bibinfo {author} {\bibfnamefont {V.}~\bibnamefont {Chang}},
  \bibinfo {author} {\bibfnamefont {K.}~\bibnamefont {Pan}}, \bibinfo {author}
  {\bibfnamefont {K.}~\bibnamefont {Lai}}, \bibinfo {author} {\bibfnamefont
  {J.~H.}\ \bibnamefont {Lu}}, \bibinfo {author} {\bibfnamefont
  {J.}~\bibnamefont {Ng}}, \bibinfo {author} {\bibfnamefont {C.}~\bibnamefont
  {Chen}}, \bibinfo {author} {\bibfnamefont {B.}~\bibnamefont {Wu}}, \bibinfo
  {author} {\bibfnamefont {C.}~\bibnamefont {Lin}}, \bibinfo {author}
  {\bibfnamefont {C.}~\bibnamefont {Liang}}, \bibinfo {author} {\bibfnamefont
  {C.}~\bibnamefont {Tsao}}, \bibinfo {author} {\bibfnamefont {Y.}~\bibnamefont
  {Mor}}, \bibinfo {author} {\bibfnamefont {C.}~\bibnamefont {Li}}, \bibinfo
  {author} {\bibfnamefont {T.}~\bibnamefont {Lin}}, \bibinfo {author}
  {\bibfnamefont {C.}~\bibnamefont {Hsieh}}, \bibinfo {author} {\bibfnamefont
  {P.}~\bibnamefont {Chen}}, \bibinfo {author} {\bibfnamefont {H.}~\bibnamefont
  {Hsu}}, \bibinfo {author} {\bibfnamefont {J.}~\bibnamefont {Chen}}, \bibinfo
  {author} {\bibfnamefont {H.}~\bibnamefont {Chen}}, \bibinfo {author}
  {\bibfnamefont {J.}~\bibnamefont {Yeh}}, \bibinfo {author} {\bibfnamefont
  {M.}~\bibnamefont {Chiang}}, \bibinfo {author} {\bibfnamefont
  {C.}~\bibnamefont {Lin}}, \bibinfo {author} {\bibfnamefont {J.}~\bibnamefont
  {Liaw}}, \bibinfo {author} {\bibfnamefont {C.}~\bibnamefont {Wang}}, \bibinfo
  {author} {\bibfnamefont {S.}~\bibnamefont {Lee}}, \bibinfo {author}
  {\bibfnamefont {C.}~\bibnamefont {Chen}}, \bibinfo {author} {\bibfnamefont
  {H.}~\bibnamefont {Lin}}, \bibinfo {author} {\bibfnamefont {R.}~\bibnamefont
  {Chen}}, \bibinfo {author} {\bibfnamefont {K.}~\bibnamefont {Chen}}, \bibinfo
  {author} {\bibfnamefont {C.}~\bibnamefont {Chui}}, \bibinfo {author}
  {\bibfnamefont {Y.}~\bibnamefont {Yeo}}, \bibinfo {author} {\bibfnamefont
  {K.}~\bibnamefont {Huang}}, \bibinfo {author} {\bibfnamefont
  {T.}~\bibnamefont {Lee}}, \bibinfo {author} {\bibfnamefont {M.}~\bibnamefont
  {Tsai}}, \bibinfo {author} {\bibfnamefont {K.}~\bibnamefont {Chen}}, \bibinfo
  {author} {\bibfnamefont {Y.}~\bibnamefont {Lu}}, \bibinfo {author}
  {\bibfnamefont {S.}~\bibnamefont {Jang}}, \ and\ \bibinfo {author}
  {\bibfnamefont {S.-Y.}\ \bibnamefont {Wu}},\ }\bibfield  {title} {\enquote
  {\bibinfo {title} {Critical {Process} {Features} {Enabling} {Aggressive}
  {Contacted} {Gate} {Pitch} {Scaling} for 3nm {CMOS} {Technology} and
  {Beyond}},}\ }in\ \href {\doibase 10.1109/IEDM45625.2022.10019565} {\emph
  {\bibinfo {booktitle} {2022 {International} {Electron} {Devices} {Meeting}
  ({IEDM})}}}\ (\bibinfo {year} {2022})\ pp.\ \bibinfo {pages}
  {27.1.1--27.1.4}\BibitemShut {NoStop}%
\bibitem [{\citenamefont {Corporation}(2021)}]{src_decadal_report}%
  \BibitemOpen
  \bibfield  {author} {\bibinfo {author} {\bibfnamefont {S.~R.}\ \bibnamefont
  {Corporation}},\ }\bibfield  {title} {\enquote {\bibinfo {title} {Decadal
  {Plan} for {Semiconductors}},}\ }\href
  {https://www.src.org/about/decadal-plan/} {\bibfield  {journal} {\bibinfo
  {journal} {https://www.src.org/about/decadal-plan/}\ } (\bibinfo {year}
  {2021})}\BibitemShut {NoStop}%
\bibitem [{\citenamefont {Rasmussen}, \citenamefont {Wilthan},\ and\
  \citenamefont {Simonds}(2023)}]{nist_sota_euv}%
  \BibitemOpen
  \bibfield  {author} {\bibinfo {author} {\bibfnamefont {E.~G.}\ \bibnamefont
  {Rasmussen}}, \bibinfo {author} {\bibfnamefont {B.}~\bibnamefont {Wilthan}},
  \ and\ \bibinfo {author} {\bibfnamefont {B.}~\bibnamefont {Simonds}},\ }\href
  {https://nvlpubs.nist.gov/nistpubs/SpecialPublications/NIST.SP.1500-208.pdf}
  {\enquote {\bibinfo {title} {Report from the {Extreme} {Ultraviolet} ({EUV})
  {Lithography} {Working} {Group} {Meeting} : current state, needs, and path
  forward},}\ }\bibinfo {type} {Tech. Rep.}\ \bibinfo {number} {NIST SP
  1500-208}\ (\bibinfo  {institution} {National Institute of Standards and
  Technology (U.S.)},\ \bibinfo {address} {Gaithersburg, MD},\ \bibinfo {year}
  {2023})\BibitemShut {NoStop}%
\bibitem [{\citenamefont {Ober}, \citenamefont {Käfer},\ and\ \citenamefont
  {Yuan}(2023)}]{duv}%
  \BibitemOpen
  \bibfield  {author} {\bibinfo {author} {\bibfnamefont {C.~K.}\ \bibnamefont
  {Ober}}, \bibinfo {author} {\bibfnamefont {F.}~\bibnamefont {Käfer}}, \ and\
  \bibinfo {author} {\bibfnamefont {C.}~\bibnamefont {Yuan}},\ }\bibfield
  {title} {\enquote {\bibinfo {title} {Recent developments in photoresists for
  extreme-ultraviolet lithography},}\ }\href {\doibase
  10.1016/j.polymer.2023.126020} {\bibfield  {journal} {\bibinfo  {journal}
  {Polymer}\ }\textbf {\bibinfo {volume} {280}},\ \bibinfo {pages} {126020}
  (\bibinfo {year} {2023})}\BibitemShut {NoStop}%
\bibitem [{\citenamefont {Haley}(2023)}]{duv_news}%
  \BibitemOpen
  \bibfield  {author} {\bibinfo {author} {\bibfnamefont {G.}~\bibnamefont
  {Haley}},\ }\href
  {https://semiengineering.com/193i-lithography-takes-center-stage-again/}
  {\enquote {\bibinfo {title} {193i {Lithography} {Takes} {Center}
  {Stage}...{Again}},}\ } (\bibinfo {year} {2023})\BibitemShut {NoStop}%
\bibitem [{\citenamefont {Mahdi}(2023)}]{dsa_euv}%
  \BibitemOpen
  \bibfield  {author} {\bibinfo {author} {\bibfnamefont {T.}~\bibnamefont
  {Mahdi}},\ }\bibfield  {title} {\enquote {\bibinfo {title} {Material and
  {Patterning} {Innovation}: {The} {Foundation} for {Moore}’s {Law}
  {Extension}},}\ }\href
  {https://www.src.org/calendar/e007840/tayseer_mahdi_ilt.pdf} {\bibfield
  {journal} {\bibinfo  {journal} {SRC Industry-led Talk:
  https://www.src.org/calendar/e007840/tayseer\_mahdi\_ilt.pdf}\ } (\bibinfo
  {year} {2023})}\BibitemShut {NoStop}%
\bibitem [{\citenamefont {Zeloof}(2018)}]{homemade_ebeam}%
  \BibitemOpen
  \bibfield  {author} {\bibinfo {author} {\bibfnamefont {S.}~\bibnamefont
  {Zeloof}},\ }\href {http://sam.zeloof.xyz/e-beam-lithography/} {\enquote
  {\bibinfo {title} {E-beam {Lithography}},}\ } (\bibinfo {year}
  {2018})\BibitemShut {NoStop}%
\bibitem [{\citenamefont {Brackmann}\ \emph {et~al.}(2023)\citenamefont
  {Brackmann}, \citenamefont {Neul}, \citenamefont {Friedrich}, \citenamefont
  {Langheinrich}, \citenamefont {Simon}, \citenamefont {Muster}, \citenamefont
  {Pregl}, \citenamefont {Demmler}, \citenamefont {Hanisch}, \citenamefont
  {Lederer}, \citenamefont {Zimmermann}, \citenamefont {Klos}, \citenamefont
  {Reichmann}, \citenamefont {Yamamoto}, \citenamefont {Wislicenus},
  \citenamefont {Dahl}, \citenamefont {Schreiber}, \citenamefont {Bluhm},\ and\
  \citenamefont {Lilienthal-Uhlig}}]{ebeam_cmos_compatible}%
  \BibitemOpen
  \bibfield  {author} {\bibinfo {author} {\bibfnamefont {V.}~\bibnamefont
  {Brackmann}}, \bibinfo {author} {\bibfnamefont {M.}~\bibnamefont {Neul}},
  \bibinfo {author} {\bibfnamefont {M.}~\bibnamefont {Friedrich}}, \bibinfo
  {author} {\bibfnamefont {W.}~\bibnamefont {Langheinrich}}, \bibinfo {author}
  {\bibfnamefont {M.}~\bibnamefont {Simon}}, \bibinfo {author} {\bibfnamefont
  {P.}~\bibnamefont {Muster}}, \bibinfo {author} {\bibfnamefont
  {S.}~\bibnamefont {Pregl}}, \bibinfo {author} {\bibfnamefont
  {A.}~\bibnamefont {Demmler}}, \bibinfo {author} {\bibfnamefont
  {N.}~\bibnamefont {Hanisch}}, \bibinfo {author} {\bibfnamefont
  {M.}~\bibnamefont {Lederer}}, \bibinfo {author} {\bibfnamefont
  {K.}~\bibnamefont {Zimmermann}}, \bibinfo {author} {\bibfnamefont
  {J.}~\bibnamefont {Klos}}, \bibinfo {author} {\bibfnamefont {F.}~\bibnamefont
  {Reichmann}}, \bibinfo {author} {\bibfnamefont {Y.}~\bibnamefont {Yamamoto}},
  \bibinfo {author} {\bibfnamefont {M.}~\bibnamefont {Wislicenus}}, \bibinfo
  {author} {\bibfnamefont {C.}~\bibnamefont {Dahl}}, \bibinfo {author}
  {\bibfnamefont {L.~R.}\ \bibnamefont {Schreiber}}, \bibinfo {author}
  {\bibfnamefont {H.}~\bibnamefont {Bluhm}}, \ and\ \bibinfo {author}
  {\bibfnamefont {B.}~\bibnamefont {Lilienthal-Uhlig}},\ }\bibfield  {title}
  {\enquote {\bibinfo {title} {Fabrication of gate electrodes for scalable
  quantum computing using {CMOS} industry compatible e-beam lithography and
  numerical simulation of the resulting quantum device},}\ }in\ \href {\doibase
  10.1117/12.2675943} {\emph {\bibinfo {booktitle} {38th {European} {Mask} and
  {Lithography} {Conference} ({EMLC} 2023)}}},\ Vol.\ \bibinfo {volume}
  {12802}\ (\bibinfo  {publisher} {SPIE},\ \bibinfo {year} {2023})\ pp.\
  \bibinfo {pages} {150--167}\BibitemShut {NoStop}%
\bibitem [{\citenamefont {Žaper}\ \emph {et~al.}(2024)\citenamefont {Žaper},
  \citenamefont {Rickhaus}, \citenamefont {Wyss}, \citenamefont {Gross},
  \citenamefont {Wagner}, \citenamefont {Poggio},\ and\ \citenamefont
  {Braakman}}]{magnet_making_techniques}%
  \BibitemOpen
  \bibfield  {author} {\bibinfo {author} {\bibfnamefont {L.}~\bibnamefont
  {Žaper}}, \bibinfo {author} {\bibfnamefont {P.}~\bibnamefont {Rickhaus}},
  \bibinfo {author} {\bibfnamefont {M.}~\bibnamefont {Wyss}}, \bibinfo {author}
  {\bibfnamefont {B.}~\bibnamefont {Gross}}, \bibinfo {author} {\bibfnamefont
  {K.}~\bibnamefont {Wagner}}, \bibinfo {author} {\bibfnamefont
  {M.}~\bibnamefont {Poggio}}, \ and\ \bibinfo {author} {\bibfnamefont
  {F.}~\bibnamefont {Braakman}},\ }\bibfield  {title} {\enquote {\bibinfo
  {title} {Scanning {Nitrogen}-{Vacancy} {Magnetometry} of
  {Focused}-{Electron}-{Beam}-{Deposited} {Cobalt} {Nanomagnets}},}\ }\href
  {\doibase 10.1021/acsanm.3c05470} {\bibfield  {journal} {\bibinfo  {journal}
  {ACS Applied Nano Materials}\ }\textbf {\bibinfo {volume} {7}},\ \bibinfo
  {pages} {3854--3860} (\bibinfo {year} {2024})}\BibitemShut {NoStop}%
\bibitem [{\citenamefont {Ha}\ \emph {et~al.}(2022)\citenamefont {Ha},
  \citenamefont {Ha}, \citenamefont {Choi}, \citenamefont {Tang}, \citenamefont
  {Schmitz}, \citenamefont {Levendorf}, \citenamefont {Lee}, \citenamefont
  {Chappell}, \citenamefont {Adams}, \citenamefont {Hulbert}, \citenamefont
  {Acuna}, \citenamefont {Noah}, \citenamefont {Matten}, \citenamefont {Jura},
  \citenamefont {Wright}, \citenamefont {Rakher},\ and\ \citenamefont
  {Borselli}}]{fc_lc3}%
  \BibitemOpen
  \bibfield  {author} {\bibinfo {author} {\bibfnamefont {W.}~\bibnamefont
  {Ha}}, \bibinfo {author} {\bibfnamefont {S.~D.}\ \bibnamefont {Ha}}, \bibinfo
  {author} {\bibfnamefont {M.~D.}\ \bibnamefont {Choi}}, \bibinfo {author}
  {\bibfnamefont {Y.}~\bibnamefont {Tang}}, \bibinfo {author} {\bibfnamefont
  {A.~E.}\ \bibnamefont {Schmitz}}, \bibinfo {author} {\bibfnamefont {M.~P.}\
  \bibnamefont {Levendorf}}, \bibinfo {author} {\bibfnamefont {K.}~\bibnamefont
  {Lee}}, \bibinfo {author} {\bibfnamefont {J.~M.}\ \bibnamefont {Chappell}},
  \bibinfo {author} {\bibfnamefont {T.~S.}\ \bibnamefont {Adams}}, \bibinfo
  {author} {\bibfnamefont {D.~R.}\ \bibnamefont {Hulbert}}, \bibinfo {author}
  {\bibfnamefont {E.}~\bibnamefont {Acuna}}, \bibinfo {author} {\bibfnamefont
  {R.~S.}\ \bibnamefont {Noah}}, \bibinfo {author} {\bibfnamefont {J.~W.}\
  \bibnamefont {Matten}}, \bibinfo {author} {\bibfnamefont {M.~P.}\
  \bibnamefont {Jura}}, \bibinfo {author} {\bibfnamefont {J.~A.}\ \bibnamefont
  {Wright}}, \bibinfo {author} {\bibfnamefont {M.~T.}\ \bibnamefont {Rakher}},
  \ and\ \bibinfo {author} {\bibfnamefont {M.~G.}\ \bibnamefont {Borselli}},\
  }\bibfield  {title} {\enquote {\bibinfo {title} {A {Flexible} {Design}
  {Platform} for {Si}/{SiGe} {Exchange}-{Only} {Qubits} with {Low}
  {Disorder}},}\ }\href {\doibase 10.1021/acs.nanolett.1c03026} {\bibfield
  {journal} {\bibinfo  {journal} {Nano Letters}\ }\textbf {\bibinfo {volume}
  {22}},\ \bibinfo {pages} {1443--1448} (\bibinfo {year} {2022})},\ \bibinfo
  {note} {publisher: American Chemical Society}\BibitemShut {NoStop}%
\bibitem [{\citenamefont {Constantinou}\ \emph {et~al.}(2024)\citenamefont
  {Constantinou}, \citenamefont {Stock}, \citenamefont {Tseng}, \citenamefont
  {Kazazis}, \citenamefont {Muntwiler}, \citenamefont {Vaz}, \citenamefont
  {Ekinci}, \citenamefont {Aeppli}, \citenamefont {Curson},\ and\ \citenamefont
  {Schofield}}]{euv_for_si_qubits}%
  \BibitemOpen
  \bibfield  {author} {\bibinfo {author} {\bibfnamefont {P.}~\bibnamefont
  {Constantinou}}, \bibinfo {author} {\bibfnamefont {T.~J.~Z.}\ \bibnamefont
  {Stock}}, \bibinfo {author} {\bibfnamefont {L.-T.}\ \bibnamefont {Tseng}},
  \bibinfo {author} {\bibfnamefont {D.}~\bibnamefont {Kazazis}}, \bibinfo
  {author} {\bibfnamefont {M.}~\bibnamefont {Muntwiler}}, \bibinfo {author}
  {\bibfnamefont {C.~A.~F.}\ \bibnamefont {Vaz}}, \bibinfo {author}
  {\bibfnamefont {Y.}~\bibnamefont {Ekinci}}, \bibinfo {author} {\bibfnamefont
  {G.}~\bibnamefont {Aeppli}}, \bibinfo {author} {\bibfnamefont {N.~J.}\
  \bibnamefont {Curson}}, \ and\ \bibinfo {author} {\bibfnamefont {S.~R.}\
  \bibnamefont {Schofield}},\ }\bibfield  {title} {\enquote {\bibinfo {title}
  {{EUV}-induced hydrogen desorption as a step towards large-scale silicon
  quantum device patterning},}\ }\href {\doibase 10.1038/s41467-024-44790-6}
  {\bibfield  {journal} {\bibinfo  {journal} {Nature Communications}\ }\textbf
  {\bibinfo {volume} {15}},\ \bibinfo {pages} {694} (\bibinfo {year}
  {2024})}\BibitemShut {NoStop}%
\bibitem [{\citenamefont {Kostic}\ \emph {et~al.}(2017)\citenamefont {Kostic},
  \citenamefont {Vutova}, \citenamefont {Bencurova}, \citenamefont {Ritomsky},\
  and\ \citenamefont {Andok}}]{ebeam_limitations}%
  \BibitemOpen
  \bibfield  {author} {\bibinfo {author} {\bibfnamefont {I.}~\bibnamefont
  {Kostic}}, \bibinfo {author} {\bibfnamefont {K.}~\bibnamefont {Vutova}},
  \bibinfo {author} {\bibfnamefont {A.}~\bibnamefont {Bencurova}}, \bibinfo
  {author} {\bibfnamefont {A.}~\bibnamefont {Ritomsky}}, \ and\ \bibinfo
  {author} {\bibfnamefont {R.}~\bibnamefont {Andok}},\ }\bibfield  {title}
  {\enquote {\bibinfo {title} {Limitations of variable shaped electron beam
  lithography for advanced research and semiconductor applications},}\ }in\
  \href {\doibase 10.1109/ISSE.2017.8000969} {\emph {\bibinfo {booktitle} {2017
  40th {International} {Spring} {Seminar} on {Electronics} {Technology}
  ({ISSE})}}}\ (\bibinfo {year} {2017})\ pp.\ \bibinfo {pages}
  {1--6}\BibitemShut {NoStop}%
\bibitem [{\citenamefont {Brackmann}\ \emph {et~al.}(2019)\citenamefont
  {Brackmann}, \citenamefont {Friedrich}, \citenamefont {Browning},
  \citenamefont {Hanisch},\ and\ \citenamefont
  {Uhlig}}]{ebeam_optimization_challenges}%
  \BibitemOpen
  \bibfield  {author} {\bibinfo {author} {\bibfnamefont {V.}~\bibnamefont
  {Brackmann}}, \bibinfo {author} {\bibfnamefont {M.}~\bibnamefont
  {Friedrich}}, \bibinfo {author} {\bibfnamefont {C.}~\bibnamefont {Browning}},
  \bibinfo {author} {\bibfnamefont {N.}~\bibnamefont {Hanisch}}, \ and\
  \bibinfo {author} {\bibfnamefont {B.}~\bibnamefont {Uhlig}},\ }\bibfield
  {title} {\enquote {\bibinfo {title} {Influence of the dose assignment and
  fracturing type on patterns exposed by a variable shaped e-beam writer:
  simulation vs experiment},}\ }in\ \href {\doibase 10.1117/12.2534642} {\emph
  {\bibinfo {booktitle} {35th {European} {Mask} and {Lithography} {Conference}
  ({EMLC} 2019)}}},\ Vol.\ \bibinfo {volume} {11177}\ (\bibinfo  {publisher}
  {SPIE},\ \bibinfo {year} {2019})\ pp.\ \bibinfo {pages}
  {159--168}\BibitemShut {NoStop}%
\bibitem [{\citenamefont {Lawrie}\ \emph {et~al.}(2020)\citenamefont {Lawrie},
  \citenamefont {Eenink}, \citenamefont {Hendrickx}, \citenamefont {Boter},
  \citenamefont {Petit}, \citenamefont {Amitonov}, \citenamefont {Lodari},
  \citenamefont {Wuetz}, \citenamefont {Volk}, \citenamefont {Philips},
  \citenamefont {Droulers}, \citenamefont {Kalhor}, \citenamefont {van
  Riggelen}, \citenamefont {Brousse}, \citenamefont {Sammak}, \citenamefont
  {Vandersypen}, \citenamefont {Scappucci},\ and\ \citenamefont
  {Veldhorst}}]{si_sige_array}%
  \BibitemOpen
  \bibfield  {author} {\bibinfo {author} {\bibfnamefont {W.}~\bibnamefont
  {Lawrie}}, \bibinfo {author} {\bibnamefont {Eenink}}, \bibinfo {author}
  {\bibnamefont {Hendrickx}}, \bibinfo {author} {\bibnamefont {Boter}},
  \bibinfo {author} {\bibnamefont {Petit}}, \bibinfo {author} {\bibnamefont
  {Amitonov}}, \bibinfo {author} {\bibnamefont {Lodari}}, \bibinfo {author}
  {\bibfnamefont {P.}~\bibnamefont {Wuetz}}, \bibinfo {author} {\bibnamefont
  {Volk}}, \bibinfo {author} {\bibnamefont {Philips}}, \bibinfo {author}
  {\bibfnamefont {G.}~\bibnamefont {Droulers}}, \bibinfo {author}
  {\bibfnamefont {N.}~\bibnamefont {Kalhor}}, \bibinfo {author} {\bibfnamefont
  {F.}~\bibnamefont {van Riggelen}}, \bibinfo {author} {\bibfnamefont
  {D.}~\bibnamefont {Brousse}}, \bibinfo {author} {\bibfnamefont
  {A.}~\bibnamefont {Sammak}}, \bibinfo {author} {\bibfnamefont
  {L.}~\bibnamefont {Vandersypen}}, \bibinfo {author} {\bibfnamefont
  {G.}~\bibnamefont {Scappucci}}, \ and\ \bibinfo {author} {\bibfnamefont
  {M.}~\bibnamefont {Veldhorst}},\ }\bibfield  {title} {\enquote {\bibinfo
  {title} {Quantum dot arrays in silicon and germanium},}\ }\href {\doibase
  10.1063/5.0002013} {\bibfield  {journal} {\bibinfo  {journal} {Applied
  Physics Letters}\ }\textbf {\bibinfo {volume} {116}},\ \bibinfo {pages}
  {080501} (\bibinfo {year} {2020})}\BibitemShut {NoStop}%
\bibitem [{\citenamefont {Philips}\ \emph {et~al.}(2022)\citenamefont
  {Philips}, \citenamefont {Madzik}, \citenamefont {Amitonov}, \citenamefont
  {de~Snoo}, \citenamefont {Russ}, \citenamefont {Kalhor}, \citenamefont
  {Volk}, \citenamefont {Lawrie}, \citenamefont {Brousse}, \citenamefont
  {Tryputen}, \citenamefont {Wuetz}, \citenamefont {Sammak}, \citenamefont
  {Veldhorst}, \citenamefont {Scappucci},\ and\ \citenamefont
  {Vandersypen}}]{si_6qubits}%
  \BibitemOpen
  \bibfield  {author} {\bibinfo {author} {\bibfnamefont {S.~G.~J.}\
  \bibnamefont {Philips}}, \bibinfo {author} {\bibfnamefont {M.~T.}\
  \bibnamefont {Madzik}}, \bibinfo {author} {\bibfnamefont {S.~V.}\
  \bibnamefont {Amitonov}}, \bibinfo {author} {\bibfnamefont {S.~L.}\
  \bibnamefont {de~Snoo}}, \bibinfo {author} {\bibfnamefont {M.}~\bibnamefont
  {Russ}}, \bibinfo {author} {\bibfnamefont {N.}~\bibnamefont {Kalhor}},
  \bibinfo {author} {\bibfnamefont {C.}~\bibnamefont {Volk}}, \bibinfo {author}
  {\bibfnamefont {W.~I.~L.}\ \bibnamefont {Lawrie}}, \bibinfo {author}
  {\bibfnamefont {D.}~\bibnamefont {Brousse}}, \bibinfo {author} {\bibfnamefont
  {L.}~\bibnamefont {Tryputen}}, \bibinfo {author} {\bibfnamefont {B.~P.}\
  \bibnamefont {Wuetz}}, \bibinfo {author} {\bibfnamefont {A.}~\bibnamefont
  {Sammak}}, \bibinfo {author} {\bibfnamefont {M.}~\bibnamefont {Veldhorst}},
  \bibinfo {author} {\bibfnamefont {G.}~\bibnamefont {Scappucci}}, \ and\
  \bibinfo {author} {\bibfnamefont {L.~M.~K.}\ \bibnamefont {Vandersypen}},\
  }\bibfield  {title} {\enquote {\bibinfo {title} {Universal control of a
  six-qubit quantum processor in silicon},}\ }\href {\doibase
  10.1038/s41586-022-05117-x} {\bibfield  {journal} {\bibinfo  {journal}
  {Nature}\ }\textbf {\bibinfo {volume} {609}},\ \bibinfo {pages} {919--924}
  (\bibinfo {year} {2022})}\BibitemShut {NoStop}%
\bibitem [{\citenamefont {Kuhlmann}\ \emph {et~al.}(2013)\citenamefont
  {Kuhlmann}, \citenamefont {Houel}, \citenamefont {Ludwig}, \citenamefont
  {Greuter}, \citenamefont {Reuter}, \citenamefont {Wieck}, \citenamefont
  {Poggio},\ and\ \citenamefont {Warburton}}]{fc_cn1}%
  \BibitemOpen
  \bibfield  {author} {\bibinfo {author} {\bibfnamefont {A.~V.}\ \bibnamefont
  {Kuhlmann}}, \bibinfo {author} {\bibfnamefont {J.}~\bibnamefont {Houel}},
  \bibinfo {author} {\bibfnamefont {A.}~\bibnamefont {Ludwig}}, \bibinfo
  {author} {\bibfnamefont {L.}~\bibnamefont {Greuter}}, \bibinfo {author}
  {\bibfnamefont {D.}~\bibnamefont {Reuter}}, \bibinfo {author} {\bibfnamefont
  {A.~D.}\ \bibnamefont {Wieck}}, \bibinfo {author} {\bibfnamefont
  {M.}~\bibnamefont {Poggio}}, \ and\ \bibinfo {author} {\bibfnamefont {R.~J.}\
  \bibnamefont {Warburton}},\ }\bibfield  {title} {\enquote {\bibinfo {title}
  {Charge noise and spin noise in a semiconductor quantum device},}\ }\href
  {\doibase 10.1038/nphys2688} {\bibfield  {journal} {\bibinfo  {journal}
  {Nature Physics}\ }\textbf {\bibinfo {volume} {9}},\ \bibinfo {pages}
  {570--575} (\bibinfo {year} {2013})},\ \bibinfo {note} {number: 9 Publisher:
  Nature Publishing Group}\BibitemShut {NoStop}%
\bibitem [{\citenamefont {Vinet}\ \emph {et~al.}(2021)\citenamefont {Vinet},
  \citenamefont {Bédécarrats}, \citenamefont {Paz}, \citenamefont {Martinez},
  \citenamefont {Chanrion}, \citenamefont {Catapano}, \citenamefont {Contamin},
  \citenamefont {Pallegoix}, \citenamefont {Venitucci}, \citenamefont
  {Mazzocchi}, \citenamefont {Niebojewski}, \citenamefont {Bertrand},
  \citenamefont {Rambal}, \citenamefont {Thomas}, \citenamefont {Charbonnier},
  \citenamefont {Mortemousque}, \citenamefont {Hartmann}, \citenamefont
  {Nowak}, \citenamefont {Thonnart}, \citenamefont {Billiot}, \citenamefont
  {Cassé}, \citenamefont {Urdampilleta}, \citenamefont {Niquet}, \citenamefont
  {Perruchot}, \citenamefont {De~Franceschi},\ and\ \citenamefont
  {Meunier}}]{material_integration_challenges_si_qubits}%
  \BibitemOpen
  \bibfield  {author} {\bibinfo {author} {\bibfnamefont {M.}~\bibnamefont
  {Vinet}}, \bibinfo {author} {\bibfnamefont {T.}~\bibnamefont
  {Bédécarrats}}, \bibinfo {author} {\bibfnamefont {B.~C.}\ \bibnamefont
  {Paz}}, \bibinfo {author} {\bibfnamefont {B.}~\bibnamefont {Martinez}},
  \bibinfo {author} {\bibfnamefont {E.}~\bibnamefont {Chanrion}}, \bibinfo
  {author} {\bibfnamefont {E.}~\bibnamefont {Catapano}}, \bibinfo {author}
  {\bibfnamefont {L.}~\bibnamefont {Contamin}}, \bibinfo {author}
  {\bibfnamefont {L.}~\bibnamefont {Pallegoix}}, \bibinfo {author}
  {\bibfnamefont {B.}~\bibnamefont {Venitucci}}, \bibinfo {author}
  {\bibfnamefont {V.}~\bibnamefont {Mazzocchi}}, \bibinfo {author}
  {\bibfnamefont {H.}~\bibnamefont {Niebojewski}}, \bibinfo {author}
  {\bibfnamefont {B.}~\bibnamefont {Bertrand}}, \bibinfo {author}
  {\bibfnamefont {N.}~\bibnamefont {Rambal}}, \bibinfo {author} {\bibfnamefont
  {C.}~\bibnamefont {Thomas}}, \bibinfo {author} {\bibfnamefont
  {J.}~\bibnamefont {Charbonnier}}, \bibinfo {author} {\bibfnamefont {P.-A.}\
  \bibnamefont {Mortemousque}}, \bibinfo {author} {\bibfnamefont {J.-M.}\
  \bibnamefont {Hartmann}}, \bibinfo {author} {\bibfnamefont {E.}~\bibnamefont
  {Nowak}}, \bibinfo {author} {\bibfnamefont {Y.}~\bibnamefont {Thonnart}},
  \bibinfo {author} {\bibfnamefont {G.}~\bibnamefont {Billiot}}, \bibinfo
  {author} {\bibfnamefont {M.}~\bibnamefont {Cassé}}, \bibinfo {author}
  {\bibfnamefont {M.}~\bibnamefont {Urdampilleta}}, \bibinfo {author}
  {\bibfnamefont {Y.-M.}\ \bibnamefont {Niquet}}, \bibinfo {author}
  {\bibfnamefont {F.}~\bibnamefont {Perruchot}}, \bibinfo {author}
  {\bibfnamefont {S.}~\bibnamefont {De~Franceschi}}, \ and\ \bibinfo {author}
  {\bibfnamefont {T.}~\bibnamefont {Meunier}},\ }\bibfield  {title} {\enquote
  {\bibinfo {title} {Material and integration challenges for large scale {Si}
  quantum computing},}\ }in\ \href {\doibase 10.1109/IEDM19574.2021.9720708}
  {\emph {\bibinfo {booktitle} {2021 {IEEE} {International} {Electron}
  {Devices} {Meeting} ({IEDM})}}}\ (\bibinfo {year} {2021})\ pp.\ \bibinfo
  {pages} {14.2.1--14.2.4}\BibitemShut {NoStop}%
\bibitem [{\citenamefont {Keith}\ \emph {et~al.}(2022)\citenamefont {Keith},
  \citenamefont {Gorman}, \citenamefont {He}, \citenamefont {Kranz},\ and\
  \citenamefont {Simmons}}]{fc_cn2}%
  \BibitemOpen
  \bibfield  {author} {\bibinfo {author} {\bibfnamefont {D.}~\bibnamefont
  {Keith}}, \bibinfo {author} {\bibfnamefont {S.~K.}\ \bibnamefont {Gorman}},
  \bibinfo {author} {\bibfnamefont {Y.}~\bibnamefont {He}}, \bibinfo {author}
  {\bibfnamefont {L.}~\bibnamefont {Kranz}}, \ and\ \bibinfo {author}
  {\bibfnamefont {M.~Y.}\ \bibnamefont {Simmons}},\ }\bibfield  {title}
  {\enquote {\bibinfo {title} {Impact of charge noise on electron exchange
  interactions in semiconductors},}\ }\href {\doibase
  10.1038/s41534-022-00523-5} {\bibfield  {journal} {\bibinfo  {journal} {npj
  Quantum Information}\ }\textbf {\bibinfo {volume} {8}},\ \bibinfo {pages}
  {1--8} (\bibinfo {year} {2022})},\ \bibinfo {note} {number: 1 Publisher:
  Nature Publishing Group}\BibitemShut {NoStop}%
\bibitem [{\citenamefont {Shehata}\ \emph {et~al.}(2023)\citenamefont
  {Shehata}, \citenamefont {Simion}, \citenamefont {Li}, \citenamefont
  {Mohiyaddin}, \citenamefont {Wan}, \citenamefont {Mongillo}, \citenamefont
  {Govoreanu}, \citenamefont {Radu}, \citenamefont {De~Greve},\ and\
  \citenamefont {Van~Dorpe}}]{fc_cn3}%
  \BibitemOpen
  \bibfield  {author} {\bibinfo {author} {\bibfnamefont {M.~M. E.~K.}\
  \bibnamefont {Shehata}}, \bibinfo {author} {\bibfnamefont {G.}~\bibnamefont
  {Simion}}, \bibinfo {author} {\bibfnamefont {R.}~\bibnamefont {Li}}, \bibinfo
  {author} {\bibfnamefont {F.~A.}\ \bibnamefont {Mohiyaddin}}, \bibinfo
  {author} {\bibfnamefont {D.}~\bibnamefont {Wan}}, \bibinfo {author}
  {\bibfnamefont {M.}~\bibnamefont {Mongillo}}, \bibinfo {author}
  {\bibfnamefont {B.}~\bibnamefont {Govoreanu}}, \bibinfo {author}
  {\bibfnamefont {I.}~\bibnamefont {Radu}}, \bibinfo {author} {\bibfnamefont
  {K.}~\bibnamefont {De~Greve}}, \ and\ \bibinfo {author} {\bibfnamefont
  {P.}~\bibnamefont {Van~Dorpe}},\ }\bibfield  {title} {\enquote {\bibinfo
  {title} {Modeling semiconductor spin qubits and their charge noise
  environment for quantum gate fidelity estimation},}\ }\href {\doibase
  10.1103/PhysRevB.108.045305} {\bibfield  {journal} {\bibinfo  {journal}
  {Physical Review B}\ }\textbf {\bibinfo {volume} {108}},\ \bibinfo {pages}
  {045305} (\bibinfo {year} {2023})},\ \bibinfo {note} {publisher: American
  Physical Society}\BibitemShut {NoStop}%
\bibitem [{\citenamefont {Lundstrom}(2017)}]{fon}%
  \BibitemOpen
  \bibfield  {author} {\bibinfo {author} {\bibfnamefont {M.}~\bibnamefont
  {Lundstrom}},\ }\href
  {https://www.worldscientific.com/worldscibooks/10.1142/9018} {\emph {\bibinfo
  {title} {Fundamentals of {Nanotransistors}}}},\ \bibinfo {series} {Lessons
  from {Nanoscience}: {A} {Lecture} {Notes} {Series}}, Vol.~\bibinfo {volume}
  {06}\ (\bibinfo  {publisher} {WORLD SCIENTIFIC},\ \bibinfo {year}
  {2017})\BibitemShut {NoStop}%
\bibitem [{\citenamefont {Zwanenburg}\ \emph {et~al.}(2013)\citenamefont
  {Zwanenburg}, \citenamefont {Dzurak}, \citenamefont {Morello}, \citenamefont
  {Simmons}, \citenamefont {Hollenberg}, \citenamefont {Klimeck}, \citenamefont
  {Rogge}, \citenamefont {Coppersmith},\ and\ \citenamefont {Eriksson}}]{sqe}%
  \BibitemOpen
  \bibfield  {author} {\bibinfo {author} {\bibfnamefont {F.~A.}\ \bibnamefont
  {Zwanenburg}}, \bibinfo {author} {\bibfnamefont {A.~S.}\ \bibnamefont
  {Dzurak}}, \bibinfo {author} {\bibfnamefont {A.}~\bibnamefont {Morello}},
  \bibinfo {author} {\bibfnamefont {M.~Y.}\ \bibnamefont {Simmons}}, \bibinfo
  {author} {\bibfnamefont {L.~C.~L.}\ \bibnamefont {Hollenberg}}, \bibinfo
  {author} {\bibfnamefont {G.}~\bibnamefont {Klimeck}}, \bibinfo {author}
  {\bibfnamefont {S.}~\bibnamefont {Rogge}}, \bibinfo {author} {\bibfnamefont
  {S.~N.}\ \bibnamefont {Coppersmith}}, \ and\ \bibinfo {author} {\bibfnamefont
  {M.~A.}\ \bibnamefont {Eriksson}},\ }\bibfield  {title} {\enquote {\bibinfo
  {title} {Silicon quantum electronics},}\ }\href {\doibase
  10.1103/RevModPhys.85.961} {\bibfield  {journal} {\bibinfo  {journal}
  {Reviews of Modern Physics}\ }\textbf {\bibinfo {volume} {85}},\ \bibinfo
  {pages} {961--1019} (\bibinfo {year} {2013})}\BibitemShut {NoStop}%
\bibitem [{\citenamefont {Buterakos}\ and\ \citenamefont
  {Das~Sarma}(2021)}]{ch2_valleys_bad}%
  \BibitemOpen
  \bibfield  {author} {\bibinfo {author} {\bibfnamefont {D.}~\bibnamefont
  {Buterakos}}\ and\ \bibinfo {author} {\bibfnamefont {S.}~\bibnamefont
  {Das~Sarma}},\ }\bibfield  {title} {\enquote {\bibinfo {title} {Spin-{Valley}
  {Qubit} {Dynamics} in {Exchange}-{Coupled} {Silicon} {Quantum} {Dots}},}\
  }\href {\doibase 10.1103/PRXQuantum.2.040358} {\bibfield  {journal} {\bibinfo
   {journal} {PRX Quantum}\ }\textbf {\bibinfo {volume} {2}},\ \bibinfo {pages}
  {040358} (\bibinfo {year} {2021})}\BibitemShut {NoStop}%
\bibitem [{\citenamefont {Penthorn}\ \emph {et~al.}(2019)\citenamefont
  {Penthorn}, \citenamefont {Schoenfield}, \citenamefont {Rooney},
  \citenamefont {Edge},\ and\ \citenamefont {Jiang}}]{ch2_valley_qubit1}%
  \BibitemOpen
  \bibfield  {author} {\bibinfo {author} {\bibfnamefont {N.~E.}\ \bibnamefont
  {Penthorn}}, \bibinfo {author} {\bibfnamefont {J.~S.}\ \bibnamefont
  {Schoenfield}}, \bibinfo {author} {\bibfnamefont {J.~D.}\ \bibnamefont
  {Rooney}}, \bibinfo {author} {\bibfnamefont {L.~F.}\ \bibnamefont {Edge}}, \
  and\ \bibinfo {author} {\bibfnamefont {H.}~\bibnamefont {Jiang}},\ }\bibfield
   {title} {\enquote {\bibinfo {title} {Two-axis quantum control of a fast
  valley qubit in silicon},}\ }\href {\doibase 10.1038/s41534-019-0212-5}
  {\bibfield  {journal} {\bibinfo  {journal} {npj Quantum Information}\
  }\textbf {\bibinfo {volume} {5}},\ \bibinfo {pages} {1--6} (\bibinfo {year}
  {2019})}\BibitemShut {NoStop}%
\bibitem [{\citenamefont {Huang}\ and\ \citenamefont
  {Hu}(2021)}]{ch2_valley_qubit2}%
  \BibitemOpen
  \bibfield  {author} {\bibinfo {author} {\bibfnamefont {P.}~\bibnamefont
  {Huang}}\ and\ \bibinfo {author} {\bibfnamefont {X.}~\bibnamefont {Hu}},\
  }\bibfield  {title} {\enquote {\bibinfo {title} {Fast spin-valley-based
  quantum gates in {Si} with micromagnets},}\ }\href {\doibase
  10.1038/s41534-021-00500-4} {\bibfield  {journal} {\bibinfo  {journal} {npj
  Quantum Information}\ }\textbf {\bibinfo {volume} {7}},\ \bibinfo {pages}
  {1--8} (\bibinfo {year} {2021})}\BibitemShut {NoStop}%
\bibitem [{\citenamefont {Jock}\ \emph {et~al.}(2022)\citenamefont {Jock},
  \citenamefont {Jacobson}, \citenamefont {Rudolph}, \citenamefont {Ward},
  \citenamefont {Carroll},\ and\ \citenamefont {Luhman}}]{ch2_valley_qubit3}%
  \BibitemOpen
  \bibfield  {author} {\bibinfo {author} {\bibfnamefont {R.~M.}\ \bibnamefont
  {Jock}}, \bibinfo {author} {\bibfnamefont {N.~T.}\ \bibnamefont {Jacobson}},
  \bibinfo {author} {\bibfnamefont {M.}~\bibnamefont {Rudolph}}, \bibinfo
  {author} {\bibfnamefont {D.~R.}\ \bibnamefont {Ward}}, \bibinfo {author}
  {\bibfnamefont {M.~S.}\ \bibnamefont {Carroll}}, \ and\ \bibinfo {author}
  {\bibfnamefont {D.~R.}\ \bibnamefont {Luhman}},\ }\bibfield  {title}
  {\enquote {\bibinfo {title} {A silicon singlet–triplet qubit driven by
  spin-valley coupling},}\ }\href {\doibase 10.1038/s41467-022-28302-y}
  {\bibfield  {journal} {\bibinfo  {journal} {Nature Communications}\ }\textbf
  {\bibinfo {volume} {13}},\ \bibinfo {pages} {641} (\bibinfo {year}
  {2022})}\BibitemShut {NoStop}%
\bibitem [{\citenamefont {Elsayed}\ \emph {et~al.}(2022)\citenamefont
  {Elsayed}, \citenamefont {Shehata}, \citenamefont {Godfrin}, \citenamefont
  {Kubicek}, \citenamefont {Massar}, \citenamefont {Canvel}, \citenamefont
  {Jussot}, \citenamefont {Simion}, \citenamefont {Mongillo}, \citenamefont
  {Wan}, \citenamefont {Govoreanu}, \citenamefont {Radu}, \citenamefont {Li},
  \citenamefont {Dorpe},\ and\ \citenamefont {De~Greve}}]{asser}%
  \BibitemOpen
  \bibfield  {author} {\bibinfo {author} {\bibfnamefont {A.}~\bibnamefont
  {Elsayed}}, \bibinfo {author} {\bibfnamefont {M.~M.~E.}\ \bibnamefont
  {Shehata}}, \bibinfo {author} {\bibfnamefont {C.}~\bibnamefont {Godfrin}},
  \bibinfo {author} {\bibfnamefont {S.}~\bibnamefont {Kubicek}}, \bibinfo
  {author} {\bibfnamefont {S.}~\bibnamefont {Massar}}, \bibinfo {author}
  {\bibfnamefont {Y.}~\bibnamefont {Canvel}}, \bibinfo {author} {\bibfnamefont
  {J.}~\bibnamefont {Jussot}}, \bibinfo {author} {\bibfnamefont
  {G.}~\bibnamefont {Simion}}, \bibinfo {author} {\bibfnamefont
  {M.}~\bibnamefont {Mongillo}}, \bibinfo {author} {\bibfnamefont
  {D.}~\bibnamefont {Wan}}, \bibinfo {author} {\bibfnamefont {B.}~\bibnamefont
  {Govoreanu}}, \bibinfo {author} {\bibfnamefont {I.~P.}\ \bibnamefont {Radu}},
  \bibinfo {author} {\bibfnamefont {R.}~\bibnamefont {Li}}, \bibinfo {author}
  {\bibfnamefont {P.~V.}\ \bibnamefont {Dorpe}}, \ and\ \bibinfo {author}
  {\bibfnamefont {K.}~\bibnamefont {De~Greve}},\ }\href {\doibase
  10.21203/rs.3.rs-2297196/v1} {\enquote {\bibinfo {title} {Low charge noise
  quantum dots with industrial {CMOS} manufacturing},}\ }\bibinfo {type}
  {preprint}\ (\bibinfo  {institution} {In Review},\ \bibinfo {year}
  {2022})\BibitemShut {NoStop}%
\bibitem [{\citenamefont {Cifuentes}\ \emph {et~al.}(2023)\citenamefont
  {Cifuentes}, \citenamefont {Tanttu}, \citenamefont {Gilbert}, \citenamefont
  {Huang}, \citenamefont {Vahapoglu}, \citenamefont {Leon}, \citenamefont
  {Serrano}, \citenamefont {Otter}, \citenamefont {Dunmore}, \citenamefont
  {Mai}, \citenamefont {Schlattner}, \citenamefont {Feng}, \citenamefont
  {Itoh}, \citenamefont {Abrosimov}, \citenamefont {Pohl}, \citenamefont
  {Thewalt}, \citenamefont {Laucht}, \citenamefont {Yang}, \citenamefont
  {Escott}, \citenamefont {Lim}, \citenamefont {Hudson}, \citenamefont
  {Rahman}, \citenamefont {Dzurak},\ and\ \citenamefont
  {Saraiva}}]{si_variability}%
  \BibitemOpen
  \bibfield  {author} {\bibinfo {author} {\bibfnamefont {J.~D.}\ \bibnamefont
  {Cifuentes}}, \bibinfo {author} {\bibfnamefont {T.}~\bibnamefont {Tanttu}},
  \bibinfo {author} {\bibfnamefont {W.}~\bibnamefont {Gilbert}}, \bibinfo
  {author} {\bibfnamefont {J.~Y.}\ \bibnamefont {Huang}}, \bibinfo {author}
  {\bibfnamefont {E.}~\bibnamefont {Vahapoglu}}, \bibinfo {author}
  {\bibfnamefont {R.~C.~C.}\ \bibnamefont {Leon}}, \bibinfo {author}
  {\bibfnamefont {S.}~\bibnamefont {Serrano}}, \bibinfo {author} {\bibfnamefont
  {D.}~\bibnamefont {Otter}}, \bibinfo {author} {\bibfnamefont
  {D.}~\bibnamefont {Dunmore}}, \bibinfo {author} {\bibfnamefont {P.~Y.}\
  \bibnamefont {Mai}}, \bibinfo {author} {\bibfnamefont {F.}~\bibnamefont
  {Schlattner}}, \bibinfo {author} {\bibfnamefont {M.}~\bibnamefont {Feng}},
  \bibinfo {author} {\bibfnamefont {K.}~\bibnamefont {Itoh}}, \bibinfo {author}
  {\bibfnamefont {N.}~\bibnamefont {Abrosimov}}, \bibinfo {author}
  {\bibfnamefont {H.-J.}\ \bibnamefont {Pohl}}, \bibinfo {author}
  {\bibfnamefont {M.}~\bibnamefont {Thewalt}}, \bibinfo {author} {\bibfnamefont
  {A.}~\bibnamefont {Laucht}}, \bibinfo {author} {\bibfnamefont {C.~H.}\
  \bibnamefont {Yang}}, \bibinfo {author} {\bibfnamefont {C.~C.}\ \bibnamefont
  {Escott}}, \bibinfo {author} {\bibfnamefont {W.~H.}\ \bibnamefont {Lim}},
  \bibinfo {author} {\bibfnamefont {F.~E.}\ \bibnamefont {Hudson}}, \bibinfo
  {author} {\bibfnamefont {R.}~\bibnamefont {Rahman}}, \bibinfo {author}
  {\bibfnamefont {A.~S.}\ \bibnamefont {Dzurak}}, \ and\ \bibinfo {author}
  {\bibfnamefont {A.}~\bibnamefont {Saraiva}},\ }\href {\doibase
  10.48550/arXiv.2303.14864} {\enquote {\bibinfo {title} {Bounds to electron
  spin qubit variability for scalable {CMOS} architectures},}\ } (\bibinfo
  {year} {2023})\BibitemShut {NoStop}%
\bibitem [{\citenamefont {Peña}\ \emph {et~al.}(2024)\citenamefont {Peña},
  \citenamefont {Koepke}, \citenamefont {Dycus}, \citenamefont {Mounce},
  \citenamefont {Baczewski}, \citenamefont {Jacobson},\ and\ \citenamefont
  {Bussmann}}]{modeling_sige_qd_variability}%
  \BibitemOpen
  \bibfield  {author} {\bibinfo {author} {\bibfnamefont {L.~F.}\ \bibnamefont
  {Peña}}, \bibinfo {author} {\bibfnamefont {J.~C.}\ \bibnamefont {Koepke}},
  \bibinfo {author} {\bibfnamefont {J.~H.}\ \bibnamefont {Dycus}}, \bibinfo
  {author} {\bibfnamefont {A.}~\bibnamefont {Mounce}}, \bibinfo {author}
  {\bibfnamefont {A.~D.}\ \bibnamefont {Baczewski}}, \bibinfo {author}
  {\bibfnamefont {N.~T.}\ \bibnamefont {Jacobson}}, \ and\ \bibinfo {author}
  {\bibfnamefont {E.}~\bibnamefont {Bussmann}},\ }\bibfield  {title} {\enquote
  {\bibinfo {title} {Modeling {Si}/{SiGe} quantum dot variability induced by
  interface disorder reconstructed from multiperspective microscopy},}\ }\href
  {\doibase 10.1038/s41534-024-00827-8} {\bibfield  {journal} {\bibinfo
  {journal} {npj Quantum Information}\ }\textbf {\bibinfo {volume} {10}},\
  \bibinfo {pages} {1--10} (\bibinfo {year} {2024})}\BibitemShut {NoStop}%
\bibitem [{\citenamefont {De~Leon}\ \emph {et~al.}(2021)\citenamefont
  {De~Leon}, \citenamefont {Itoh}, \citenamefont {Kim}, \citenamefont {Mehta},
  \citenamefont {Northup}, \citenamefont {Paik}, \citenamefont {Palmer},
  \citenamefont {Samarth}, \citenamefont {Sangtawesin},\ and\ \citenamefont
  {Steuerman}}]{material_challenges_for_qc}%
  \BibitemOpen
  \bibfield  {author} {\bibinfo {author} {\bibfnamefont {N.~P.}\ \bibnamefont
  {De~Leon}}, \bibinfo {author} {\bibfnamefont {K.~M.}\ \bibnamefont {Itoh}},
  \bibinfo {author} {\bibfnamefont {D.}~\bibnamefont {Kim}}, \bibinfo {author}
  {\bibfnamefont {K.~K.}\ \bibnamefont {Mehta}}, \bibinfo {author}
  {\bibfnamefont {T.~E.}\ \bibnamefont {Northup}}, \bibinfo {author}
  {\bibfnamefont {H.}~\bibnamefont {Paik}}, \bibinfo {author} {\bibfnamefont
  {B.~S.}\ \bibnamefont {Palmer}}, \bibinfo {author} {\bibfnamefont
  {N.}~\bibnamefont {Samarth}}, \bibinfo {author} {\bibfnamefont
  {S.}~\bibnamefont {Sangtawesin}}, \ and\ \bibinfo {author} {\bibfnamefont
  {D.~W.}\ \bibnamefont {Steuerman}},\ }\bibfield  {title} {\enquote {\bibinfo
  {title} {Materials challenges and opportunities for quantum computing
  hardware},}\ }\href {\doibase 10.1126/science.abb2823} {\bibfield  {journal}
  {\bibinfo  {journal} {Science}\ }\textbf {\bibinfo {volume} {372}},\ \bibinfo
  {pages} {eabb2823} (\bibinfo {year} {2021})}\BibitemShut {NoStop}%
\bibitem [{\citenamefont {Degli~Esposti}\ \emph {et~al.}(2024)\citenamefont
  {Degli~Esposti}, \citenamefont {Stehouwer}, \citenamefont {Gul},
  \citenamefont {Samkharadze}, \citenamefont {Deprez}, \citenamefont {Meyer},
  \citenamefont {Meijer}, \citenamefont {Tryputen}, \citenamefont {Karwal},
  \citenamefont {Botifoll}, \citenamefont {Arbiol}, \citenamefont {Amitonov},
  \citenamefont {Vandersypen}, \citenamefont {Sammak}, \citenamefont
  {Veldhorst},\ and\ \citenamefont
  {Scappucci}}]{sige_disorder_valley_mitigation}%
  \BibitemOpen
  \bibfield  {author} {\bibinfo {author} {\bibfnamefont {D.}~\bibnamefont
  {Degli~Esposti}}, \bibinfo {author} {\bibfnamefont {L.~E.~A.}\ \bibnamefont
  {Stehouwer}}, \bibinfo {author} {\bibfnamefont {O.}~\bibnamefont {Gul}},
  \bibinfo {author} {\bibfnamefont {N.}~\bibnamefont {Samkharadze}}, \bibinfo
  {author} {\bibfnamefont {C.}~\bibnamefont {Deprez}}, \bibinfo {author}
  {\bibfnamefont {M.}~\bibnamefont {Meyer}}, \bibinfo {author} {\bibfnamefont
  {I.~N.}\ \bibnamefont {Meijer}}, \bibinfo {author} {\bibfnamefont
  {L.}~\bibnamefont {Tryputen}}, \bibinfo {author} {\bibfnamefont
  {S.}~\bibnamefont {Karwal}}, \bibinfo {author} {\bibfnamefont
  {M.}~\bibnamefont {Botifoll}}, \bibinfo {author} {\bibfnamefont
  {J.}~\bibnamefont {Arbiol}}, \bibinfo {author} {\bibfnamefont {S.~V.}\
  \bibnamefont {Amitonov}}, \bibinfo {author} {\bibfnamefont {L.~M.~K.}\
  \bibnamefont {Vandersypen}}, \bibinfo {author} {\bibfnamefont
  {A.}~\bibnamefont {Sammak}}, \bibinfo {author} {\bibfnamefont
  {M.}~\bibnamefont {Veldhorst}}, \ and\ \bibinfo {author} {\bibfnamefont
  {G.}~\bibnamefont {Scappucci}},\ }\bibfield  {title} {\enquote {\bibinfo
  {title} {Low disorder and high valley splitting in silicon},}\ }\href
  {\doibase 10.1038/s41534-024-00826-9} {\bibfield  {journal} {\bibinfo
  {journal} {npj Quantum Information}\ }\textbf {\bibinfo {volume} {10}},\
  \bibinfo {pages} {1--9} (\bibinfo {year} {2024})}\BibitemShut {NoStop}%
\bibitem [{\citenamefont {Reed}\ \emph {et~al.}(2016)\citenamefont {Reed},
  \citenamefont {Maune}, \citenamefont {Andrews}, \citenamefont {Borselli},
  \citenamefont {Eng}, \citenamefont {Jura}, \citenamefont {Kiselev},
  \citenamefont {Ladd}, \citenamefont {Merkel}, \citenamefont {Milosavljevic},
  \citenamefont {Pritchett}, \citenamefont {Rakher}, \citenamefont {Ross},
  \citenamefont {Schmitz}, \citenamefont {Smith}, \citenamefont {Wright},
  \citenamefont {Gyure},\ and\ \citenamefont {Hunter}}]{sweet_eo}%
  \BibitemOpen
  \bibfield  {author} {\bibinfo {author} {\bibfnamefont {M.}~\bibnamefont
  {Reed}}, \bibinfo {author} {\bibfnamefont {B.}~\bibnamefont {Maune}},
  \bibinfo {author} {\bibfnamefont {R.}~\bibnamefont {Andrews}}, \bibinfo
  {author} {\bibfnamefont {M.}~\bibnamefont {Borselli}}, \bibinfo {author}
  {\bibfnamefont {K.}~\bibnamefont {Eng}}, \bibinfo {author} {\bibfnamefont
  {M.}~\bibnamefont {Jura}}, \bibinfo {author} {\bibfnamefont {A.}~\bibnamefont
  {Kiselev}}, \bibinfo {author} {\bibfnamefont {T.}~\bibnamefont {Ladd}},
  \bibinfo {author} {\bibfnamefont {S.}~\bibnamefont {Merkel}}, \bibinfo
  {author} {\bibfnamefont {I.}~\bibnamefont {Milosavljevic}}, \bibinfo {author}
  {\bibfnamefont {E.}~\bibnamefont {Pritchett}}, \bibinfo {author}
  {\bibfnamefont {M.}~\bibnamefont {Rakher}}, \bibinfo {author} {\bibfnamefont
  {R.}~\bibnamefont {Ross}}, \bibinfo {author} {\bibfnamefont {A.}~\bibnamefont
  {Schmitz}}, \bibinfo {author} {\bibfnamefont {A.}~\bibnamefont {Smith}},
  \bibinfo {author} {\bibfnamefont {J.}~\bibnamefont {Wright}}, \bibinfo
  {author} {\bibfnamefont {M.}~\bibnamefont {Gyure}}, \ and\ \bibinfo {author}
  {\bibfnamefont {A.}~\bibnamefont {Hunter}},\ }\bibfield  {title} {\enquote
  {\bibinfo {title} {Reduced {Sensitivity} to {Charge} {Noise} in
  {Semiconductor} {Spin} {Qubits} via {Symmetric} {Operation}},}\ }\href
  {\doibase 10.1103/PhysRevLett.116.110402} {\bibfield  {journal} {\bibinfo
  {journal} {Physical Review Letters}\ }\textbf {\bibinfo {volume} {116}},\
  \bibinfo {pages} {110402} (\bibinfo {year} {2016})}\BibitemShut {NoStop}%
\bibitem [{\citenamefont {Piot}\ \emph {et~al.}(2022)\citenamefont {Piot},
  \citenamefont {Brun}, \citenamefont {Schmitt}, \citenamefont {Zihlmann},
  \citenamefont {Michal}, \citenamefont {Apra}, \citenamefont {Abadillo-Uriel},
  \citenamefont {Jehl}, \citenamefont {Bertrand}, \citenamefont {Niebojewski},
  \citenamefont {Hutin}, \citenamefont {Vinet}, \citenamefont {Urdampilleta},
  \citenamefont {Meunier}, \citenamefont {Niquet}, \citenamefont {Maurand},\
  and\ \citenamefont {Franceschi}}]{silvano_6}%
  \BibitemOpen
  \bibfield  {author} {\bibinfo {author} {\bibfnamefont {N.}~\bibnamefont
  {Piot}}, \bibinfo {author} {\bibfnamefont {B.}~\bibnamefont {Brun}}, \bibinfo
  {author} {\bibfnamefont {V.}~\bibnamefont {Schmitt}}, \bibinfo {author}
  {\bibfnamefont {S.}~\bibnamefont {Zihlmann}}, \bibinfo {author}
  {\bibfnamefont {V.~P.}\ \bibnamefont {Michal}}, \bibinfo {author}
  {\bibfnamefont {A.}~\bibnamefont {Apra}}, \bibinfo {author} {\bibfnamefont
  {J.~C.}\ \bibnamefont {Abadillo-Uriel}}, \bibinfo {author} {\bibfnamefont
  {X.}~\bibnamefont {Jehl}}, \bibinfo {author} {\bibfnamefont {B.}~\bibnamefont
  {Bertrand}}, \bibinfo {author} {\bibfnamefont {H.}~\bibnamefont
  {Niebojewski}}, \bibinfo {author} {\bibfnamefont {L.}~\bibnamefont {Hutin}},
  \bibinfo {author} {\bibfnamefont {M.}~\bibnamefont {Vinet}}, \bibinfo
  {author} {\bibfnamefont {M.}~\bibnamefont {Urdampilleta}}, \bibinfo {author}
  {\bibfnamefont {T.}~\bibnamefont {Meunier}}, \bibinfo {author} {\bibfnamefont
  {Y.-M.}\ \bibnamefont {Niquet}}, \bibinfo {author} {\bibfnamefont
  {R.}~\bibnamefont {Maurand}}, \ and\ \bibinfo {author} {\bibfnamefont
  {S.~D.}\ \bibnamefont {Franceschi}},\ }\bibfield  {title} {\enquote {\bibinfo
  {title} {A single hole spin with enhanced coherence in natural silicon},}\
  }\href {\doibase 10.1038/s41565-022-01196-z} {\bibfield  {journal} {\bibinfo
  {journal} {Nature Nanotechnology}\ }\textbf {\bibinfo {volume} {17}},\
  \bibinfo {pages} {1072--1077} (\bibinfo {year} {2022})}\BibitemShut {NoStop}%
\bibitem [{\citenamefont {Camenzind}\ \emph {et~al.}(2021)\citenamefont
  {Camenzind}, \citenamefont {Elsayed}, \citenamefont {Mohiyaddin},
  \citenamefont {Li}, \citenamefont {Kubicek}, \citenamefont {Jussot},
  \citenamefont {Dorpe}, \citenamefont {Govoreanu}, \citenamefont {Radu},\ and\
  \citenamefont {Zumbühl}}]{gate_stack_disorder}%
  \BibitemOpen
  \bibfield  {author} {\bibinfo {author} {\bibfnamefont {T.~N.}\ \bibnamefont
  {Camenzind}}, \bibinfo {author} {\bibfnamefont {A.}~\bibnamefont {Elsayed}},
  \bibinfo {author} {\bibfnamefont {F.~A.}\ \bibnamefont {Mohiyaddin}},
  \bibinfo {author} {\bibfnamefont {R.}~\bibnamefont {Li}}, \bibinfo {author}
  {\bibfnamefont {S.}~\bibnamefont {Kubicek}}, \bibinfo {author} {\bibfnamefont
  {J.}~\bibnamefont {Jussot}}, \bibinfo {author} {\bibfnamefont {P.~V.}\
  \bibnamefont {Dorpe}}, \bibinfo {author} {\bibfnamefont {B.}~\bibnamefont
  {Govoreanu}}, \bibinfo {author} {\bibfnamefont {I.}~\bibnamefont {Radu}}, \
  and\ \bibinfo {author} {\bibfnamefont {D.~M.}\ \bibnamefont {Zumbühl}},\
  }\bibfield  {title} {\enquote {\bibinfo {title} {High mobility {SiMOSFETs}
  fabricated in a full 300 mm {CMOS} process},}\ }\href {\doibase
  10.1088/2633-4356/ac40f4} {\bibfield  {journal} {\bibinfo  {journal}
  {Materials for Quantum Technology}\ }\textbf {\bibinfo {volume} {1}},\
  \bibinfo {pages} {041001} (\bibinfo {year} {2021})}\BibitemShut {NoStop}%
\bibitem [{\citenamefont {Iyer}(2022)}]{no_more_poly}%
  \BibitemOpen
  \bibfield  {author} {\bibinfo {author} {\bibfnamefont {S.}~\bibnamefont
  {Iyer}},\ }\bibfield  {title} {\enquote {\bibinfo {title} {Lemons are for
  {Lemonade}},}\ }\href
  {https://www.src.org/calendar/e007658/e007658_presentation_iyer.pdf}
  {\bibfield  {journal} {\bibinfo  {journal} {SRC Industry-led Talk:
  https://www.src.org/calendar/e007658/e007658\_presentation\_iyer.pdf}\ }
  (\bibinfo {year} {2022})}\BibitemShut {NoStop}%
\bibitem [{\citenamefont {Penny}\ \emph {et~al.}(2022)\citenamefont {Penny},
  \citenamefont {Motoyama}, \citenamefont {Ghosh}, \citenamefont {Bae},
  \citenamefont {Lanzillo}, \citenamefont {Sieg}, \citenamefont {Park},
  \citenamefont {Zou}, \citenamefont {Lee}, \citenamefont {Metzler},
  \citenamefont {Lee}, \citenamefont {Cho}, \citenamefont {Shoudy},
  \citenamefont {Nguyen}, \citenamefont {Simon}, \citenamefont {Park},
  \citenamefont {Clevenger}, \citenamefont {Anderson}, \citenamefont {Child},
  \citenamefont {Yamashita}, \citenamefont {Arnold}, \citenamefont {Wu},
  \citenamefont {Spooner}, \citenamefont {Choi}, \citenamefont {Seo},\ and\
  \citenamefont {Guo}}]{ruthenium}%
  \BibitemOpen
  \bibfield  {author} {\bibinfo {author} {\bibfnamefont {C.}~\bibnamefont
  {Penny}}, \bibinfo {author} {\bibfnamefont {K.}~\bibnamefont {Motoyama}},
  \bibinfo {author} {\bibfnamefont {S.}~\bibnamefont {Ghosh}}, \bibinfo
  {author} {\bibfnamefont {T.}~\bibnamefont {Bae}}, \bibinfo {author}
  {\bibfnamefont {N.}~\bibnamefont {Lanzillo}}, \bibinfo {author}
  {\bibfnamefont {S.}~\bibnamefont {Sieg}}, \bibinfo {author} {\bibfnamefont
  {C.}~\bibnamefont {Park}}, \bibinfo {author} {\bibfnamefont {L.}~\bibnamefont
  {Zou}}, \bibinfo {author} {\bibfnamefont {H.}~\bibnamefont {Lee}}, \bibinfo
  {author} {\bibfnamefont {D.}~\bibnamefont {Metzler}}, \bibinfo {author}
  {\bibfnamefont {J.}~\bibnamefont {Lee}}, \bibinfo {author} {\bibfnamefont
  {S.}~\bibnamefont {Cho}}, \bibinfo {author} {\bibfnamefont {M.}~\bibnamefont
  {Shoudy}}, \bibinfo {author} {\bibfnamefont {S.}~\bibnamefont {Nguyen}},
  \bibinfo {author} {\bibfnamefont {A.}~\bibnamefont {Simon}}, \bibinfo
  {author} {\bibfnamefont {K.}~\bibnamefont {Park}}, \bibinfo {author}
  {\bibfnamefont {L.}~\bibnamefont {Clevenger}}, \bibinfo {author}
  {\bibfnamefont {B.}~\bibnamefont {Anderson}}, \bibinfo {author}
  {\bibfnamefont {C.}~\bibnamefont {Child}}, \bibinfo {author} {\bibfnamefont
  {T.}~\bibnamefont {Yamashita}}, \bibinfo {author} {\bibfnamefont
  {J.}~\bibnamefont {Arnold}}, \bibinfo {author} {\bibfnamefont
  {T.}~\bibnamefont {Wu}}, \bibinfo {author} {\bibfnamefont {T.}~\bibnamefont
  {Spooner}}, \bibinfo {author} {\bibfnamefont {K.}~\bibnamefont {Choi}},
  \bibinfo {author} {\bibfnamefont {K.-I.}\ \bibnamefont {Seo}}, \ and\
  \bibinfo {author} {\bibfnamefont {D.}~\bibnamefont {Guo}},\ }\bibfield
  {title} {\enquote {\bibinfo {title} {Subtractive {Ru} {Interconnect}
  {Enabled} by {Novel} {Patterning} {Solution} for {EUV} {Double} {Patterning}
  and {TopVia} with {Embedded} {Airgap} {Integration} for {Post} {Cu}
  {Interconnect} {Scaling}},}\ }in\ \href {\doibase
  10.1109/IEDM45625.2022.10019479} {\emph {\bibinfo {booktitle} {2022
  {International} {Electron} {Devices} {Meeting} ({IEDM})}}}\ (\bibinfo {year}
  {2022})\ pp.\ \bibinfo {pages} {12.1.1--12.1.4}\BibitemShut {NoStop}%
\bibitem [{\citenamefont {Patel}()}]{gate_stack_sota_materials}%
  \BibitemOpen
  \bibfield  {author} {\bibinfo {author} {\bibfnamefont {D.}~\bibnamefont
  {Patel}},\ }\href {https://www.semianalysis.com/p/iedm2022p1} {\enquote
  {\bibinfo {title} {{TSMC} 3nm {FinFlex} + {Self}-{Aligned} {Contacts},
  {Intel} {EMIB} 3 + {Foveros} {Direct}, {AMD} {Yield} {Issues}, {IBM}
  {Vertical} {Transport} {FET} ({VTFET}) + {RU} {Interconnects}, {CFET},
  {Sequential} {Stacking}, {Samsung} {Yield}, and more},}\ }\BibitemShut
  {NoStop}%
\bibitem [{\citenamefont {Robertson}\ and\ \citenamefont
  {Wallace}(2015)}]{dielectric_and_metal_materials}%
  \BibitemOpen
  \bibfield  {author} {\bibinfo {author} {\bibfnamefont {J.}~\bibnamefont
  {Robertson}}\ and\ \bibinfo {author} {\bibfnamefont {R.~M.}\ \bibnamefont
  {Wallace}},\ }\bibfield  {title} {\enquote {\bibinfo {title} {High-{K}
  materials and metal gates for {CMOS} applications},}\ }\href {\doibase
  10.1016/j.mser.2014.11.001} {\bibfield  {journal} {\bibinfo  {journal}
  {Materials Science and Engineering: R: Reports}\ }\textbf {\bibinfo {volume}
  {88}},\ \bibinfo {pages} {1--41} (\bibinfo {year} {2015})}\BibitemShut
  {NoStop}%
\bibitem [{\citenamefont {Verberk}\ \emph {et~al.}(2022)\citenamefont
  {Verberk}, \citenamefont {Michalak}, \citenamefont {Versluis}, \citenamefont
  {Polinder}, \citenamefont {Samkharadze}, \citenamefont {Amitonov},
  \citenamefont {Sammak}, \citenamefont {Tryputen}, \citenamefont {Brousse},\
  and\ \citenamefont {Hanfoug}}]{qutech_whitepaper}%
  \BibitemOpen
  \bibfield  {author} {\bibinfo {author} {\bibfnamefont {R.}~\bibnamefont
  {Verberk}}, \bibinfo {author} {\bibfnamefont {D.}~\bibnamefont {Michalak}},
  \bibinfo {author} {\bibfnamefont {R.}~\bibnamefont {Versluis}}, \bibinfo
  {author} {\bibfnamefont {H.}~\bibnamefont {Polinder}}, \bibinfo {author}
  {\bibfnamefont {N.}~\bibnamefont {Samkharadze}}, \bibinfo {author}
  {\bibfnamefont {S.}~\bibnamefont {Amitonov}}, \bibinfo {author}
  {\bibfnamefont {A.}~\bibnamefont {Sammak}}, \bibinfo {author} {\bibfnamefont
  {L.}~\bibnamefont {Tryputen}}, \bibinfo {author} {\bibfnamefont
  {D.}~\bibnamefont {Brousse}}, \ and\ \bibinfo {author} {\bibfnamefont
  {R.}~\bibnamefont {Hanfoug}},\ }\bibfield  {title} {\enquote {\bibinfo
  {title} {Synergy between quantum computing and semiconductor technology},}\
  }in\ \href {\doibase 10.1117/12.2639994} {\emph {\bibinfo {booktitle} {37th
  {European} {Mask} and {Lithography} {Conference}}}},\ \bibinfo {editor}
  {edited by\ \bibinfo {editor} {\bibfnamefont {U.}~\bibnamefont {Behringer}}}\
  (\bibinfo  {publisher} {SPIE},\ \bibinfo {address} {Leuven, Belgium},\
  \bibinfo {year} {2022})\ p.~\bibinfo {pages} {27}\BibitemShut {NoStop}%
\bibitem [{\citenamefont {Dumoulin~Stuyck}\ \emph {et~al.}(2021)\citenamefont
  {Dumoulin~Stuyck}, \citenamefont {Mohiyaddin}, \citenamefont {Li},
  \citenamefont {Heyns}, \citenamefont {Govoreanu},\ and\ \citenamefont
  {Radu}}]{optimizing_magnets}%
  \BibitemOpen
  \bibfield  {author} {\bibinfo {author} {\bibfnamefont {N.~I.}\ \bibnamefont
  {Dumoulin~Stuyck}}, \bibinfo {author} {\bibfnamefont {F.~A.}\ \bibnamefont
  {Mohiyaddin}}, \bibinfo {author} {\bibfnamefont {R.}~\bibnamefont {Li}},
  \bibinfo {author} {\bibfnamefont {M.}~\bibnamefont {Heyns}}, \bibinfo
  {author} {\bibfnamefont {B.}~\bibnamefont {Govoreanu}}, \ and\ \bibinfo
  {author} {\bibfnamefont {I.~P.}\ \bibnamefont {Radu}},\ }\bibfield  {title}
  {\enquote {\bibinfo {title} {Low dephasing and robust micromagnet designs for
  silicon spin qubits},}\ }\href {\doibase 10.1063/5.0059939} {\bibfield
  {journal} {\bibinfo  {journal} {Applied Physics Letters}\ }\textbf {\bibinfo
  {volume} {119}},\ \bibinfo {pages} {094001} (\bibinfo {year}
  {2021})}\BibitemShut {NoStop}%
\bibitem [{\citenamefont {Hu}\ \emph {et~al.}(2021)\citenamefont {Hu},
  \citenamefont {Ma}, \citenamefont {Ni}, \citenamefont {Zhang}, \citenamefont
  {Zhou}, \citenamefont {Wang}, \citenamefont {Luo}, \citenamefont {Cao},
  \citenamefont {Kong}, \citenamefont {Wang}, \citenamefont {Li},\ and\
  \citenamefont {Guo}}]{fc_op_guide_simos_qubits}%
  \BibitemOpen
  \bibfield  {author} {\bibinfo {author} {\bibfnamefont {R.-Z.}\ \bibnamefont
  {Hu}}, \bibinfo {author} {\bibfnamefont {R.-L.}\ \bibnamefont {Ma}}, \bibinfo
  {author} {\bibfnamefont {M.}~\bibnamefont {Ni}}, \bibinfo {author}
  {\bibfnamefont {X.}~\bibnamefont {Zhang}}, \bibinfo {author} {\bibfnamefont
  {Y.}~\bibnamefont {Zhou}}, \bibinfo {author} {\bibfnamefont {K.}~\bibnamefont
  {Wang}}, \bibinfo {author} {\bibfnamefont {G.}~\bibnamefont {Luo}}, \bibinfo
  {author} {\bibfnamefont {G.}~\bibnamefont {Cao}}, \bibinfo {author}
  {\bibfnamefont {Z.-Z.}\ \bibnamefont {Kong}}, \bibinfo {author}
  {\bibfnamefont {G.-L.}\ \bibnamefont {Wang}}, \bibinfo {author}
  {\bibfnamefont {H.-O.}\ \bibnamefont {Li}}, \ and\ \bibinfo {author}
  {\bibfnamefont {G.-P.}\ \bibnamefont {Guo}},\ }\bibfield  {title} {\enquote
  {\bibinfo {title} {An {Operation} {Guide} of {Si}-{MOS} {Quantum} {Dots} for
  {Spin} {Qubits}},}\ }\href {\doibase 10.3390/nano11102486} {\bibfield
  {journal} {\bibinfo  {journal} {Nanomaterials}\ }\textbf {\bibinfo {volume}
  {11}},\ \bibinfo {pages} {2486} (\bibinfo {year} {2021})}\BibitemShut
  {NoStop}%
\bibitem [{\citenamefont {Denisov}\ \emph {et~al.}(2023)\citenamefont
  {Denisov}, \citenamefont {Fuchs}, \citenamefont {Oh},\ and\ \citenamefont
  {Petta}}]{sige_dispersive_readout}%
  \BibitemOpen
  \bibfield  {author} {\bibinfo {author} {\bibfnamefont {A.~O.}\ \bibnamefont
  {Denisov}}, \bibinfo {author} {\bibfnamefont {G.}~\bibnamefont {Fuchs}},
  \bibinfo {author} {\bibfnamefont {S.~W.}\ \bibnamefont {Oh}}, \ and\ \bibinfo
  {author} {\bibfnamefont {J.~R.}\ \bibnamefont {Petta}},\ }\bibfield  {title}
  {\enquote {\bibinfo {title} {Dispersive readout of a silicon quantum device
  using an atomic force microscope-based rf gate sensor},}\ }\href {\doibase
  10.1063/5.0158196} {\bibfield  {journal} {\bibinfo  {journal} {Applied
  Physics Letters}\ }\textbf {\bibinfo {volume} {123}},\ \bibinfo {pages}
  {093502} (\bibinfo {year} {2023})}\BibitemShut {NoStop}%
\bibitem [{\citenamefont {Vigneau}\ \emph {et~al.}(2023)\citenamefont
  {Vigneau}, \citenamefont {Fedele}, \citenamefont {Chatterjee}, \citenamefont
  {Reilly}, \citenamefont {Kuemmeth}, \citenamefont {Gonzalez-Zalba},
  \citenamefont {Laird},\ and\ \citenamefont {Ares}}]{fc_s13}%
  \BibitemOpen
  \bibfield  {author} {\bibinfo {author} {\bibfnamefont {F.}~\bibnamefont
  {Vigneau}}, \bibinfo {author} {\bibfnamefont {F.}~\bibnamefont {Fedele}},
  \bibinfo {author} {\bibfnamefont {A.}~\bibnamefont {Chatterjee}}, \bibinfo
  {author} {\bibfnamefont {D.}~\bibnamefont {Reilly}}, \bibinfo {author}
  {\bibfnamefont {F.}~\bibnamefont {Kuemmeth}}, \bibinfo {author}
  {\bibfnamefont {M.~F.}\ \bibnamefont {Gonzalez-Zalba}}, \bibinfo {author}
  {\bibfnamefont {E.}~\bibnamefont {Laird}}, \ and\ \bibinfo {author}
  {\bibfnamefont {N.}~\bibnamefont {Ares}},\ }\bibfield  {title} {\enquote
  {\bibinfo {title} {Probing quantum devices with radio-frequency
  reflectometry},}\ }\href {\doibase 10.1063/5.0088229} {\bibfield  {journal}
  {\bibinfo  {journal} {Applied Physics Reviews}\ }\textbf {\bibinfo {volume}
  {10}},\ \bibinfo {pages} {021305} (\bibinfo {year} {2023})}\BibitemShut
  {NoStop}%
\bibitem [{\citenamefont {Hutin}\ \emph {et~al.}(2019)\citenamefont {Hutin},
  \citenamefont {Lundberg}, \citenamefont {Chatterjee}, \citenamefont {Crippa},
  \citenamefont {Li}, \citenamefont {Maurand}, \citenamefont {Jehl},
  \citenamefont {Sanquer}, \citenamefont {Gonzalez-Zalba}, \citenamefont
  {Kuemmeth}, \citenamefont {Niquet}, \citenamefont {Bertrand}, \citenamefont
  {De~Franceschi}, \citenamefont {Urdampilleta}, \citenamefont {Meunier},
  \citenamefont {Vinet}, \citenamefont {Chanrion}, \citenamefont
  {Bohuslavskyi}, \citenamefont {Ansaloni}, \citenamefont {Yang}, \citenamefont
  {Michniewicz}, \citenamefont {Niegemann},\ and\ \citenamefont
  {Spence}}]{mine_many}%
  \BibitemOpen
  \bibfield  {author} {\bibinfo {author} {\bibfnamefont {L.}~\bibnamefont
  {Hutin}}, \bibinfo {author} {\bibfnamefont {T.}~\bibnamefont {Lundberg}},
  \bibinfo {author} {\bibfnamefont {A.}~\bibnamefont {Chatterjee}}, \bibinfo
  {author} {\bibfnamefont {A.}~\bibnamefont {Crippa}}, \bibinfo {author}
  {\bibfnamefont {J.}~\bibnamefont {Li}}, \bibinfo {author} {\bibfnamefont
  {R.}~\bibnamefont {Maurand}}, \bibinfo {author} {\bibfnamefont
  {X.}~\bibnamefont {Jehl}}, \bibinfo {author} {\bibfnamefont {M.}~\bibnamefont
  {Sanquer}}, \bibinfo {author} {\bibfnamefont {M.~F.}\ \bibnamefont
  {Gonzalez-Zalba}}, \bibinfo {author} {\bibfnamefont {F.}~\bibnamefont
  {Kuemmeth}}, \bibinfo {author} {\bibfnamefont {Y.-M.}\ \bibnamefont
  {Niquet}}, \bibinfo {author} {\bibfnamefont {B.}~\bibnamefont {Bertrand}},
  \bibinfo {author} {\bibfnamefont {S.}~\bibnamefont {De~Franceschi}}, \bibinfo
  {author} {\bibfnamefont {M.}~\bibnamefont {Urdampilleta}}, \bibinfo {author}
  {\bibfnamefont {T.}~\bibnamefont {Meunier}}, \bibinfo {author} {\bibfnamefont
  {M.}~\bibnamefont {Vinet}}, \bibinfo {author} {\bibfnamefont
  {E.}~\bibnamefont {Chanrion}}, \bibinfo {author} {\bibfnamefont
  {H.}~\bibnamefont {Bohuslavskyi}}, \bibinfo {author} {\bibfnamefont
  {F.}~\bibnamefont {Ansaloni}}, \bibinfo {author} {\bibfnamefont {T.-Y.}\
  \bibnamefont {Yang}}, \bibinfo {author} {\bibfnamefont {J.}~\bibnamefont
  {Michniewicz}}, \bibinfo {author} {\bibfnamefont {D.~J.}\ \bibnamefont
  {Niegemann}}, \ and\ \bibinfo {author} {\bibfnamefont {C.}~\bibnamefont
  {Spence}},\ }\bibfield  {title} {\enquote {\bibinfo {title} {Gate
  reflectometry for probing charge and spin states in linear {Si} {MOS}
  split-gate arrays},}\ }in\ \href {\doibase 10.1109/IEDM19573.2019.8993580}
  {\emph {\bibinfo {booktitle} {2019 {IEEE} {International} {Electron}
  {Devices} {Meeting} ({IEDM})}}}\ (\bibinfo  {publisher} {IEEE},\ \bibinfo
  {address} {San Francisco, CA, USA},\ \bibinfo {year} {2019})\ pp.\ \bibinfo
  {pages} {37.7.1--37.7.4}\BibitemShut {NoStop}%
\bibitem [{\citenamefont {Ciriano-Tejel}\ \emph {et~al.}(2021)\citenamefont
  {Ciriano-Tejel}, \citenamefont {Fogarty}, \citenamefont {Schaal},
  \citenamefont {Hutin}, \citenamefont {Bertrand}, \citenamefont {Ibberson},
  \citenamefont {Gonzalez-Zalba}, \citenamefont {Li}, \citenamefont {Niquet},
  \citenamefont {Vinet},\ and\ \citenamefont
  {Morton}}]{dispersive_lever-arm_theory}%
  \BibitemOpen
  \bibfield  {author} {\bibinfo {author} {\bibfnamefont {V.~N.}\ \bibnamefont
  {Ciriano-Tejel}}, \bibinfo {author} {\bibfnamefont {M.~A.}\ \bibnamefont
  {Fogarty}}, \bibinfo {author} {\bibfnamefont {S.}~\bibnamefont {Schaal}},
  \bibinfo {author} {\bibfnamefont {L.}~\bibnamefont {Hutin}}, \bibinfo
  {author} {\bibfnamefont {B.}~\bibnamefont {Bertrand}}, \bibinfo {author}
  {\bibfnamefont {L.}~\bibnamefont {Ibberson}}, \bibinfo {author}
  {\bibfnamefont {M.~F.}\ \bibnamefont {Gonzalez-Zalba}}, \bibinfo {author}
  {\bibfnamefont {J.}~\bibnamefont {Li}}, \bibinfo {author} {\bibfnamefont
  {Y.-M.}\ \bibnamefont {Niquet}}, \bibinfo {author} {\bibfnamefont
  {M.}~\bibnamefont {Vinet}}, \ and\ \bibinfo {author} {\bibfnamefont {J.~J.}\
  \bibnamefont {Morton}},\ }\bibfield  {title} {\enquote {\bibinfo {title}
  {Spin {Readout} of a {CMOS} {Quantum} {Dot} by {Gate} {Reflectometry} and
  {Spin}-{Dependent} {Tunneling}},}\ }\href {\doibase
  10.1103/PRXQuantum.2.010353} {\bibfield  {journal} {\bibinfo  {journal} {PRX
  Quantum}\ }\textbf {\bibinfo {volume} {2}},\ \bibinfo {pages} {010353}
  (\bibinfo {year} {2021})}\BibitemShut {NoStop}%
\bibitem [{\citenamefont {von Horstig}\ \emph {et~al.}(2024)\citenamefont {von
  Horstig}, \citenamefont {Ibberson}, \citenamefont {Oakes}, \citenamefont
  {Cochrane}, \citenamefont {Wise}, \citenamefont {Stelmashenko}, \citenamefont
  {Barraud}, \citenamefont {Robinson}, \citenamefont {Martins},\ and\
  \citenamefont {Gonzalez-Zalba}}]{dispersive_platform_comparison}%
  \BibitemOpen
  \bibfield  {author} {\bibinfo {author} {\bibfnamefont {F.-E.}\ \bibnamefont
  {von Horstig}}, \bibinfo {author} {\bibfnamefont {D.~J.}\ \bibnamefont
  {Ibberson}}, \bibinfo {author} {\bibfnamefont {G.~A.}\ \bibnamefont {Oakes}},
  \bibinfo {author} {\bibfnamefont {L.}~\bibnamefont {Cochrane}}, \bibinfo
  {author} {\bibfnamefont {D.~F.}\ \bibnamefont {Wise}}, \bibinfo {author}
  {\bibfnamefont {N.}~\bibnamefont {Stelmashenko}}, \bibinfo {author}
  {\bibfnamefont {S.}~\bibnamefont {Barraud}}, \bibinfo {author} {\bibfnamefont
  {J.~A.}\ \bibnamefont {Robinson}}, \bibinfo {author} {\bibfnamefont
  {F.}~\bibnamefont {Martins}}, \ and\ \bibinfo {author} {\bibfnamefont
  {M.~F.}\ \bibnamefont {Gonzalez-Zalba}},\ }\bibfield  {title} {\enquote
  {\bibinfo {title} {Multimodule microwave assembly for fast readout and
  charge-noise characterization of silicon quantum dots},}\ }\href {\doibase
  10.1103/PhysRevApplied.21.044016} {\bibfield  {journal} {\bibinfo  {journal}
  {Physical Review Applied}\ }\textbf {\bibinfo {volume} {21}},\ \bibinfo
  {pages} {044016} (\bibinfo {year} {2024})}\BibitemShut {NoStop}%
\bibitem [{\citenamefont {Crippa}\ \emph {et~al.}(2019)\citenamefont {Crippa},
  \citenamefont {Ezzouch}, \citenamefont {Aprá}, \citenamefont {Amisse},
  \citenamefont {Laviéville}, \citenamefont {Hutin}, \citenamefont {Bertrand},
  \citenamefont {Vinet}, \citenamefont {Urdampilleta}, \citenamefont {Meunier},
  \citenamefont {Sanquer}, \citenamefont {Jehl}, \citenamefont {Maurand},\ and\
  \citenamefont {De~Franceschi}}]{silvano_5}%
  \BibitemOpen
  \bibfield  {author} {\bibinfo {author} {\bibfnamefont {A.}~\bibnamefont
  {Crippa}}, \bibinfo {author} {\bibfnamefont {R.}~\bibnamefont {Ezzouch}},
  \bibinfo {author} {\bibfnamefont {A.}~\bibnamefont {Aprá}}, \bibinfo
  {author} {\bibfnamefont {A.}~\bibnamefont {Amisse}}, \bibinfo {author}
  {\bibfnamefont {R.}~\bibnamefont {Laviéville}}, \bibinfo {author}
  {\bibfnamefont {L.}~\bibnamefont {Hutin}}, \bibinfo {author} {\bibfnamefont
  {B.}~\bibnamefont {Bertrand}}, \bibinfo {author} {\bibfnamefont
  {M.}~\bibnamefont {Vinet}}, \bibinfo {author} {\bibfnamefont
  {M.}~\bibnamefont {Urdampilleta}}, \bibinfo {author} {\bibfnamefont
  {T.}~\bibnamefont {Meunier}}, \bibinfo {author} {\bibfnamefont
  {M.}~\bibnamefont {Sanquer}}, \bibinfo {author} {\bibfnamefont
  {X.}~\bibnamefont {Jehl}}, \bibinfo {author} {\bibfnamefont {R.}~\bibnamefont
  {Maurand}}, \ and\ \bibinfo {author} {\bibfnamefont {S.}~\bibnamefont
  {De~Franceschi}},\ }\bibfield  {title} {\enquote {\bibinfo {title}
  {Gate-reflectometry dispersive readout and coherent control of a spin qubit
  in silicon},}\ }\href {\doibase 10.1038/s41467-019-10848-z} {\bibfield
  {journal} {\bibinfo  {journal} {Nature Communications}\ }\textbf {\bibinfo
  {volume} {10}},\ \bibinfo {pages} {2776} (\bibinfo {year}
  {2019})}\BibitemShut {NoStop}%
\bibitem [{\citenamefont {Neyens}\ \emph {et~al.}(2023)\citenamefont {Neyens},
  \citenamefont {Zietz}, \citenamefont {Watson}, \citenamefont {Luthi},
  \citenamefont {Nethwewala}, \citenamefont {George}, \citenamefont {Henry},
  \citenamefont {Wagner}, \citenamefont {Islam}, \citenamefont {Pillarisetty},
  \citenamefont {Kotlyar}, \citenamefont {Millard}, \citenamefont {Pellerano},
  \citenamefont {Bishop}, \citenamefont {Bojarski}, \citenamefont {Roberts},\
  and\ \citenamefont {Clarke}}]{fc_lc4}%
  \BibitemOpen
  \bibfield  {author} {\bibinfo {author} {\bibfnamefont {S.}~\bibnamefont
  {Neyens}}, \bibinfo {author} {\bibfnamefont {O.}~\bibnamefont {Zietz}},
  \bibinfo {author} {\bibfnamefont {T.}~\bibnamefont {Watson}}, \bibinfo
  {author} {\bibfnamefont {F.}~\bibnamefont {Luthi}}, \bibinfo {author}
  {\bibfnamefont {A.}~\bibnamefont {Nethwewala}}, \bibinfo {author}
  {\bibfnamefont {H.}~\bibnamefont {George}}, \bibinfo {author} {\bibfnamefont
  {E.}~\bibnamefont {Henry}}, \bibinfo {author} {\bibfnamefont
  {A.}~\bibnamefont {Wagner}}, \bibinfo {author} {\bibfnamefont
  {M.}~\bibnamefont {Islam}}, \bibinfo {author} {\bibfnamefont
  {R.}~\bibnamefont {Pillarisetty}}, \bibinfo {author} {\bibfnamefont
  {R.}~\bibnamefont {Kotlyar}}, \bibinfo {author} {\bibfnamefont
  {K.}~\bibnamefont {Millard}}, \bibinfo {author} {\bibfnamefont
  {S.}~\bibnamefont {Pellerano}}, \bibinfo {author} {\bibfnamefont
  {N.}~\bibnamefont {Bishop}}, \bibinfo {author} {\bibfnamefont
  {S.}~\bibnamefont {Bojarski}}, \bibinfo {author} {\bibfnamefont
  {J.}~\bibnamefont {Roberts}}, \ and\ \bibinfo {author} {\bibfnamefont
  {J.~S.}\ \bibnamefont {Clarke}},\ }\href {\doibase 10.48550/arXiv.2307.04812}
  {\enquote {\bibinfo {title} {Probing single electrons across 300 mm spin
  qubit wafers},}\ } (\bibinfo {year} {2023}),\ \bibinfo {note}
  {arXiv:2307.04812 [cond-mat, physics:quant-ph]}\BibitemShut {NoStop}%
\bibitem [{\citenamefont {Sebastiano}\ \emph {et~al.}(2020)\citenamefont
  {Sebastiano}, \citenamefont {van Dijk}, \citenamefont {Hart}, \citenamefont
  {Patra}, \citenamefont {van Staveren}, \citenamefont {Xue}, \citenamefont
  {Almudever}, \citenamefont {Scappucci}, \citenamefont {Veldhorst},
  \citenamefont {Vandersypen}, \citenamefont {Vladimirescu}, \citenamefont
  {Pellerano}, \citenamefont {Babaie},\ and\ \citenamefont
  {Charbon}}]{ch6_lr_cryodem}%
  \BibitemOpen
  \bibfield  {author} {\bibinfo {author} {\bibfnamefont {F.}~\bibnamefont
  {Sebastiano}}, \bibinfo {author} {\bibfnamefont {J.}~\bibnamefont {van
  Dijk}}, \bibinfo {author} {\bibfnamefont {P.}~\bibnamefont {Hart}}, \bibinfo
  {author} {\bibfnamefont {B.}~\bibnamefont {Patra}}, \bibinfo {author}
  {\bibfnamefont {J.}~\bibnamefont {van Staveren}}, \bibinfo {author}
  {\bibfnamefont {X.}~\bibnamefont {Xue}}, \bibinfo {author} {\bibfnamefont
  {C.}~\bibnamefont {Almudever}}, \bibinfo {author} {\bibfnamefont
  {G.}~\bibnamefont {Scappucci}}, \bibinfo {author} {\bibfnamefont
  {M.}~\bibnamefont {Veldhorst}}, \bibinfo {author} {\bibfnamefont
  {L.}~\bibnamefont {Vandersypen}}, \bibinfo {author} {\bibfnamefont
  {A.}~\bibnamefont {Vladimirescu}}, \bibinfo {author} {\bibfnamefont
  {S.}~\bibnamefont {Pellerano}}, \bibinfo {author} {\bibfnamefont
  {M.}~\bibnamefont {Babaie}}, \ and\ \bibinfo {author} {\bibfnamefont
  {E.}~\bibnamefont {Charbon}},\ }\bibfield  {title} {\enquote {\bibinfo
  {title} {Cryo-{CMOS} {Interfaces} for {Large}-{Scale} {Quantum}
  {Computers}},}\ }in\ \href {\doibase 10.1109/IEDM13553.2020.9372075} {\emph
  {\bibinfo {booktitle} {2020 {IEEE} {International} {Electron} {Devices}
  {Meeting} ({IEDM})}}}\ (\bibinfo {year} {2020})\ pp.\ \bibinfo {pages}
  {25.2.1--25.2.4}\BibitemShut {NoStop}%
\bibitem [{\citenamefont {Patra}\ \emph {et~al.}(2020)\citenamefont {Patra},
  \citenamefont {van Dijk}, \citenamefont {Subramanian}, \citenamefont {Corna},
  \citenamefont {Xue}, \citenamefont {Jeon}, \citenamefont {Sheikh},
  \citenamefont {Juarez-Hernandez}, \citenamefont {Esparza}, \citenamefont
  {Rampurawala}, \citenamefont {Carlton}, \citenamefont {Samkharadze},
  \citenamefont {Ravikumar}, \citenamefont {Nieva}, \citenamefont {Kim},
  \citenamefont {Lee}, \citenamefont {Sammak}, \citenamefont {Scappucci},
  \citenamefont {Veldhorst}, \citenamefont {Vandersypen}, \citenamefont
  {Babaie}, \citenamefont {Sebastiano}, \citenamefont {Charbon},\ and\
  \citenamefont {Pellerano}}]{ch6_lr_cryodem2}%
  \BibitemOpen
  \bibfield  {author} {\bibinfo {author} {\bibfnamefont {B.}~\bibnamefont
  {Patra}}, \bibinfo {author} {\bibfnamefont {J.~P.~G.}\ \bibnamefont {van
  Dijk}}, \bibinfo {author} {\bibfnamefont {S.}~\bibnamefont {Subramanian}},
  \bibinfo {author} {\bibfnamefont {A.}~\bibnamefont {Corna}}, \bibinfo
  {author} {\bibfnamefont {X.}~\bibnamefont {Xue}}, \bibinfo {author}
  {\bibfnamefont {C.}~\bibnamefont {Jeon}}, \bibinfo {author} {\bibfnamefont
  {F.}~\bibnamefont {Sheikh}}, \bibinfo {author} {\bibfnamefont
  {E.}~\bibnamefont {Juarez-Hernandez}}, \bibinfo {author} {\bibfnamefont
  {B.~P.}\ \bibnamefont {Esparza}}, \bibinfo {author} {\bibfnamefont
  {H.}~\bibnamefont {Rampurawala}}, \bibinfo {author} {\bibfnamefont
  {B.}~\bibnamefont {Carlton}}, \bibinfo {author} {\bibfnamefont
  {N.}~\bibnamefont {Samkharadze}}, \bibinfo {author} {\bibfnamefont
  {S.}~\bibnamefont {Ravikumar}}, \bibinfo {author} {\bibfnamefont
  {C.}~\bibnamefont {Nieva}}, \bibinfo {author} {\bibfnamefont
  {S.}~\bibnamefont {Kim}}, \bibinfo {author} {\bibfnamefont {H.-J.}\
  \bibnamefont {Lee}}, \bibinfo {author} {\bibfnamefont {A.}~\bibnamefont
  {Sammak}}, \bibinfo {author} {\bibfnamefont {G.}~\bibnamefont {Scappucci}},
  \bibinfo {author} {\bibfnamefont {M.}~\bibnamefont {Veldhorst}}, \bibinfo
  {author} {\bibfnamefont {L.~M.~K.}\ \bibnamefont {Vandersypen}}, \bibinfo
  {author} {\bibfnamefont {M.}~\bibnamefont {Babaie}}, \bibinfo {author}
  {\bibfnamefont {F.}~\bibnamefont {Sebastiano}}, \bibinfo {author}
  {\bibfnamefont {E.}~\bibnamefont {Charbon}}, \ and\ \bibinfo {author}
  {\bibfnamefont {S.}~\bibnamefont {Pellerano}},\ }\bibfield  {title} {\enquote
  {\bibinfo {title} {19.1 {A} {Scalable} {Cryo}-{CMOS} 2-to-{20GHz} {Digitally}
  {Intensive} {Controller} for 4×32 {Frequency} {Multiplexed} {Spin}
  {Qubits}/{Transmons} in 22nm {FinFET} {Technology} for {Quantum}
  {Computers}},}\ }in\ \href {\doibase 10.1109/ISSCC19947.2020.9063109} {\emph
  {\bibinfo {booktitle} {2020 {IEEE} {International} {Solid}-{State} {Circuits}
  {Conference} - ({ISSCC})}}}\ (\bibinfo {year} {2020})\ pp.\ \bibinfo {pages}
  {304--306}\BibitemShut {NoStop}%
\bibitem [{\citenamefont {Xue}\ \emph {et~al.}(2021)\citenamefont {Xue},
  \citenamefont {Patra}, \citenamefont {van Dijk}, \citenamefont {Samkharadze},
  \citenamefont {Subramanian}, \citenamefont {Corna}, \citenamefont
  {Paquelet~Wuetz}, \citenamefont {Jeon}, \citenamefont {Sheikh}, \citenamefont
  {Juarez-Hernandez}, \citenamefont {Esparza}, \citenamefont {Rampurawala},
  \citenamefont {Carlton}, \citenamefont {Ravikumar}, \citenamefont {Nieva},
  \citenamefont {Kim}, \citenamefont {Lee}, \citenamefont {Sammak},
  \citenamefont {Scappucci}, \citenamefont {Veldhorst}, \citenamefont
  {Sebastiano}, \citenamefont {Babaie}, \citenamefont {Pellerano},
  \citenamefont {Charbon},\ and\ \citenamefont {Vandersypen}}]{cryo2}%
  \BibitemOpen
  \bibfield  {author} {\bibinfo {author} {\bibfnamefont {X.}~\bibnamefont
  {Xue}}, \bibinfo {author} {\bibfnamefont {B.}~\bibnamefont {Patra}}, \bibinfo
  {author} {\bibfnamefont {J.~P.~G.}\ \bibnamefont {van Dijk}}, \bibinfo
  {author} {\bibfnamefont {N.}~\bibnamefont {Samkharadze}}, \bibinfo {author}
  {\bibfnamefont {S.}~\bibnamefont {Subramanian}}, \bibinfo {author}
  {\bibfnamefont {A.}~\bibnamefont {Corna}}, \bibinfo {author} {\bibfnamefont
  {B.}~\bibnamefont {Paquelet~Wuetz}}, \bibinfo {author} {\bibfnamefont
  {C.}~\bibnamefont {Jeon}}, \bibinfo {author} {\bibfnamefont {F.}~\bibnamefont
  {Sheikh}}, \bibinfo {author} {\bibfnamefont {E.}~\bibnamefont
  {Juarez-Hernandez}}, \bibinfo {author} {\bibfnamefont {B.~P.}\ \bibnamefont
  {Esparza}}, \bibinfo {author} {\bibfnamefont {H.}~\bibnamefont
  {Rampurawala}}, \bibinfo {author} {\bibfnamefont {B.}~\bibnamefont
  {Carlton}}, \bibinfo {author} {\bibfnamefont {S.}~\bibnamefont {Ravikumar}},
  \bibinfo {author} {\bibfnamefont {C.}~\bibnamefont {Nieva}}, \bibinfo
  {author} {\bibfnamefont {S.}~\bibnamefont {Kim}}, \bibinfo {author}
  {\bibfnamefont {H.-J.}\ \bibnamefont {Lee}}, \bibinfo {author} {\bibfnamefont
  {A.}~\bibnamefont {Sammak}}, \bibinfo {author} {\bibfnamefont
  {G.}~\bibnamefont {Scappucci}}, \bibinfo {author} {\bibfnamefont
  {M.}~\bibnamefont {Veldhorst}}, \bibinfo {author} {\bibfnamefont
  {F.}~\bibnamefont {Sebastiano}}, \bibinfo {author} {\bibfnamefont
  {M.}~\bibnamefont {Babaie}}, \bibinfo {author} {\bibfnamefont
  {S.}~\bibnamefont {Pellerano}}, \bibinfo {author} {\bibfnamefont
  {E.}~\bibnamefont {Charbon}}, \ and\ \bibinfo {author} {\bibfnamefont
  {L.~M.~K.}\ \bibnamefont {Vandersypen}},\ }\bibfield  {title} {\enquote
  {\bibinfo {title} {{CMOS}-based cryogenic control of silicon quantum
  circuits},}\ }\href {\doibase 10.1038/s41586-021-03469-4} {\bibfield
  {journal} {\bibinfo  {journal} {Nature}\ }\textbf {\bibinfo {volume} {593}},\
  \bibinfo {pages} {205--210} (\bibinfo {year} {2021})}\BibitemShut {NoStop}%
\bibitem [{\citenamefont {Yang}\ \emph {et~al.}(2020)\citenamefont {Yang},
  \citenamefont {Ruffino}, \citenamefont {Michniewicz}, \citenamefont {Peng},
  \citenamefont {Charbon},\ and\ \citenamefont {Gonzalez-Zalba}}]{mine_ieee}%
  \BibitemOpen
  \bibfield  {author} {\bibinfo {author} {\bibfnamefont {T.-Y.}\ \bibnamefont
  {Yang}}, \bibinfo {author} {\bibfnamefont {A.}~\bibnamefont {Ruffino}},
  \bibinfo {author} {\bibfnamefont {J.}~\bibnamefont {Michniewicz}}, \bibinfo
  {author} {\bibfnamefont {Y.}~\bibnamefont {Peng}}, \bibinfo {author}
  {\bibfnamefont {E.}~\bibnamefont {Charbon}}, \ and\ \bibinfo {author}
  {\bibfnamefont {M.~F.}\ \bibnamefont {Gonzalez-Zalba}},\ }\bibfield  {title}
  {\enquote {\bibinfo {title} {Quantum {Transport} in 40-nm {MOSFETs} at
  {Deep}-{Cryogenic} {Temperatures}},}\ }\href {\doibase
  10.1109/LED.2020.2995645} {\bibfield  {journal} {\bibinfo  {journal} {IEEE
  Electron Device Letters}\ ,\ \bibinfo {pages} {1--1}} (\bibinfo {year}
  {2020})}\BibitemShut {NoStop}%
\bibitem [{\citenamefont {Ruffino}\ \emph {et~al.}(2021)\citenamefont
  {Ruffino}, \citenamefont {Peng}, \citenamefont {Yang}, \citenamefont
  {Michniewicz}, \citenamefont {Gonzalez-Zalba},\ and\ \citenamefont
  {Charbon}}]{mine_ieee2}%
  \BibitemOpen
  \bibfield  {author} {\bibinfo {author} {\bibfnamefont {A.}~\bibnamefont
  {Ruffino}}, \bibinfo {author} {\bibfnamefont {Y.}~\bibnamefont {Peng}},
  \bibinfo {author} {\bibfnamefont {T.-Y.}\ \bibnamefont {Yang}}, \bibinfo
  {author} {\bibfnamefont {J.}~\bibnamefont {Michniewicz}}, \bibinfo {author}
  {\bibfnamefont {M.~F.}\ \bibnamefont {Gonzalez-Zalba}}, \ and\ \bibinfo
  {author} {\bibfnamefont {E.}~\bibnamefont {Charbon}},\ }\bibfield  {title}
  {\enquote {\bibinfo {title} {13.2 {A} {Fully}-{Integrated} 40-nm 5-6.5 {GHz}
  {Cryo}-{CMOS} {System}-on-{Chip} with {I}/{Q} {Receiver} and {Frequency}
  {Synthesizer} for {Scalable} {Multiplexed} {Readout} of {Quantum} {Dots}},}\
  }in\ \href {\doibase 10.1109/ISSCC42613.2021.9365758} {\emph {\bibinfo
  {booktitle} {2021 {IEEE} {International} {Solid}- {State} {Circuits}
  {Conference} ({ISSCC})}}}\ (\bibinfo  {publisher} {IEEE},\ \bibinfo {address}
  {San Francisco, CA, USA},\ \bibinfo {year} {2021})\ pp.\ \bibinfo {pages}
  {210--212}\BibitemShut {NoStop}%
\bibitem [{\citenamefont {Ruffino}\ \emph {et~al.}(2022)\citenamefont
  {Ruffino}, \citenamefont {Yang}, \citenamefont {Michniewicz}, \citenamefont
  {Peng}, \citenamefont {Charbon},\ and\ \citenamefont
  {Gonzalez-Zalba}}]{my_nat_true}%
  \BibitemOpen
  \bibfield  {author} {\bibinfo {author} {\bibfnamefont {A.}~\bibnamefont
  {Ruffino}}, \bibinfo {author} {\bibfnamefont {T.-Y.}\ \bibnamefont {Yang}},
  \bibinfo {author} {\bibfnamefont {J.}~\bibnamefont {Michniewicz}}, \bibinfo
  {author} {\bibfnamefont {Y.}~\bibnamefont {Peng}}, \bibinfo {author}
  {\bibfnamefont {E.}~\bibnamefont {Charbon}}, \ and\ \bibinfo {author}
  {\bibfnamefont {M.~F.}\ \bibnamefont {Gonzalez-Zalba}},\ }\bibfield  {title}
  {\enquote {\bibinfo {title} {A cryo-{CMOS} chip that integrates silicon
  quantum dots and multiplexed dispersive readout electronics},}\ }\href
  {\doibase 10.1038/s41928-021-00687-6} {\bibfield  {journal} {\bibinfo
  {journal} {Nature Electronics}\ }\textbf {\bibinfo {volume} {5}},\ \bibinfo
  {pages} {53--59} (\bibinfo {year} {2022})},\ \bibinfo {note} {number: 1
  Publisher: Nature Publishing Group}\BibitemShut {NoStop}%
\bibitem [{\citenamefont {Borsoi}\ \emph {et~al.}(2023)\citenamefont {Borsoi},
  \citenamefont {Hendrickx}, \citenamefont {John}, \citenamefont {Meyer},
  \citenamefont {Motz}, \citenamefont {van Riggelen}, \citenamefont {Sammak},
  \citenamefont {de~Snoo}, \citenamefont {Scappucci},\ and\ \citenamefont
  {Veldhorst}}]{fc_s12}%
  \BibitemOpen
  \bibfield  {author} {\bibinfo {author} {\bibfnamefont {F.}~\bibnamefont
  {Borsoi}}, \bibinfo {author} {\bibfnamefont {N.~W.}\ \bibnamefont
  {Hendrickx}}, \bibinfo {author} {\bibfnamefont {V.}~\bibnamefont {John}},
  \bibinfo {author} {\bibfnamefont {M.}~\bibnamefont {Meyer}}, \bibinfo
  {author} {\bibfnamefont {S.}~\bibnamefont {Motz}}, \bibinfo {author}
  {\bibfnamefont {F.}~\bibnamefont {van Riggelen}}, \bibinfo {author}
  {\bibfnamefont {A.}~\bibnamefont {Sammak}}, \bibinfo {author} {\bibfnamefont
  {S.~L.}\ \bibnamefont {de~Snoo}}, \bibinfo {author} {\bibfnamefont
  {G.}~\bibnamefont {Scappucci}}, \ and\ \bibinfo {author} {\bibfnamefont
  {M.}~\bibnamefont {Veldhorst}},\ }\bibfield  {title} {\enquote {\bibinfo
  {title} {Shared control of a 16 semiconductor quantum dot crossbar array},}\
  }\href {\doibase 10.1038/s41565-023-01491-3} {\bibfield  {journal} {\bibinfo
  {journal} {Nature Nanotechnology}\ ,\ \bibinfo {pages} {1--7}} (\bibinfo
  {year} {2023})},\ \bibinfo {note} {publisher: Nature Publishing
  Group}\BibitemShut {NoStop}%
\bibitem [{\citenamefont {Krinner}\ \emph {et~al.}(2019)\citenamefont
  {Krinner}, \citenamefont {Storz}, \citenamefont {Kurpiers}, \citenamefont
  {Magnard}, \citenamefont {Heinsoo}, \citenamefont {Keller}, \citenamefont
  {Lütolf}, \citenamefont {Eichler},\ and\ \citenamefont {Wallraff}}]{cryo1}%
  \BibitemOpen
  \bibfield  {author} {\bibinfo {author} {\bibfnamefont {S.}~\bibnamefont
  {Krinner}}, \bibinfo {author} {\bibfnamefont {S.}~\bibnamefont {Storz}},
  \bibinfo {author} {\bibfnamefont {P.}~\bibnamefont {Kurpiers}}, \bibinfo
  {author} {\bibfnamefont {P.}~\bibnamefont {Magnard}}, \bibinfo {author}
  {\bibfnamefont {J.}~\bibnamefont {Heinsoo}}, \bibinfo {author} {\bibfnamefont
  {R.}~\bibnamefont {Keller}}, \bibinfo {author} {\bibfnamefont
  {J.}~\bibnamefont {Lütolf}}, \bibinfo {author} {\bibfnamefont
  {C.}~\bibnamefont {Eichler}}, \ and\ \bibinfo {author} {\bibfnamefont
  {A.}~\bibnamefont {Wallraff}},\ }\bibfield  {title} {\enquote {\bibinfo
  {title} {Engineering cryogenic setups for 100-qubit scale superconducting
  circuit systems},}\ }\href {\doibase 10.1140/epjqt/s40507-019-0072-0}
  {\bibfield  {journal} {\bibinfo  {journal} {EPJ Quantum Technology}\ }\textbf
  {\bibinfo {volume} {6}},\ \bibinfo {pages} {1--29} (\bibinfo {year}
  {2019})},\ \bibinfo {note} {number: 1 Publisher: SpringerOpen}\BibitemShut
  {NoStop}%
\bibitem [{\citenamefont {Fogarty}\ \emph {et~al.}(2018)\citenamefont
  {Fogarty}, \citenamefont {Chan}, \citenamefont {Hensen}, \citenamefont
  {Huang}, \citenamefont {Tanttu}, \citenamefont {Yang}, \citenamefont
  {Laucht}, \citenamefont {Veldhorst}, \citenamefont {Hudson}, \citenamefont
  {Itoh}, \citenamefont {Culcer}, \citenamefont {Morello},\ and\ \citenamefont
  {Dzurak}}]{gfactor}%
  \BibitemOpen
  \bibfield  {author} {\bibinfo {author} {\bibfnamefont {M.~A.}\ \bibnamefont
  {Fogarty}}, \bibinfo {author} {\bibfnamefont {K.~W.}\ \bibnamefont {Chan}},
  \bibinfo {author} {\bibfnamefont {B.}~\bibnamefont {Hensen}}, \bibinfo
  {author} {\bibfnamefont {W.}~\bibnamefont {Huang}}, \bibinfo {author}
  {\bibfnamefont {T.}~\bibnamefont {Tanttu}}, \bibinfo {author} {\bibfnamefont
  {C.~H.}\ \bibnamefont {Yang}}, \bibinfo {author} {\bibfnamefont
  {A.}~\bibnamefont {Laucht}}, \bibinfo {author} {\bibfnamefont
  {M.}~\bibnamefont {Veldhorst}}, \bibinfo {author} {\bibfnamefont {F.~E.}\
  \bibnamefont {Hudson}}, \bibinfo {author} {\bibfnamefont {K.~M.}\
  \bibnamefont {Itoh}}, \bibinfo {author} {\bibfnamefont {D.}~\bibnamefont
  {Culcer}}, \bibinfo {author} {\bibfnamefont {A.}~\bibnamefont {Morello}}, \
  and\ \bibinfo {author} {\bibfnamefont {A.~S.}\ \bibnamefont {Dzurak}},\
  }\bibfield  {title} {\enquote {\bibinfo {title} {Integrated silicon qubit
  platform with single-spin addressability, exchange control and robust
  single-shot singlet-triplet readout},}\ }\href {\doibase
  10.1038/s41467-018-06039-x} {\bibfield  {journal} {\bibinfo  {journal}
  {Nature Communications}\ }\textbf {\bibinfo {volume} {9}},\ \bibinfo {pages}
  {4370} (\bibinfo {year} {2018})},\ \bibinfo {note} {number: 1
  arXiv:1708.03445 [quant-ph]}\BibitemShut {NoStop}%
\bibitem [{fc_(2023)}]{fc_lc2}%
  \BibitemOpen
  \href
  {https://spectrum.ieee.org/scalable-qubits-quantum-computer-news-silicon-wafer}
  {\enquote {\bibinfo {title} {Building a {Quantum} {Computer} {From}
  {Off}-the-{Shelf} {Parts} - {IEEE} {Spectrum}},}\ } (\bibinfo {year}
  {2023})\BibitemShut {NoStop}%
\bibitem [{\citenamefont {Zwerver}\ \emph {et~al.}(2023)\citenamefont
  {Zwerver}, \citenamefont {Amitonov}, \citenamefont {de~Snoo}, \citenamefont
  {Madzik}, \citenamefont {Rimbach-Russ}, \citenamefont {Sammak}, \citenamefont
  {Scappucci},\ and\ \citenamefont {Vandersypen}}]{fc_s1}%
  \BibitemOpen
  \bibfield  {author} {\bibinfo {author} {\bibfnamefont {A.}~\bibnamefont
  {Zwerver}}, \bibinfo {author} {\bibfnamefont {S.}~\bibnamefont {Amitonov}},
  \bibinfo {author} {\bibfnamefont {S.}~\bibnamefont {de~Snoo}}, \bibinfo
  {author} {\bibfnamefont {M.}~\bibnamefont {Madzik}}, \bibinfo {author}
  {\bibfnamefont {M.}~\bibnamefont {Rimbach-Russ}}, \bibinfo {author}
  {\bibfnamefont {A.}~\bibnamefont {Sammak}}, \bibinfo {author} {\bibfnamefont
  {G.}~\bibnamefont {Scappucci}}, \ and\ \bibinfo {author} {\bibfnamefont
  {L.}~\bibnamefont {Vandersypen}},\ }\bibfield  {title} {\enquote {\bibinfo
  {title} {Shuttling an {Electron} {Spin} through a {Silicon} {Quantum} {Dot}
  {Array}},}\ }\href {\doibase 10.1103/PRXQuantum.4.030303} {\bibfield
  {journal} {\bibinfo  {journal} {PRX Quantum}\ }\textbf {\bibinfo {volume}
  {4}},\ \bibinfo {pages} {030303} (\bibinfo {year} {2023})},\ \bibinfo {note}
  {publisher: American Physical Society}\BibitemShut {NoStop}%
\bibitem [{\citenamefont {Utsugi}\ \emph {et~al.}(2023)\citenamefont {Utsugi},
  \citenamefont {Kuno}, \citenamefont {Lee}, \citenamefont {Tsuchiya},
  \citenamefont {Mine}, \citenamefont {Hisamoto}, \citenamefont {Saito},\ and\
  \citenamefont {Mizuno}}]{fc_s2}%
  \BibitemOpen
  \bibfield  {author} {\bibinfo {author} {\bibfnamefont {T.}~\bibnamefont
  {Utsugi}}, \bibinfo {author} {\bibfnamefont {T.}~\bibnamefont {Kuno}},
  \bibinfo {author} {\bibfnamefont {N.}~\bibnamefont {Lee}}, \bibinfo {author}
  {\bibfnamefont {R.}~\bibnamefont {Tsuchiya}}, \bibinfo {author}
  {\bibfnamefont {T.}~\bibnamefont {Mine}}, \bibinfo {author} {\bibfnamefont
  {D.}~\bibnamefont {Hisamoto}}, \bibinfo {author} {\bibfnamefont
  {S.}~\bibnamefont {Saito}}, \ and\ \bibinfo {author} {\bibfnamefont
  {H.}~\bibnamefont {Mizuno}},\ }\bibfield  {title} {\enquote {\bibinfo {title}
  {Single-electron routing in a silicon quantum-dot array},}\ }\href {\doibase
  10.1103/PhysRevB.108.235308} {\bibfield  {journal} {\bibinfo  {journal}
  {Physical Review B}\ }\textbf {\bibinfo {volume} {108}},\ \bibinfo {pages}
  {235308} (\bibinfo {year} {2023})},\ \bibinfo {note} {publisher: American
  Physical Society}\BibitemShut {NoStop}%
\bibitem [{\citenamefont {Seidler}\ \emph {et~al.}(2022)\citenamefont
  {Seidler}, \citenamefont {Struck}, \citenamefont {Xue}, \citenamefont
  {Focke}, \citenamefont {Trellenkamp}, \citenamefont {Bluhm},\ and\
  \citenamefont {Schreiber}}]{fc_s3}%
  \BibitemOpen
  \bibfield  {author} {\bibinfo {author} {\bibfnamefont {I.}~\bibnamefont
  {Seidler}}, \bibinfo {author} {\bibfnamefont {T.}~\bibnamefont {Struck}},
  \bibinfo {author} {\bibfnamefont {R.}~\bibnamefont {Xue}}, \bibinfo {author}
  {\bibfnamefont {N.}~\bibnamefont {Focke}}, \bibinfo {author} {\bibfnamefont
  {S.}~\bibnamefont {Trellenkamp}}, \bibinfo {author} {\bibfnamefont
  {H.}~\bibnamefont {Bluhm}}, \ and\ \bibinfo {author} {\bibfnamefont {L.~R.}\
  \bibnamefont {Schreiber}},\ }\bibfield  {title} {\enquote {\bibinfo {title}
  {Conveyor-mode single-electron shuttling in {Si}/{SiGe} for a scalable
  quantum computing architecture},}\ }\href {\doibase
  10.1038/s41534-022-00615-2} {\bibfield  {journal} {\bibinfo  {journal} {npj
  Quantum Information}\ }\textbf {\bibinfo {volume} {8}},\ \bibinfo {pages}
  {1--7} (\bibinfo {year} {2022})},\ \bibinfo {note} {number: 1 Publisher:
  Nature Publishing Group}\BibitemShut {NoStop}%
\bibitem [{\citenamefont {Tadokoro}\ \emph {et~al.}(2021)\citenamefont
  {Tadokoro}, \citenamefont {Nakajima}, \citenamefont {Kobayashi},
  \citenamefont {Takeda}, \citenamefont {Noiri}, \citenamefont {Tomari},
  \citenamefont {Yoneda}, \citenamefont {Tarucha},\ and\ \citenamefont
  {Kodera}}]{2d_better1}%
  \BibitemOpen
  \bibfield  {author} {\bibinfo {author} {\bibfnamefont {M.}~\bibnamefont
  {Tadokoro}}, \bibinfo {author} {\bibfnamefont {T.}~\bibnamefont {Nakajima}},
  \bibinfo {author} {\bibfnamefont {T.}~\bibnamefont {Kobayashi}}, \bibinfo
  {author} {\bibfnamefont {K.}~\bibnamefont {Takeda}}, \bibinfo {author}
  {\bibfnamefont {A.}~\bibnamefont {Noiri}}, \bibinfo {author} {\bibfnamefont
  {K.}~\bibnamefont {Tomari}}, \bibinfo {author} {\bibfnamefont
  {J.}~\bibnamefont {Yoneda}}, \bibinfo {author} {\bibfnamefont
  {S.}~\bibnamefont {Tarucha}}, \ and\ \bibinfo {author} {\bibfnamefont
  {T.}~\bibnamefont {Kodera}},\ }\bibfield  {title} {\enquote {\bibinfo {title}
  {Designs for a two-dimensional {Si} quantum dot array with spin qubit
  addressability},}\ }\href {\doibase 10.1038/s41598-021-98212-4} {\bibfield
  {journal} {\bibinfo  {journal} {Scientific Reports}\ }\textbf {\bibinfo
  {volume} {11}},\ \bibinfo {pages} {19406} (\bibinfo {year} {2021})},\
  \bibinfo {note} {number: 1 Publisher: Nature Publishing Group}\BibitemShut
  {NoStop}%
\bibitem [{\citenamefont {Swayne}(2020)}]{2d_better2}%
  \BibitemOpen
  \bibfield  {author} {\bibinfo {author} {\bibfnamefont {M.}~\bibnamefont
  {Swayne}},\ }\href
  {https://thequantuminsider.com/2020/12/31/niels-bohr-institute-leti-researchers-say-two-dimensional-quantum-dot-array-is-a-step-toward-practical-quantum-computing/}
  {\enquote {\bibinfo {title} {Niels {Bohr} {Institute}, {Leti} {Researchers}
  {Say} {Two}-{Dimensional} {Quantum} {Dot} {Array} {Is} {A} {Step} {Toward}
  {Practical} {Quantum} {Computing}},}\ } (\bibinfo {year} {2020})\BibitemShut
  {NoStop}%
\bibitem [{\citenamefont {Cai}, \citenamefont {Siegel},\ and\ \citenamefont
  {Benjamin}(2023)}]{fc_s4}%
  \BibitemOpen
  \bibfield  {author} {\bibinfo {author} {\bibfnamefont {Z.}~\bibnamefont
  {Cai}}, \bibinfo {author} {\bibfnamefont {A.}~\bibnamefont {Siegel}}, \ and\
  \bibinfo {author} {\bibfnamefont {S.}~\bibnamefont {Benjamin}},\ }\bibfield
  {title} {\enquote {\bibinfo {title} {Looped {Pipelines} {Enabling}
  {Effective} {3D} {Qubit} {Lattices} in a {Strictly} {2D} {Device}},}\ }\href
  {\doibase 10.1103/PRXQuantum.4.020345} {\bibfield  {journal} {\bibinfo
  {journal} {PRX Quantum}\ }\textbf {\bibinfo {volume} {4}},\ \bibinfo {pages}
  {020345} (\bibinfo {year} {2023})},\ \bibinfo {note} {publisher: American
  Physical Society}\BibitemShut {NoStop}%
\bibitem [{\citenamefont {Veldhorst}\ \emph {et~al.}(2017)\citenamefont
  {Veldhorst}, \citenamefont {Eenink}, \citenamefont {Yang},\ and\
  \citenamefont {Dzurak}}]{fc_s5}%
  \BibitemOpen
  \bibfield  {author} {\bibinfo {author} {\bibfnamefont {M.}~\bibnamefont
  {Veldhorst}}, \bibinfo {author} {\bibfnamefont {H.~G.~J.}\ \bibnamefont
  {Eenink}}, \bibinfo {author} {\bibfnamefont {C.~H.}\ \bibnamefont {Yang}}, \
  and\ \bibinfo {author} {\bibfnamefont {A.~S.}\ \bibnamefont {Dzurak}},\
  }\bibfield  {title} {\enquote {\bibinfo {title} {Silicon {CMOS} architecture
  for a spin-based quantum computer},}\ }\href {\doibase
  10.1038/s41467-017-01905-6} {\bibfield  {journal} {\bibinfo  {journal}
  {Nature Communications}\ }\textbf {\bibinfo {volume} {8}},\ \bibinfo {pages}
  {1766} (\bibinfo {year} {2017})},\ \bibinfo {note} {number: 1 Publisher:
  Nature Publishing Group}\BibitemShut {NoStop}%
\bibitem [{\citenamefont {Chanrion}\ \emph {et~al.}(2020)\citenamefont
  {Chanrion}, \citenamefont {Niegemann}, \citenamefont {Bertrand},
  \citenamefont {Spence}, \citenamefont {Jadot}, \citenamefont {Li},
  \citenamefont {Mortemousque}, \citenamefont {Hutin}, \citenamefont {Maurand},
  \citenamefont {Jehl}, \citenamefont {Sanquer}, \citenamefont {De~Franceschi},
  \citenamefont {Bäuerle}, \citenamefont {Balestro}, \citenamefont {Niquet},
  \citenamefont {Vinet}, \citenamefont {Meunier},\ and\ \citenamefont
  {Urdampilleta}}]{fc_s6}%
  \BibitemOpen
  \bibfield  {author} {\bibinfo {author} {\bibfnamefont {E.}~\bibnamefont
  {Chanrion}}, \bibinfo {author} {\bibfnamefont {D.~J.}\ \bibnamefont
  {Niegemann}}, \bibinfo {author} {\bibfnamefont {B.}~\bibnamefont {Bertrand}},
  \bibinfo {author} {\bibfnamefont {C.}~\bibnamefont {Spence}}, \bibinfo
  {author} {\bibfnamefont {B.}~\bibnamefont {Jadot}}, \bibinfo {author}
  {\bibfnamefont {J.}~\bibnamefont {Li}}, \bibinfo {author} {\bibfnamefont
  {P.-A.}\ \bibnamefont {Mortemousque}}, \bibinfo {author} {\bibfnamefont
  {L.}~\bibnamefont {Hutin}}, \bibinfo {author} {\bibfnamefont
  {R.}~\bibnamefont {Maurand}}, \bibinfo {author} {\bibfnamefont
  {X.}~\bibnamefont {Jehl}}, \bibinfo {author} {\bibfnamefont {M.}~\bibnamefont
  {Sanquer}}, \bibinfo {author} {\bibfnamefont {S.}~\bibnamefont
  {De~Franceschi}}, \bibinfo {author} {\bibfnamefont {C.}~\bibnamefont
  {Bäuerle}}, \bibinfo {author} {\bibfnamefont {F.}~\bibnamefont {Balestro}},
  \bibinfo {author} {\bibfnamefont {Y.-M.}\ \bibnamefont {Niquet}}, \bibinfo
  {author} {\bibfnamefont {M.}~\bibnamefont {Vinet}}, \bibinfo {author}
  {\bibfnamefont {T.}~\bibnamefont {Meunier}}, \ and\ \bibinfo {author}
  {\bibfnamefont {M.}~\bibnamefont {Urdampilleta}},\ }\bibfield  {title}
  {\enquote {\bibinfo {title} {Charge {Detection} in an {Array} of {CMOS}
  {Quantum} {Dots}},}\ }\href {\doibase 10.1103/PhysRevApplied.14.024066}
  {\bibfield  {journal} {\bibinfo  {journal} {Physical Review Applied}\
  }\textbf {\bibinfo {volume} {14}},\ \bibinfo {pages} {024066} (\bibinfo
  {year} {2020})},\ \bibinfo {note} {publisher: American Physical
  Society}\BibitemShut {NoStop}%
\bibitem [{\citenamefont {Gilbert}\ \emph {et~al.}(2020)\citenamefont
  {Gilbert}, \citenamefont {Saraiva}, \citenamefont {Lim}, \citenamefont
  {Yang}, \citenamefont {Laucht}, \citenamefont {Bertrand}, \citenamefont
  {Rambal}, \citenamefont {Hutin}, \citenamefont {Escott}, \citenamefont
  {Vinet},\ and\ \citenamefont {Dzurak}}]{fc_s7}%
  \BibitemOpen
  \bibfield  {author} {\bibinfo {author} {\bibfnamefont {W.}~\bibnamefont
  {Gilbert}}, \bibinfo {author} {\bibfnamefont {A.}~\bibnamefont {Saraiva}},
  \bibinfo {author} {\bibfnamefont {W.~H.}\ \bibnamefont {Lim}}, \bibinfo
  {author} {\bibfnamefont {C.~H.}\ \bibnamefont {Yang}}, \bibinfo {author}
  {\bibfnamefont {A.}~\bibnamefont {Laucht}}, \bibinfo {author} {\bibfnamefont
  {B.}~\bibnamefont {Bertrand}}, \bibinfo {author} {\bibfnamefont
  {N.}~\bibnamefont {Rambal}}, \bibinfo {author} {\bibfnamefont
  {L.}~\bibnamefont {Hutin}}, \bibinfo {author} {\bibfnamefont {C.~C.}\
  \bibnamefont {Escott}}, \bibinfo {author} {\bibfnamefont {M.}~\bibnamefont
  {Vinet}}, \ and\ \bibinfo {author} {\bibfnamefont {A.~S.}\ \bibnamefont
  {Dzurak}},\ }\bibfield  {title} {\enquote {\bibinfo {title}
  {Single-{Electron} {Operation} of a {Silicon}-{CMOS} 2 × 2 {Quantum} {Dot}
  {Array} with {Integrated} {Charge} {Sensing}},}\ }\href {\doibase
  10.1021/acs.nanolett.0c02397} {\bibfield  {journal} {\bibinfo  {journal}
  {Nano Letters}\ }\textbf {\bibinfo {volume} {20}},\ \bibinfo {pages}
  {7882--7888} (\bibinfo {year} {2020})}\BibitemShut {NoStop}%
\bibitem [{\citenamefont {Unseld}\ \emph {et~al.}(2023)\citenamefont {Unseld},
  \citenamefont {Meyer}, \citenamefont {Madzik}, \citenamefont {Borsoi},
  \citenamefont {de~Snoo}, \citenamefont {Amitonov}, \citenamefont {Sammak},
  \citenamefont {Scappucci}, \citenamefont {Veldhorst},\ and\ \citenamefont
  {Vandersypen}}]{fc_s8}%
  \BibitemOpen
  \bibfield  {author} {\bibinfo {author} {\bibfnamefont {F.~K.}\ \bibnamefont
  {Unseld}}, \bibinfo {author} {\bibfnamefont {M.}~\bibnamefont {Meyer}},
  \bibinfo {author} {\bibfnamefont {M.~T.}\ \bibnamefont {Madzik}}, \bibinfo
  {author} {\bibfnamefont {F.}~\bibnamefont {Borsoi}}, \bibinfo {author}
  {\bibfnamefont {S.~L.}\ \bibnamefont {de~Snoo}}, \bibinfo {author}
  {\bibfnamefont {S.~V.}\ \bibnamefont {Amitonov}}, \bibinfo {author}
  {\bibfnamefont {A.}~\bibnamefont {Sammak}}, \bibinfo {author} {\bibfnamefont
  {G.}~\bibnamefont {Scappucci}}, \bibinfo {author} {\bibfnamefont
  {M.}~\bibnamefont {Veldhorst}}, \ and\ \bibinfo {author} {\bibfnamefont
  {L.~M.~K.}\ \bibnamefont {Vandersypen}},\ }\href
  {http://arxiv.org/abs/2305.19681} {\enquote {\bibinfo {title} {A {2D} quantum
  dot array in planar 28 {Si}/{SiGe}},}\ } (\bibinfo {year} {2023}),\ \bibinfo
  {note} {arXiv:2305.19681 [cond-mat, physics:quant-ph]}\BibitemShut {NoStop}%
\bibitem [{\citenamefont {Crawford}\ \emph {et~al.}(2023)\citenamefont
  {Crawford}, \citenamefont {Cruise}, \citenamefont {Mertig},\ and\
  \citenamefont {Gonzalez-Zalba}}]{fc_s9}%
  \BibitemOpen
  \bibfield  {author} {\bibinfo {author} {\bibfnamefont {O.}~\bibnamefont
  {Crawford}}, \bibinfo {author} {\bibfnamefont {J.~R.}\ \bibnamefont
  {Cruise}}, \bibinfo {author} {\bibfnamefont {N.}~\bibnamefont {Mertig}}, \
  and\ \bibinfo {author} {\bibfnamefont {M.~F.}\ \bibnamefont
  {Gonzalez-Zalba}},\ }\bibfield  {title} {\enquote {\bibinfo {title}
  {Compilation and scaling strategies for a silicon quantum processor with
  sparse two-dimensional connectivity},}\ }\href {\doibase
  10.1038/s41534-023-00679-8} {\bibfield  {journal} {\bibinfo  {journal} {npj
  Quantum Information}\ }\textbf {\bibinfo {volume} {9}},\ \bibinfo {pages}
  {1--11} (\bibinfo {year} {2023})},\ \bibinfo {note} {number: 1 Publisher:
  Nature Publishing Group}\BibitemShut {NoStop}%
\bibitem [{\citenamefont {Jnane}\ \emph {et~al.}(2022)\citenamefont {Jnane},
  \citenamefont {Undseth}, \citenamefont {Cai}, \citenamefont {Benjamin},\ and\
  \citenamefont {Koczor}}]{fc_s10}%
  \BibitemOpen
  \bibfield  {author} {\bibinfo {author} {\bibfnamefont {H.}~\bibnamefont
  {Jnane}}, \bibinfo {author} {\bibfnamefont {B.}~\bibnamefont {Undseth}},
  \bibinfo {author} {\bibfnamefont {Z.}~\bibnamefont {Cai}}, \bibinfo {author}
  {\bibfnamefont {S.~C.}\ \bibnamefont {Benjamin}}, \ and\ \bibinfo {author}
  {\bibfnamefont {B.}~\bibnamefont {Koczor}},\ }\bibfield  {title} {\enquote
  {\bibinfo {title} {Multicore {Quantum} {Computing}},}\ }\href {\doibase
  10.1103/PhysRevApplied.18.044064} {\bibfield  {journal} {\bibinfo  {journal}
  {Physical Review Applied}\ }\textbf {\bibinfo {volume} {18}},\ \bibinfo
  {pages} {044064} (\bibinfo {year} {2022})},\ \bibinfo {note} {publisher:
  American Physical Society}\BibitemShut {NoStop}%
\bibitem [{\citenamefont {Li}\ and\ \citenamefont {Chen}(2020)}]{fc_s11}%
  \BibitemOpen
  \bibfield  {author} {\bibinfo {author} {\bibfnamefont {M.-C.}\ \bibnamefont
  {Li}}\ and\ \bibinfo {author} {\bibfnamefont {A.-X.}\ \bibnamefont {Chen}},\
  }\bibfield  {title} {\enquote {\bibinfo {title} {Elementary quantum gates
  between long-distance qubits mediated by a resonator},}\ }\href {\doibase
  10.1007/s11128-020-02858-4} {\bibfield  {journal} {\bibinfo  {journal}
  {Quantum Information Processing}\ }\textbf {\bibinfo {volume} {19}},\
  \bibinfo {pages} {365} (\bibinfo {year} {2020})}\BibitemShut {NoStop}%
\bibitem [{\citenamefont {Zhang}\ \emph {et~al.}(2024)\citenamefont {Zhang},
  \citenamefont {Yin}, \citenamefont {Wu}, \citenamefont {Shapourian},
  \citenamefont {Shabani},\ and\ \citenamefont {Ding}}]{chiplets}%
  \BibitemOpen
  \bibfield  {author} {\bibinfo {author} {\bibfnamefont {H.}~\bibnamefont
  {Zhang}}, \bibinfo {author} {\bibfnamefont {K.}~\bibnamefont {Yin}}, \bibinfo
  {author} {\bibfnamefont {A.}~\bibnamefont {Wu}}, \bibinfo {author}
  {\bibfnamefont {H.}~\bibnamefont {Shapourian}}, \bibinfo {author}
  {\bibfnamefont {A.}~\bibnamefont {Shabani}}, \ and\ \bibinfo {author}
  {\bibfnamefont {Y.}~\bibnamefont {Ding}},\ }\bibfield  {title} {\enquote
  {\bibinfo {title} {{MECH}: {Multi}-{Entry} {Communication} {Highway} for
  {Superconducting} {Quantum} {Chiplets}},}\ }\href {\doibase
  10.48550/arXiv.2305.05149} {\  (\bibinfo {year} {2024}),\
  10.48550/arXiv.2305.05149}\BibitemShut {NoStop}%
\bibitem [{\citenamefont {Pauka}\ \emph {et~al.}(2021)\citenamefont {Pauka},
  \citenamefont {Das}, \citenamefont {Kalra}, \citenamefont {Moini},
  \citenamefont {Yang}, \citenamefont {Trainer}, \citenamefont {Bousquet},
  \citenamefont {Cantaloube}, \citenamefont {Dick}, \citenamefont {Gardner},
  \citenamefont {Manfra},\ and\ \citenamefont {Reilly}}]{all_stack1}%
  \BibitemOpen
  \bibfield  {author} {\bibinfo {author} {\bibfnamefont {S.~J.}\ \bibnamefont
  {Pauka}}, \bibinfo {author} {\bibfnamefont {K.}~\bibnamefont {Das}}, \bibinfo
  {author} {\bibfnamefont {R.}~\bibnamefont {Kalra}}, \bibinfo {author}
  {\bibfnamefont {A.}~\bibnamefont {Moini}}, \bibinfo {author} {\bibfnamefont
  {Y.}~\bibnamefont {Yang}}, \bibinfo {author} {\bibfnamefont {M.}~\bibnamefont
  {Trainer}}, \bibinfo {author} {\bibfnamefont {A.}~\bibnamefont {Bousquet}},
  \bibinfo {author} {\bibfnamefont {C.}~\bibnamefont {Cantaloube}}, \bibinfo
  {author} {\bibfnamefont {N.}~\bibnamefont {Dick}}, \bibinfo {author}
  {\bibfnamefont {G.~C.}\ \bibnamefont {Gardner}}, \bibinfo {author}
  {\bibfnamefont {M.~J.}\ \bibnamefont {Manfra}}, \ and\ \bibinfo {author}
  {\bibfnamefont {D.~J.}\ \bibnamefont {Reilly}},\ }\bibfield  {title}
  {\enquote {\bibinfo {title} {A cryogenic {CMOS} chip for generating control
  signals for multiple qubits},}\ }\href {\doibase 10.1038/s41928-020-00528-y}
  {\bibfield  {journal} {\bibinfo  {journal} {Nature Electronics}\ }\textbf
  {\bibinfo {volume} {4}},\ \bibinfo {pages} {64--70} (\bibinfo {year}
  {2021})},\ \bibinfo {note} {number: 1 Publisher: Nature Publishing
  Group}\BibitemShut {NoStop}%
\bibitem [{\citenamefont {Gonzalez-Zalba}\ \emph {et~al.}(2021)\citenamefont
  {Gonzalez-Zalba}, \citenamefont {de~Franceschi}, \citenamefont {Charbon},
  \citenamefont {Meunier}, \citenamefont {Vinet},\ and\ \citenamefont
  {Dzurak}}]{all_stack2}%
  \BibitemOpen
  \bibfield  {author} {\bibinfo {author} {\bibfnamefont {M.~F.}\ \bibnamefont
  {Gonzalez-Zalba}}, \bibinfo {author} {\bibfnamefont {S.}~\bibnamefont
  {de~Franceschi}}, \bibinfo {author} {\bibfnamefont {E.}~\bibnamefont
  {Charbon}}, \bibinfo {author} {\bibfnamefont {T.}~\bibnamefont {Meunier}},
  \bibinfo {author} {\bibfnamefont {M.}~\bibnamefont {Vinet}}, \ and\ \bibinfo
  {author} {\bibfnamefont {A.~S.}\ \bibnamefont {Dzurak}},\ }\bibfield  {title}
  {\enquote {\bibinfo {title} {Scaling silicon-based quantum computing using
  {CMOS} technology},}\ }\href {\doibase 10.1038/s41928-021-00681-y} {\bibfield
   {journal} {\bibinfo  {journal} {Nature Electronics}\ }\textbf {\bibinfo
  {volume} {4}},\ \bibinfo {pages} {872--884} (\bibinfo {year} {2021})},\
  \bibinfo {note} {number: 12 Publisher: Nature Publishing Group}\BibitemShut
  {NoStop}%
\bibitem [{\citenamefont {Michniewicz}(2022)}]{nawa_myphd}%
  \BibitemOpen
  \bibfield  {author} {\bibinfo {author} {\bibfnamefont {J.}~\bibnamefont
  {Michniewicz}},\ }\bibfield  {title} {\enquote {\bibinfo {title}
  {Expansibility {Evaluation} of a {Two}-dimensional {Access} {Array} for
  {Quantum} {Computing}},}\ }\href
  {https://www.repository.cam.ac.uk/handle/1810/336062} {\  (\bibinfo {year}
  {2022})}\BibitemShut {NoStop}%
\bibitem [{\citenamefont {Stefanazzi}\ \emph {et~al.}(2022)\citenamefont
  {Stefanazzi}, \citenamefont {Treptow}, \citenamefont {Wilcer}, \citenamefont
  {Stoughton}, \citenamefont {Bradford}, \citenamefont {Uemura}, \citenamefont
  {Zorzetti}, \citenamefont {Montella}, \citenamefont {Cancelo}, \citenamefont
  {Sussman}, \citenamefont {Houck}, \citenamefont {Saxena}, \citenamefont
  {Arnaldi}, \citenamefont {Agrawal}, \citenamefont {Zhang}, \citenamefont
  {Ding},\ and\ \citenamefont {Schuster}}]{nawa_fpga}%
  \BibitemOpen
  \bibfield  {author} {\bibinfo {author} {\bibfnamefont {L.}~\bibnamefont
  {Stefanazzi}}, \bibinfo {author} {\bibfnamefont {K.}~\bibnamefont {Treptow}},
  \bibinfo {author} {\bibfnamefont {N.}~\bibnamefont {Wilcer}}, \bibinfo
  {author} {\bibfnamefont {C.}~\bibnamefont {Stoughton}}, \bibinfo {author}
  {\bibfnamefont {C.}~\bibnamefont {Bradford}}, \bibinfo {author}
  {\bibfnamefont {S.}~\bibnamefont {Uemura}}, \bibinfo {author} {\bibfnamefont
  {S.}~\bibnamefont {Zorzetti}}, \bibinfo {author} {\bibfnamefont
  {S.}~\bibnamefont {Montella}}, \bibinfo {author} {\bibfnamefont
  {G.}~\bibnamefont {Cancelo}}, \bibinfo {author} {\bibfnamefont
  {S.}~\bibnamefont {Sussman}}, \bibinfo {author} {\bibfnamefont
  {A.}~\bibnamefont {Houck}}, \bibinfo {author} {\bibfnamefont
  {S.}~\bibnamefont {Saxena}}, \bibinfo {author} {\bibfnamefont
  {H.}~\bibnamefont {Arnaldi}}, \bibinfo {author} {\bibfnamefont
  {A.}~\bibnamefont {Agrawal}}, \bibinfo {author} {\bibfnamefont
  {H.}~\bibnamefont {Zhang}}, \bibinfo {author} {\bibfnamefont
  {C.}~\bibnamefont {Ding}}, \ and\ \bibinfo {author} {\bibfnamefont {D.~I.}\
  \bibnamefont {Schuster}},\ }\bibfield  {title} {\enquote {\bibinfo {title}
  {The {QICK} ({Quantum} {Instrumentation} {Control} {Kit}): {Readout} and
  control for qubits and detectors},}\ }\href {\doibase 10.1063/5.0076249}
  {\bibfield  {journal} {\bibinfo  {journal} {Review of Scientific
  Instruments}\ }\textbf {\bibinfo {volume} {93}},\ \bibinfo {pages} {044709}
  (\bibinfo {year} {2022})}\BibitemShut {NoStop}%
\bibitem [{\citenamefont {Bandic}, \citenamefont {Feld},\ and\ \citenamefont
  {Almudever}(2022)}]{nawa_fullstack}%
  \BibitemOpen
  \bibfield  {author} {\bibinfo {author} {\bibfnamefont {M.}~\bibnamefont
  {Bandic}}, \bibinfo {author} {\bibfnamefont {S.}~\bibnamefont {Feld}}, \ and\
  \bibinfo {author} {\bibfnamefont {C.~G.}\ \bibnamefont {Almudever}},\
  }\bibfield  {title} {\enquote {\bibinfo {title} {Full-stack quantum computing
  systems in the {NISQ} era: algorithm-driven and hardware-aware compilation
  techniques},}\ }in\ \href {\doibase 10.23919/DATE54114.2022.9774643} {\emph
  {\bibinfo {booktitle} {2022 {Design}, {Automation} \& {Test} in {Europe}
  {Conference} \& {Exhibition} ({DATE})}}}\ (\bibinfo  {publisher} {IEEE},\
  \bibinfo {address} {Antwerp, Belgium},\ \bibinfo {year} {2022})\ pp.\
  \bibinfo {pages} {1--6}\BibitemShut {NoStop}%
\bibitem [{\citenamefont {Moon}\ \emph {et~al.}(2020)\citenamefont {Moon},
  \citenamefont {Lennon}, \citenamefont {Kirkpatrick}, \citenamefont {van
  Esbroeck}, \citenamefont {Camenzind}, \citenamefont {Yu}, \citenamefont
  {Vigneau}, \citenamefont {Zumbühl}, \citenamefont {Briggs}, \citenamefont
  {Osborne}, \citenamefont {Sejdinovic}, \citenamefont {Laird},\ and\
  \citenamefont {Ares}}]{nawa_ares_ml}%
  \BibitemOpen
  \bibfield  {author} {\bibinfo {author} {\bibfnamefont {H.}~\bibnamefont
  {Moon}}, \bibinfo {author} {\bibfnamefont {D.~T.}\ \bibnamefont {Lennon}},
  \bibinfo {author} {\bibfnamefont {J.}~\bibnamefont {Kirkpatrick}}, \bibinfo
  {author} {\bibfnamefont {N.~M.}\ \bibnamefont {van Esbroeck}}, \bibinfo
  {author} {\bibfnamefont {L.~C.}\ \bibnamefont {Camenzind}}, \bibinfo {author}
  {\bibfnamefont {L.}~\bibnamefont {Yu}}, \bibinfo {author} {\bibfnamefont
  {F.}~\bibnamefont {Vigneau}}, \bibinfo {author} {\bibfnamefont {D.~M.}\
  \bibnamefont {Zumbühl}}, \bibinfo {author} {\bibfnamefont {G.~a.~D.}\
  \bibnamefont {Briggs}}, \bibinfo {author} {\bibfnamefont {M.~A.}\
  \bibnamefont {Osborne}}, \bibinfo {author} {\bibfnamefont {D.}~\bibnamefont
  {Sejdinovic}}, \bibinfo {author} {\bibfnamefont {E.~A.}\ \bibnamefont
  {Laird}}, \ and\ \bibinfo {author} {\bibfnamefont {N.}~\bibnamefont {Ares}},\
  }\bibfield  {title} {\enquote {\bibinfo {title} {Machine learning enables
  completely automatic tuning of a quantum device faster than human experts},}\
  }\href {\doibase 10.1038/s41467-020-17835-9} {\bibfield  {journal} {\bibinfo
  {journal} {Nature Communications}\ }\textbf {\bibinfo {volume} {11}},\
  \bibinfo {pages} {4161} (\bibinfo {year} {2020})}\BibitemShut {NoStop}%
\bibitem [{\citenamefont {Zwolak}\ and\ \citenamefont {Taylor}(2023)}]{zwolak}%
  \BibitemOpen
  \bibfield  {author} {\bibinfo {author} {\bibfnamefont {J.~P.}\ \bibnamefont
  {Zwolak}}\ and\ \bibinfo {author} {\bibfnamefont {J.~M.}\ \bibnamefont
  {Taylor}},\ }\bibfield  {title} {\enquote {\bibinfo {title} {Colloquium:
  {Advances} in automation of quantum dot devices control},}\ }\href {\doibase
  10.1103/RevModPhys.95.011006} {\bibfield  {journal} {\bibinfo  {journal}
  {Reviews of Modern Physics}\ }\textbf {\bibinfo {volume} {95}},\ \bibinfo
  {pages} {011006} (\bibinfo {year} {2023})},\ \bibinfo {note} {publisher:
  American Physical Society}\BibitemShut {NoStop}%
\bibitem [{\citenamefont {Stano}\ and\ \citenamefont
  {Loss}(2022)}]{spinqubit_database}%
  \BibitemOpen
  \bibfield  {author} {\bibinfo {author} {\bibfnamefont {P.}~\bibnamefont
  {Stano}}\ and\ \bibinfo {author} {\bibfnamefont {D.}~\bibnamefont {Loss}},\
  }\bibfield  {title} {\enquote {\bibinfo {title} {Review of performance
  metrics of spin qubits in gated semiconducting nanostructures},}\ }\href
  {\doibase 10.1038/s42254-022-00484-w} {\bibfield  {journal} {\bibinfo
  {journal} {Nature Reviews Physics}\ }\textbf {\bibinfo {volume} {4}},\
  \bibinfo {pages} {672--688} (\bibinfo {year} {2022})}\BibitemShut {NoStop}%
\end{thebibliography}%

\end{document}